\pdfoutput=1
\documentclass[12pt,preprint]{aastex}
\usepackage[margin=0.75in]{geometry}
\usepackage{graphicx}
\usepackage{natbib}
\usepackage{aas_macros}
\usepackage[section] {placeins}
\usepackage{subfigure}
\usepackage{float}
\usepackage{color}

\def\msun{{\rm\,M_\odot}}

\def\msun{{\rm\,M_\odot}} 

\def\zsun{{\rm\,Z_\odot}}

\newcommand{\kms}{\, {\rm km\, s}^{-1}}

\newcommand{\mpc}{\, {\rm Mpc}}
\newcommand{\kpc}{\, {\rm kpc}}
\newcommand{\hmpc}{\, h^{-1} \mpc}

\newcommand{\hi}{\mbox{H\,{\scriptsize I}\ }}
\newcommand{\heii}{\mbox{He\,{\scriptsize II}\ }}

\def\h2{${\rm\,H_2}$}

\makeatletter 

\def\hi{\noindent \hangindent=2.5em}

\def\kpc{{\rm\,kpc}}

\def\mpc{{\rm\,Mpc}}

\def\kms{{\rm\,km/s}}
\def\msun{{\rm\,M_\odot}}

\def\vol#1  {{{#1}{\rm,}\ }}

\def\hi{{H\thinspace I}\ }

\def\heii{{He\thinspace II}\ }

\newcount\refno
\newcount\rfno
\def\eq{$^{\the\refno\ }$\advance\refno by 1}
\def\ad{\advance\rfno by 1}

\def\clock{\count0=\time \divide\count0 by 60
     \count1=\count0 \multiply\count1 by -60 \advance\count1 by \time
     \number\count0:\ifnum\count1<10{0\number\count1}\else\number\count1\fi}

\def\myputfigure#1#2#3#4#5%
{\hskip0.03\textwidth\vskip#5pt
\makebox[0pt]{\hskip#2in
\includegraphics[width=#3\textwidth]{#1}}\vskip#4pt\hfill}

\definecolor{burntorange}{rgb}{1,0.4,0.2}

\newcount\refno
\newcount\rfno
\def\eq{$^{\the\refno\ }$\advance\refno by 1}
\def\ad{\advance\rfno by 1}

\begin{document}

\title{On the Origin of the Hubble Sequence: I. Insights on Galaxy Color Migration from Cosmological Simulations}
 
\author{
Renyue Cen$^{1}$
} 

\footnotetext[1]{Princeton University Observatory, Princeton, NJ 08544;
 cen@astro.princeton.edu}

\begin{abstract} 

An analysis of more than $3000$ galaxies resolved at better than $114~$pc/h at $z=0.62$ 
in a ``{\color{red}\bf LAOZI}" cosmological adaptive mesh refinement hydrodynamic 
simulation is performed and insights gained on star formation quenching and color migration.
The vast majority of red galaxies are found to be within 
three virial radii of a larger galaxy, at the onset of quenching when the specific star formation rate experiences
the sharpest decline to fall below $\sim 10^{-2}-10^{-1}$Gyr$^{-1}$ (depending on the redshift).
We shall thus call this mechanism ``environment quenching", which encompasses satellite quenching. 
Two physical processes are largely responsible: 
ram-pressure stripping first disconnects the galaxy from the cold gas supply on large scales,
followed by a longer period of cold gas starvation taking place in high velocity dispersion environment,
during the early part of which the existing dense cold gas in the central region ($\le 10$kpc) 
is consumed by {\it in situ} star formation.  
Quenching is found to be {\it more efficient} (i.e., a larger fraction of galaxies being quenched), {\it but not faster} (i.e., duration being weakly dependent on environment), 
on average, in denser environment.
Throughout this quenching period and the ensuing one in the red sequence
galaxies follow nearly vertical tracks in the color-stellar-mass diagram.
In contrast, individual galaxies of all masses grow most of their stellar masses in the blue cloud, prior to the onset of quenching,  
and progressively more massive blue galaxies 
{\it 
with already relatively older mean stellar ages} continue to enter the red sequence.
Consequently, correlations among observables of red galaxies - such as the age-mass relation - 
are largely inherited from their blue progenitors at the onset of quenching.
While the color makeup of the entire galaxy population strongly depends on environment,
which is a direct result of environment quenching,
physical properties of blue galaxies as a sub-population show little dependence on environment.
A variety of predictions from the simulation are shown to be in accord with extant observations.

\end{abstract}
 

\section{Introduction}

The bimodal distribution of galaxy colors at low redshift is well established
\citep[e.g.,][]{2001Strateva, 2003bBlanton, 2003Kauffmann, 2004Baldry}.
The ``blue cloud", sometimes referred to as the ``star formation sequence" \citep[][]{2007Salim},
is occupied by star-forming galaxies, while the ``red sequence" galaxies appear to have little ongoing star formation (SF).
It has been argued that this bimodality suggests that SF 
of the blue cloud galaxies en route to the red sequence must be turned off promptly 
to prevent them from lingering in the green valley between the blue and red peaks.
A number of physical mechanisms have been proposed to cause this apparent ``quenching" of SF.
Galaxy mergers have been suggested to trigger strong and rapid SF that subsequently
drives gas away and shuts down further SF activities in a sudden fashion. 
However,
recent observations show that galaxies in the green valley do not 
show merger signatures, perhaps disfavoring the merger scenario \citep[e.g.,][]{2011Mendez}. 
Feedback from active galactic nuclei (AGN) has also been suggested to provide quenching, but
observational evidence for this scenario has been 
either inconclusive or at best circumstantial \citep[e.g.,][]{2008Bundy, 2012Santini, 2012Bongiorno, 2013Rosario}.
Some recent studies based on large data sets do not find evidence
for AGN feedback playing a role in galaxy color migration \citep[e.g.,][]{2007Zheng, 2010Xue, 2012Aird, 2012Harrison, 2012Swinbank, 2013Mendel}.

External, environmental effects may have played an important role in shaping galaxy colors.
High density environments are observed to be occupied primarily by early type (elliptical and S0) red galaxies
- the ``density-morphology relation" \citep[e.g.,][]{1974Oemler, 1980Dressler,1984Postman} -  
with giant elliptical galaxies anchoring the centers of rich clusters of galaxies \citep[e.g.,][]{2009Kormendy}. 
This relation is consistent with the larger trend of the galaxy population appearing bluer in more 
underdense regions in the local universe \citep[e.g.,][]{2003Goto,2003Gomez,2004Tanaka,2004Rojas}.
Different types of galaxies are seen to cluster differently and have different
environment-dependencies, in the same sense as the density-morphology relation
\citep[e.g.,][]{1976Davis, 2003Hogg, 2004Balogh, 2004Kauffmann, 2007Park, 2008Coil, 2011Zehavi}. 

Recent quantitative studies have yielded richer details on SF dependence on halo mass
and environment, probing their relationships at higher redshifts. 
For example, using a large group catalog from the Sloan Digital Sky Survey (SDSS) Data Release 2, 
\citet[][]{2006Weinmann} find that 
at fixed luminosity the fraction of early-type galaxies increases with increasing halo mass
and this mass dependence is smooth and persists over the entire mass range probed without any break or feature at any mass-scale.
From a spectral analysis of galaxies at $z=0.4-0.8$ based on the ESO Distant Cluster Survey,
\citet[][]{2009Poggianti} find that the incidence of K+A galaxies increases strongly with increasing velocity dispersion 
of the environment from groups to clusters.
\citet[][]{2011McGee}, examining the SF properties of group and field galaxies from SDSS at $z\sim 0.08$
and from ultraviolet imaging with GALEX at $z\sim 0.4$, 
find that the fraction of passive galaxies is higher in groups than the field at both redshifts, 
with the difference between the group and field growing with time and  larger at low masses. 
With the NOAO Extremely Wide-Field Infrared Imager (NEWFIRM) Survey 
of the All-wavelength Extended Groth strip International Survey (AEGIS) and Cosmic Evolution Survey (COSMOS) fields,
\citet[][]{2011Whitaker}
show evidence for a bimodal color distribution between quiescent and star-forming galaxies that persists to 
$z\sim 3$.
\citet[][]{2012Presotto} study the evolution of galaxies located within groups 
using the group catalog obtained from zCOSMOS spectroscopic data and the complementary photometric data from the COSMOS survey at $z=0.2-0.8$
and find the rate of SF quenching to be faster in groups than in the field. 
\citet[][]{2012Muzzin} analyze galaxy properties at $z=0.85-1.20$
using a spectroscopic sample of 797 cluster and field galaxies drawn from the Gemini Cluster Astrophysics Spectroscopic Survey,
finding that post starburst galaxies with $M^*=10^{9.3-10.7}\msun$ are three times more common in high-density regions compared to low-density regions. 
Based on data from the zCOSMOS survey \citet[][]{2012Tanaka} perform
an environment study  and find that quiescent galaxies prefer more massive systems at $z=0.5-1$.
\citet[][]{2012Rasmussen}, analyzing GALEX imaging of a statistically representative sample of 23 galaxy groups at $z \sim 0.06$,
suggest an average quenching timescale of $\ge 2$Gyr.
\citet[][]{2013Mok}, with deep GMOS-S spectroscopy for 11 galaxy groups at $z=0.8-1$,
show that the strongest environmental dependence is observed in the fraction of passive galaxies, 
which make up only $\sim 20$ per cent of the field in the mass range $M_{\rm star}=10^{10.3}-10^{11.0}\msun$ 
but are the dominant component of groups. 
Using SDSS ($z\sim 0.1$) and the All-Wavelength Extended Groth Strip International Survey (AEGIS; $z\sim 1$) data,
\citet[][]{2013Woo} 
find a strong environmental dependence of quenching in terms of halo mass and 
distance to the centrals 
at both redshifts.

The widespread observational evidence of environment quenching is unsurprising theoretically. 
In regions of overdensity, whether around a large collapsed halo or unvirialized 
structure (e.g., a Zel'dovich pancake or a filament), gas is gravitationally shock heated when converging flows meet.
In regions filled with hot shock-heated gas, 
multiple gasdynamical processes would occur. 
One of the most important gasdynamical processes is ram-pressure stripping of gas,
when a galaxy moves through the ambient hot gas at a significant speed, which includes, but is not limited to,
the infall velocity of a satellite galaxy.
The theoretical basis for the ram-pressure stripping process is laid down in the seminal work of \citet[][]{1972Gunn}. 
Recent works with detailed simulations of this effect on galaxies (in non-cosmological settings) 
include those of 
\citet[][]{2000Mori},
\citet[][]{2000Quilis},
\citet[][]{2008Kronberger}, \citet[][]{2009Bekki} and \citet[][]{2009Tonnesen}.

Even in the absence of ram-pressure stripping, ubiquitous supersonic and transonic motions of galaxies of complex acceleration patterns
through ambient medium (intergalactic or circumgalactic medium) subject them to the Raleigh-Taylor and Richtmyer-Meshkov instabilities.
Large shear velocities at the interfaces between galaxies and the ambient medium
allow the Kelvin-Helmholtz (KH) instability to play an important role.
When these processes work in tandem with ram-pressure displacements,
the disruptive effects are amplified.
For example, the KH instability time scale is substantially shorter 
for a non-self-gravitating gas cloud \citep[e.g.,][]{1993Murray} than for one sitting inside a virialized
dark matter halo \citep[e.g.,][]{2008Cen}.

Another important process in hot environments is starvation of cold gas that is fuel for SF \citep[e.g.,][]{1980Larson,2000Balogh, 2006Dekel}.
In regions with high-temperature and high-entropy 
cooling of hot gas is an inefficient process for fueling SF, 
an important point noted long ago to account for the basic 
properties (mass, size) of galaxies \citep[e.g.,][]{1977Binney, 1977Rees, 1977Silk}.
This phenomenon may be understood by considering  the dependence of cooling time on the entropy of the gas:
the gas cooling time can be written as
$t_{\rm cool}(T,S) = S^{3/2} \left[{3 \over 2}\left({\mu_e\over \mu}\right)^2 {k_B\over T^{1/2}\Lambda(T)}\right]$ \citep[][]{2004Scannapieco}
\footnote[1]{
where 
$k_B$ is Boltzmann's constant, $T$ temperature and $\Lambda$ cooling function,
$\mu=0.62$ and $\mu_e=1.18$ for ionized gas that we are concerned with,
$S$ is the gas entropy defined as $S \equiv {T\over n^{2/3}}$ in units of K~cm$^2$ ($n$ is gas number density).
If one conservatively adopts the lowest value of the term inside the bracket at the cooling peak at temperature $T_{\rm min}\sim 10^{5.3}$K,
it follows that the minimum cooling time of a gas parcel just scales with $S^{3/2}$.}.
It follows that the minimum cooling time of a gas parcel just scales with $S^{3/2}$.
As a numerical example, for a gas parcel of entropy $S=10^9$K~cm$^{2}$ (say, for temperature $10^7$K and  density $10^{-3}$cm$^{-3}$) 
and metallicity $0.1\zsun$,
its cooling time is no shorter than the Hubble time at $z=1$ hence 
the gas can no longer cool efficiently to fuel SF.

It may be that the combination of cold gas removal and dispersal by ram-pressure stripping, hydrodynamic instabilities, 
and cold gas starvation, all of which are expected to become increasingly important in more massive environments,
plays a primary role in driving the color migration from the blue cloud to the red sequence.
In dense environments, gravitational tidal (stripping and shock) effects 
and relatively close fly-bys between galaxies \citep[e.g.,][]{1996Moore} also become important.
To understand the overall effect on SF quenching by these external processes in the context of the 
standard cold dark matter model, a realistic cosmological setting is imperative,
in order to capture complex external processes 
that are likely intertwined with large variations of internal properties of galaxies. 
In this paper we perform {\it ab initio}
{\color{red}\bf L}arge-scale {\color{red}\bf A}daptive-mesh-refinement {\color{red}\bf O}mniscient {\color{red}\bf Z}oom-{\color{red}\bf I}n cosmological hydrodynamic simulations,
called {\color{red}\bf LAOZI Simulations}, 
to obtain a large sample of galaxies to, for the first time, perform a chronological and statistical investigation 
on a very large scale.
The large simulated galaxy sample size and very high resolution of LAOZI simulations
provide an unprecedented opportunity to undertake the study presented.
Our study shares the spirit of the work by \citet[][]{2011Feldmann}, who 
examine the evolution of a dozen galaxies falling onto a forming group of galaxies,
with a substantial improvement in the statistical treatment, the simulation resolution, the range of environment probed, 
and the analysis scope.
Feedback from AGN is not included in this simulation, partly because of its large uncertainties and present lack of definitive driving sources
and primarily due to our intention to focus on external effects.
Internal effects due to SF are automatically included and 
we find no evidence that SF or merger triggered SF plays a primary role in quenching from our study.
The outline of this paper is as follows.
In \S 2 we detail our simulations (\S 2.1), method of making galaxy catalogs (\S 2.2),
construction of histories of galaxies (\S 2.3), tests and validation of simulations (\S 2.4)
and the produced bimodal distribution of galaxies colors (\S 2.5). 
Results are presented in \S 3, organized in an approximately chronological order,
starting with the ram-pressure stripping effects in \S 3.1,
followed by the ensuing period of gas starvation in hot environment in \S 3.2.
In \S 3.3 we discuss stellar mass growth,
evolution of stellar mass function of red galaxies
and present galaxy color migration tracks.
\S 3.4 gives an example of consequences of the found color migration picture - galaxy age-mass relation.
We present observable environmental dependence of galaxy makeup at $z=0.62$ in \S 3.5.
Conclusions are given in \S 4.

\section{Simulations}\label{sec: sims}

\subsection{Hydrocode and Simulation Parameters}

We perform cosmological simulations with the AMR 
Eulerian hydro code, Enzo 
\citep[][]{1999bBryan, 2009Joung}.  
First we run a low resolution simulation with a periodic box of $120~h^{-1}$Mpc 
(comoving) on a side.
We identify a region centered on a cluster of mass of $\sim 3\times 10^{14}\msun$ at $z=0$.
We then resimulate with high resolution of the chosen region embedded
in the outer $120h^{-1}$Mpc box to properly take into account the large-scale tidal field
and appropriate boundary conditions at the surface of a refined region.
The refined region 
has a comoving size of $21\times 24\times 20h^{-3}$Mpc$^3$ 
and represents a $+1.8\sigma$ matter density fluctuation on that volume.
The dark matter particle mass in the refined region is $1.3\times 10^7h^{-1}\msun$.
The refined region is surrounded by three layers (each of $\sim 1h^{-1}$Mpc) of buffer zones with 
particle masses successively larger by a factor of $8$ for each layer, 
which then connects with
the outer root grid that has a dark matter particle mass $8^4$ times that in the refined region.
We choose the mesh refinement criterion such that the resolution is 
always smaller than $111h^{-1}$pc (physical), corresponding to a maximum mesh refinement level of $13$ at $z=0$.
An identical comparison run that has four times better resolution of $29$pc/h
was also run down to $z=3$ and some relevant comparisons between the two simulations are made
to understand effects of limited resolution on our results.
The simulations include
a metagalactic UV background
\citep[][]{1996Haardt},  
and a model for self-shielding of UV radiation \citep[][]{2005Cen}.
They include metallicity-dependent radiative cooling \citep[][]{1995Cen}.
Our simulations also solve relevant gas chemistry
chains for molecular hydrogen formation \citep[][]{1997Abel},
molecular formation on dust grains \citep[][]{2009Joung},
and metal cooling extended down to $10~$K \citep[][]{1972Dalgarno}.
Star particles are created in cells that satisfy a set of criteria for 
SF proposed by \citet[][]{1992CenOstriker}.
Each star particle is tagged with its initial mass, creation time, and metallicity; 
star particles typically have masses of $\sim$$10^6\msun$.

Supernova feedback from SF is modeled following \citet[][]{2005Cen}.
Feedback energy and ejected metal-enriched mass are distributed into 
27 local gas cells centered at the star particle in question, 
weighted by the specific volume of each cell, which is to mimic the physical process of supernova
blastwave propagation that tends to channel energy, momentum and mass into the least dense regions
(with the least resistance and cooling).
The primary advantages of this supernova energy based feedback mechanism are three-fold.
First, nature does drive winds in this way and energy input is realistic.
Second, it has only one free parameter $e_{SN}$, namely, the fraction of the rest mass energy of stars formed
that is deposited as thermal energy on the cell scale at the location of supernovae.
Third, the processes are treated physically, obeying their respective conservation laws (where they apply),
allowing transport of metals, mass, energy and momentum to be treated self-consistently
and taking into account relevant heating/cooling processes at all times.
We allow the entire feedback processes to be hydrodynamically coupled to surroundings
and subject to relevant physical processes, such as cooling and heating. 
The total amount of explosion kinetic energy from Type II supernovae
with a Chabrier initial mass function (IMF) is $6.6\times 10^{-6} M_* c^2$ (where $c$ is the speed of light),
for an amount $M_{*}$ of star formed.
Taking into account the contribution of prompt Type I supernovae,
we use $e_{SN}=1\times 10^{-5}$ in our simulations.
Observations of local starburst galaxies indicate
that nearly all of the SF produced kinetic energy 
is used to power galactic superwinds \citep[e.g.,][]{2001Heckman}. 
Supernova feedback is important primarily for regulating SF
and for transporting energy and metals into the intergalactic medium.
The extremely inhomogeneous metal enrichment process
demands that both metals and energy (and momentum) are correctly modeled so that they
are transported in a physically sound (albeit still approximate at the current resolution) way.

We use the following cosmological parameters that are consistent with 
the WMAP7-normalized \citep[][]{2010Komatsu} $\Lambda$CDM model:
$\Omega_M=0.28$, $\Omega_b=0.046$, $\Omega_{\Lambda}=0.72$, $\sigma_8=0.82$,
$H_0=100 h \,{\rm km\, s}^{-1} {\rm Mpc}^{-1} = 70 \,{\rm km\, s}^{-1} {\rm Mpc}^{-1}$ and $n=0.96$.
These parameters are consistent with those from Planck first-year data \citep[][]{2013Planck}
if we average Planck derived $H_0$ with SN Ia and HST based $H_0$.

We note that the size of the refined region, $21\times 24\times 20h^{-3}$Mpc$^3$, is still relatively small and the region biased.
This is, of course, designed on purpose.
Because of that, however, we are not able to cover all possible environment, such as the center of a void.
Also because of that, we have avoided addressing any measures that requires a precise
characterization of the abundance of any large galaxy systems, such as the
mass function or luminosity function of massive galaxies or groups.
Despite that, measures that are characterized as a function of environment/system masses 
should still be valid.
Our environment coverage is substantially larger than probed in, for example, \citet[][]{2011Feldmann}.
In \citet[][]{2012Tonnesen} we show that 
the present simulation box (C box) (run to $z=0$ with a lower resolution previously)
spans a wide range in environment from rich clusters to the field, and 
there is a substantial overlap in the field environment with another simulation centered on a void (V box).
It is the density peaks higher than we model here (i.e., more massive clusters of galaxies) that we fail to probe.
As it should be clear later, this shortcoming should not affect any of our conclusions,
which may be appropriately extrapolated.

\subsection{Simulated Galaxy Catalogs}

We identify galaxies in our high resolution simulations using the HOP algorithm 
\citep[][]{1999Eisenstein} operating on the stellar particles, which is tested to be robust
and insensitive to specific choices of concerned parameters within reasonable ranges.
Satellites within a galaxy down to mass of $\sim 10^9\msun$ are clearly identified separately in most cases.
The luminosity of each stellar particle in each of the Sloan Digital Sky Survey (SDSS) five bands 
is computed using the GISSEL stellar synthesis code \citep[][]{Bruzual03}, 
by supplying the formation time, metallicity and stellar mass.
Collecting luminosity and other quantities of member stellar particles, gas cells and dark matter 
particles yields
the following physical parameters for each galaxy:
position, velocity, total mass, stellar mass, gas mass, 
mean formation time, 
mean stellar metallicity, mean gas metallicity, SFR,
luminosities in five SDSS bands (and various colors) and others.
At a spatial resolution of $159$pc (physical) with thousands of well resolved galaxies at $z\sim 0.6-6$,
the simulated galaxy catalogs present an excellent (by far, the best available) tool to study galaxy formation and evolution.

\subsection{Construction of Histories of Simulated Galaxies}
\label{ssec:mergertrees}

When we start the analysis for this paper, the simulation has reached $z=0.62$.
For each galaxy at $z=0.62$ a genealogical line is constructed from
$z=0.62$ to $z=6$ by connecting galaxy catalogs at a series of redshifts.
Galaxy catalogs are constructed from $z=0.62$ to $z=1.40$ at a redshift increment of $\Delta z=0.02$ and 
from $z=1.40$ to $z=6$ at a redshift increment of $\Delta z=0.05$.
The parent of each galaxy is identified with the one at the next higher redshift
catalog that has the most overlap in stellar mass.

We call galaxies with {\color{blue} $g-r<0.55$ ``blue"}, 
those with {\color{green} $g-r=0.55-0.65$ ``green"} and those with {\color{red} $g-r>0.65$ ``red"},
in accord with the bimodal color distribution that we will show below
and with that of observed galaxies \citep[e.g.,][]{2003Blanton},
where $g$ and $r$ are magnitudes of SDSS $g$ and $r$ bands.

In subsequent analysis, we will examine gasdynamic processes, e.g., cold gas loss or lack of cold gas accretion,
under the working hypothesis that ram-pressure stripping and gas starvation 
are the primary detrimental processes to star formation.
We should assume that other processes, such as hydrodynamical instabilities (e.g., RT, KH, tidal shocks, etc),
may either be ``lumped together" with ram-pressure stripping or 
play some role to enhance cold gas destruction that is initiated by ram-pressure stripping. 
A word on tidal stripping may be instructive.
It is noted that while tidal stripping would affect both stars and gas, ram-pressure operates only on the latter. 
As one will see later, in some cases the stellar masses of 
galaxies decrease with time, which are likely due to tidal effects. 
A simple argument suggests that ram-pressure effects
are likely to be more far-reaching spatially and are more consistent with
the environment effects becoming effective at 2-3 virial radii than tidal stripping that we will show later.
Let us take a specific example to illustrate this.

Let us assume that the primary and infalling galaxies have
a velocity dispersion of $\sigma_1$ and $\sigma_2$, respectively, and that they both
have isothermal sphere density profiles for both dark matter and baryons.
The virial radius is proportional to its velocity dispersion in each case.
Under such a configuration, we find that the tidal radius for 
the satellite galaxy at its virial radius is equal to the virial radius
of the primary galaxy. On the other hand, the ram-pressure force on
the gas in the satellite at its virial radius is already equal to the
gravitational restoring by the satellite, when the satellite is
($\sigma_1/\sigma_2$) virial radii away from the primary galaxy.
In reality, of course, the density profiles for dark matter and baryons are different
and neither is isothermal, and the gas may display a varying degree of non-sphericity.
But the relative importance of ram-pressure and tidal strippings is likely to
remain the same for relatively diffuse gas. The relative situation is unchanged,
if one allows the gas to cool and condense. As an example, if the gas within the virial
radius of the satellite and the primary galaxies in the above example is allowed to shrink
spherically by a factor of 10 in radius (we will continue to assume that the velocity
dispersion or rotation velocity remains flat and at the same amplitude), we find
that the tidal stripping radius is now a factor of 10 smaller than before (equal to 0.1 times
the virial radius of the primary galaxy), while the new ram-pressure stripping radius is
$\sigma_1/\sigma_2$ times the new tidal stripping radius.
As a third example, if the gas within the virial radius of the satellite galaxy in the above
example is allowed to shrink spherically by a factor of 10 in radius but the gas in the
primary galaxy does not shrink in size, it can be shown that in this case the
tidal stripping radius is equal to 0.1 times the virial radius of the primary galaxy,
while the new ram-pressure stripping radius is now 0.1 times (sigma1/sigma2) times the
virial radius of the primary galaxy.
As the last example, if the gas within the virial radius of the satellite galaxy in the above
example is allowed to shrink by a factor of 10 in radius to become a disk but the gas in the
primary galaxy does not shrink in size, it can be shown that in this case the
tidal stripping radius is equal to 0.1 times the virial radius of the primary galaxy.
The new ram-pressure stripping radius depends on the orientation of the motion vector
and the normal of the disk: if the motion vector is normal to the disk, the tidal stripping
radius is 0.1 times $\sigma_1/\sigma_2$ times the virial radius of the primary galaxy; if the motion
vector is in the plane to the disk, the tidal stripping radius is zero.

\begin{figure}[ht]
\centering
\vskip -0.0cm
\resizebox{6.0in}{!}{\includegraphics[angle=0]{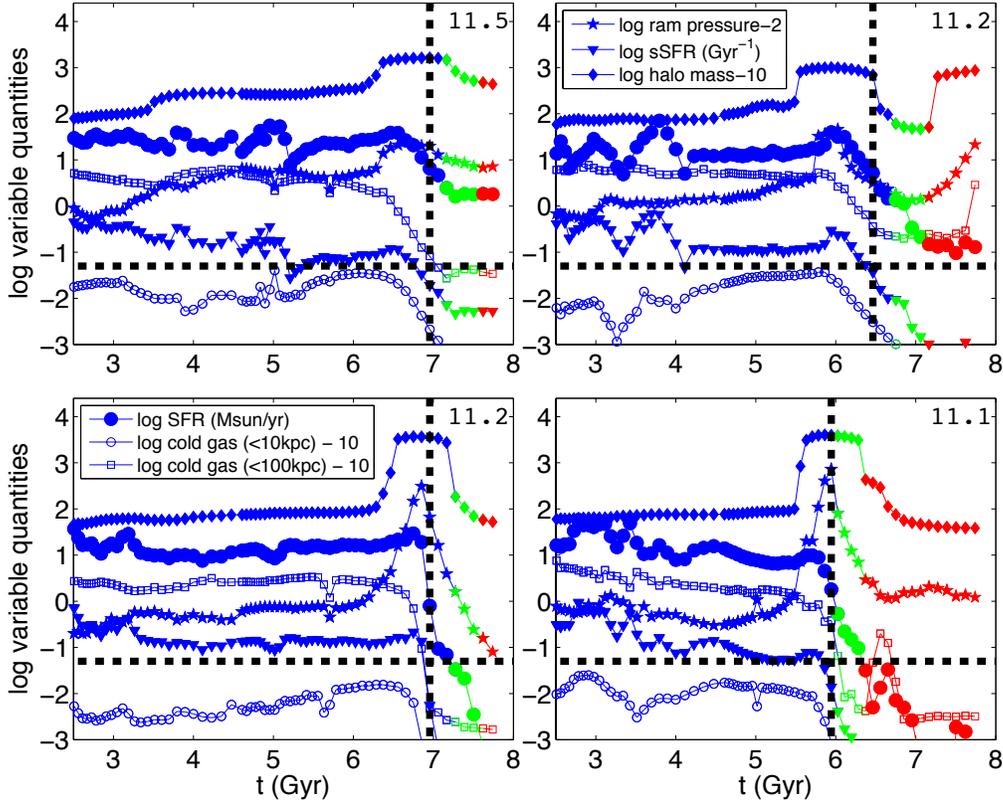}}   
\vskip -0.5cm
\caption{
four panels show the histories of six variables for four randomly selected red galaxies with stellar mass 
of $\sim 10^{11}\msun$ at $z=0.62$:
log SFR (in $\msun$/yr) (solid dots), 
log ram pressure (in Kelvin cm$^{-3}$) - 2 (stars),
log cold gas within 10kpc (in $\msun$) -10 (open circles),
log cold gas within 100kpc (in $\msun$) -10 (open squares),
log sSFR (in Gyr$^{-1}$) (down-pointing triangles)
and 
log halo mass (in $\msun$) -10 (solid diamonds);
The color of symbols at any given time corresponds the color of the galaxy at that time.
The logarithm of the stellar mass at $z=0.62$ is indicated in the upper-right corner in each panel.
The vertical dashed line in each panel indicates the location of $t_{\rm q}$.
}
\label{fig:red0}
\end{figure}

\begin{figure}[ht]
\centering
\vskip -0.0cm
\resizebox{6.0in}{!}{\includegraphics[angle=0]{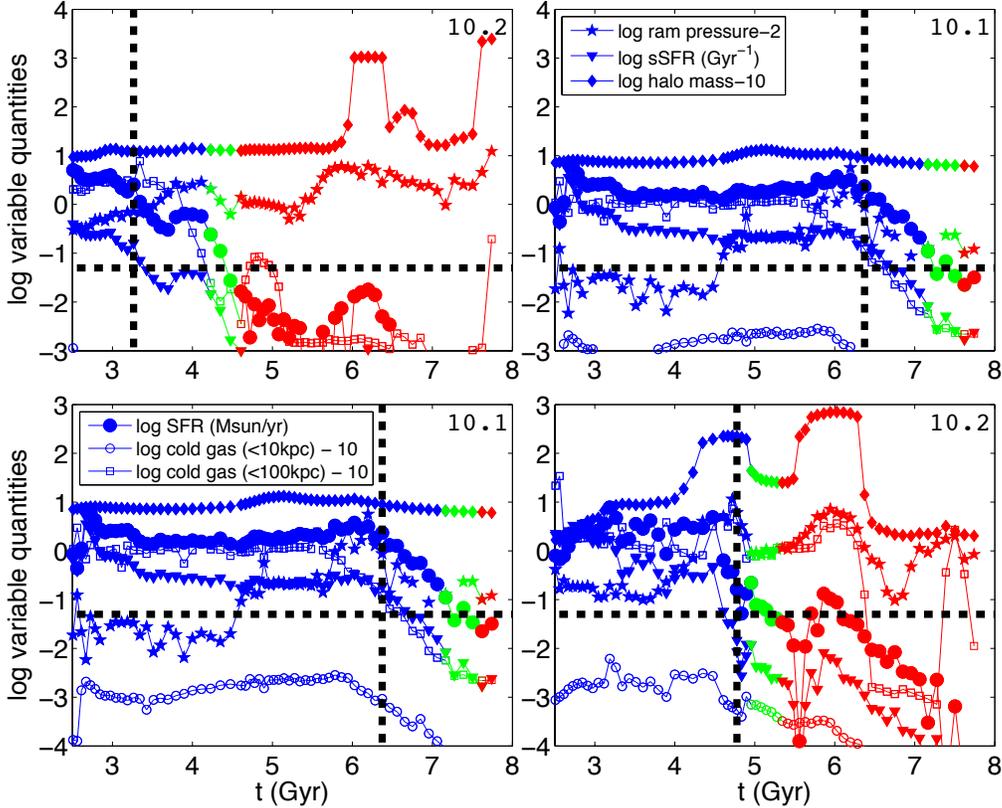}}   
\vskip -0.5cm
\caption{
four panels show the histories of six variables for four randomly selected red galaxies with stellar mass 
of $\sim 10^{10}\msun$ at $z=0.62$:
log SFR (in $\msun$/yr) (solid dots), 
log ram pressure (in Kelvin cm$^{-3}$) - 2 (stars),
log cold gas within 10kpc (in $\msun$) -10 (open circles),
log cold gas within 100kpc (in $\msun$) -10 (open squares),
log sSFR (in Gyr$^{-1}$) (down-pointing triangles)
and 
log halo mass (in $\msun$) -10 (solid diamonds);
The color of symbols at any given time corresponds the color of the galaxy at that time.
The logarithm of the stellar mass at $z=0.62$ is indicated in the upper-right corner in each panel.
The vertical dashed line in each panel indicates the location of $t_{\rm q}$.
}
\label{fig:red1}
\end{figure}

We denote a point in time when the galaxy turns from blue to green as $t_{\rm g}$,
a point in time when the galaxy turns from green to red as $t_{\rm r}$.
Convention for time is that the Big Bang occurs at $t=0$.  
We identify a point in time, searched over the range $t_g-$2Gyr to $t_g+$1Gyr, 
when the derivative of SFR with respect to time, ${\rm dSFR/dt}$, is most negative,
as $t_{\rm q}$ (${\rm q}$ stands for quenching); 
in practice, to reduce uncertainties due to temporal fluctuations in SFR,
$t_{\rm q}$ is set to equal to $t(n+1)$ when the sliding-window difference ${\rm (SFR(n+3)-SFR(n)/(t(n+3)-t(n))}$ 
is most negative, where $t(1), t(2), ..., t(n), ...$ are the times of our data outputs, as noted earlier.
Galaxies at $t_{\rm q}$ are collectively called SFQs for star formation quenching galaxies.
To demonstrate the reliability and accuracy of identification
of $t_{\rm q}$ we show in Figure~\ref{fig:red0}
the histories for a set of four randomly selected red galaxies at $z=0.62$ of stellar mass $\sim 10^{11}\msun$. 
The vertical dashed line in each panel shows
$t_{\rm q}$, which is the location of steepest drop of SFR (solid dots).
In all four cases, our method identifies the location accurately.
Figure~\ref{fig:red1} is similar to Figure~\ref{fig:red0} 
but for galaxies of stellar mass $\sim 10^{10}\msun$, 
where we see our method identifies $t_{\rm q}$ with a similar accuracy.

Similarly, we identify a point in time in the range $t_g-$2Gyr to $t_g+$1Gyr, 
when the derivative of the amount of cold gas, 
$(M_{10},M_{30},M_{100})$ within radial ranges $(0-10,0-30,0-100)$kpc, 
with respect to time is most negative as ($t_{\rm 30}, t_{\rm 100}, t_{\rm 300})$, respectively.
We define cold gas as gas with temperature less than $10^5$K.
The exponential decay time scale of SFR at $t_{q}$ is defined by $\tau_{\rm q}\equiv (d\ln~{\rm SFR}/dt)^{-1}$.
The exponential decay time scale of $(M_{10},M_{30},M_{100})$ at ($t_{\rm 30}, t_{\rm 100}, t_{\rm 300})$
are defined by $[\tau_{\rm 10}\equiv (d\ln M_{10}/dt)^{-1},\tau_{\rm 30}\equiv (d\ln M_{30}/dt)^{-1}, \tau_{\rm 100}\equiv (d\ln M_{100}/dt)^{-1}]$.
The time interval between $t_{\rm q}$ and $t_{\rm r}$ is denoted as $t_{\rm qr}$,
The time duration that the galaxy spends in the green valley before turning red is called $t_{\rm green}$.
The time duration the galaxy has spent in the red sequence by $z=0.62$ is denoted as $t_{\rm red}$.

We make a needed simplification by approximating the ram-pressure, denoted as $p_{300}$, 
by ${\rm p_{300}}\equiv \rho_6(300) T(300)$, where $\rho_6(300)$ and $T(300)$, respectively, are the mean density of gas with temperature $\ge 10^6$K
and $T(300)$ the mean mass-weighted gas temperature within a proper radius of $300$kpc centered on the galaxy in question.
This tradeoff is made thanks chiefly to 
the difficulty of defining precisely the motion of a galaxy relative to its ambient gas environment,
where the latter often has complex density and velocity structures, and the former has complex, 
generally non-spherical gas distribution geometry.
In a gravitationally shock heated medium, this approximation should be reasonably good,
because the ram-pressure is approximately equal to thermal pressure in post-shock regions. 
We define a point in time 
searched over the time interval between $t_{\rm q}-$2Gyr and $t_{\rm q}+$1Gyrs, 
when the derivative of $p_{300}$ with respect to time is maximum as $t_{\rm ram}$, 
intended to serve as the point in time when ram-pressure has the steepest rise,

As stated in the introduction, it is convenient to express gas cooling time that is proportional 
to gas entropy to the power ${3/2}$, $S^{3/2}$.
Thus, we approximate gas starvation from large scales by the value of environmental entropy $S_{300}$,
defined to be the average gas entropy within a top-hat sphere of proper radius $300$kpc.

For convenience, frequently used symbols and their definitions are given in Table ~\ref{tab:table1}.


\placetable{tab:table1}

\begin{deluxetable}{ll}
\tabletypesize{\small}
\tablecolumns{2}
\tablewidth{0pc}
\tablecaption{Definitions of symbols and names \label{tab:table1}}
\tablehead{
\colhead{symbol/name} & \colhead{definition/meaning}}
\startdata
$t_{\rm g}$ & a point in time when galaxy has $g-r=0.55$ \\
$t_{\rm r}$ & a point in time when galaxy has $g-r=0.65$ \\
$M_{\rm 10}$ & amount of cold gas within a radius of 10kpc \\
$M_{\rm 30}$ & amount of cold gas within a radius of 30kpc \\
$M_{\rm 100}$ & amount of cold gas within a radius of 100kpc \\
$t_{\rm q}$ & a point in time of quenching for SFR \\
$t_{\rm 10}$ & a point time of quenching for $M_{10}$ \\
$t_{\rm 30}$ & a point time of quenching for $M_{30}$ \\
$t_{\rm 100}$ & a point time of quenching for $M_{100}$ \\
$t_{\rm ram}$ & a point in time of largest first derivative of ram-pressure w.r.t time\\
$\tau_{\rm q}$ & exponential decay time of SFR at $t_{\rm q}$ \\
$\tau_{\rm 10}$ & exponential decay time of $M_{10}$ at $t_{\rm 10}$ \\
$\tau_{\rm 30}$ & exponential decay time of $M_{30}$ at $t_{\rm 30}$ \\
$\tau_{\rm 100}$ & exponential decay time of $M_{100}$ at $t_{\rm 100}$ \\
$\Delta M_*$ & stellar mass change between $t_{\rm q}$ and $t_{\rm r}$ \\
$\Delta M_{10}$ & $M_{10}$ mass change between $t_{\rm q}$ and $t_{\rm r}$ \\
$\Delta M_{30}$ & $M_{30}$ mass change between $t_{\rm q}$ and $t_{\rm r}$ \\
$\Delta M_{100}$ & $M_{100}$ mass between $t_{\rm q}$ and $t_{\rm r}$ \\
$r_{\rm e}^{\rm SFR}$ & effective radius of young stars formed within the past 100Myr \\
$T_{300}$ & environmental temperature within physical radius of $300$kpc \\
$S_{300}$ & environmental entropy within physical radius of $300$kpc \\
$p_{300}$ & environmental pressure within physical radius of $300$kpc \\
$\delta_2$ & environmental overdensity within comoving radius of $2h^{-1}$Mpc \\
$d/r_{\rm v}^{\rm c}$ & distance to primary galaxy in units of virial radius of primary galaxy \\
$t_{\rm qr}$ & time duration from $t_{\rm q}$ to $t_{\rm r}$ \\
$t_{\rm green}$ & time spent in green valley \\
$t_{\rm red}$ & time spent in red sequence \\
$M_{\rm h}^{\rm c}$ & halo mass of primary galaxy \\
$M_*^{\rm s}/M_*^{\rm c}$ & stellar mass ratio of satellite to primary galaxy 
\enddata
\end{deluxetable}

\subsection{Tests and Validation of Simulation}

The galaxy formation simulation in a cosmological setting used here
includes sophisticated physical treatment, ultra-high resolution and a very large galaxy 
sample to statistically address cosmological and astrophysical questions.
While this simulation represents the current state-of-the-art in these respects,
feedback from SF is still far from being treated from first principles.
Thus, it is necessary that we validate the feedback prescription empirically.

In \citet[][]{2012Cen} 
we presented an examination of the damped Lyman alpha systems (DLAs)
and found that the simulations, for the first time, are able to match
all observed properties of DLAs, including abundance, size, metallicity and kinematics.
In particular, the metal distribution in and around galaxies over a wide range of redshift
($z=0-5$) is shown to be in excellent agreement with observations \citep[][]{2012Rafelski}.
The scales probed by DLAs range from stellar disks at low redshift to about one half of the virial radius at high redshift.
In \citet[][]{2012bCen} we further show that
the properties of O~VI absorption lines at low redshift, including their abundance, Doppler-column density distribution,
temperature range, metallicity and coincidence between O~VII and O~VI lines,
are all in good agreement with observations \citep[][]{2008Danforth,2008Tripp, 2009Yao}.
The agreement between simulations and observations with respect to O~VI lines 
is recently shown to extend to the correlation between galaxies and O~VI lines,
the relative incidence ratio of O~VI around red to blue galaxies,
the amount of oxygen mass around red and blue galaxies as well as 
cold gas around red galaxies \citep[][]{2013Cen}.

In addition to agreements with observations with respect to circumgalactic and intergalactic medium,
we find that our simulations are able to match the 
global SFR history (the Madau plot) and galaxy evolution \citep{2011bCen}, 
the luminosity function of galaxies at high \citep[][]{2011cCen} and low redshift \citep[][]{2011bCen}, 
and the galaxy color distribution \citep[][]{2011bCen, 2012Tonnesen}, within observational uncertainties.
In \citet{2011bCen} we show  
that our simulations reproduce many trends in the global evolution of galaxies and various manifestations of 
the cosmic downsizing phenomenon.
Specifically, our simulations show that, at any redshift, the specific star formation rate of galaxies, on average, 
correlates negatively with galaxy stellar mass,
which seems to be the primary physical process for driving the cosmic downsizing phenomena observed.
Smoothed particle hydrodynamic (SPH) simulations and semi-analytic methods, in comparison, appear to produce a positive 
correlation between the specific star formation rate of galaxies and galaxy stellar mass,
which is opposite to what we find \citep[e.g.,][]{2012Weinmann}.
These broad agreements between our simulations and observations indicate that, among others, 
our treatment of feedback processes from SF, i.e., the transport of metals and energy from galaxies,
from SF sites to megaparsec scale (i.e., from interstellar to intergalactic medium)
are realistically modeled as a function of distance and environment, at least in a statistical sense,
and it is meaningful to employ our simulated galaxies, circumgalactic and intergalactic medium 
for understanding physical processes and for confrontations with other, independent observations.

\begin{figure}[!ht]
\centering
\vskip -1.0cm
\resizebox{6.0in}{!}{\includegraphics[angle=0]{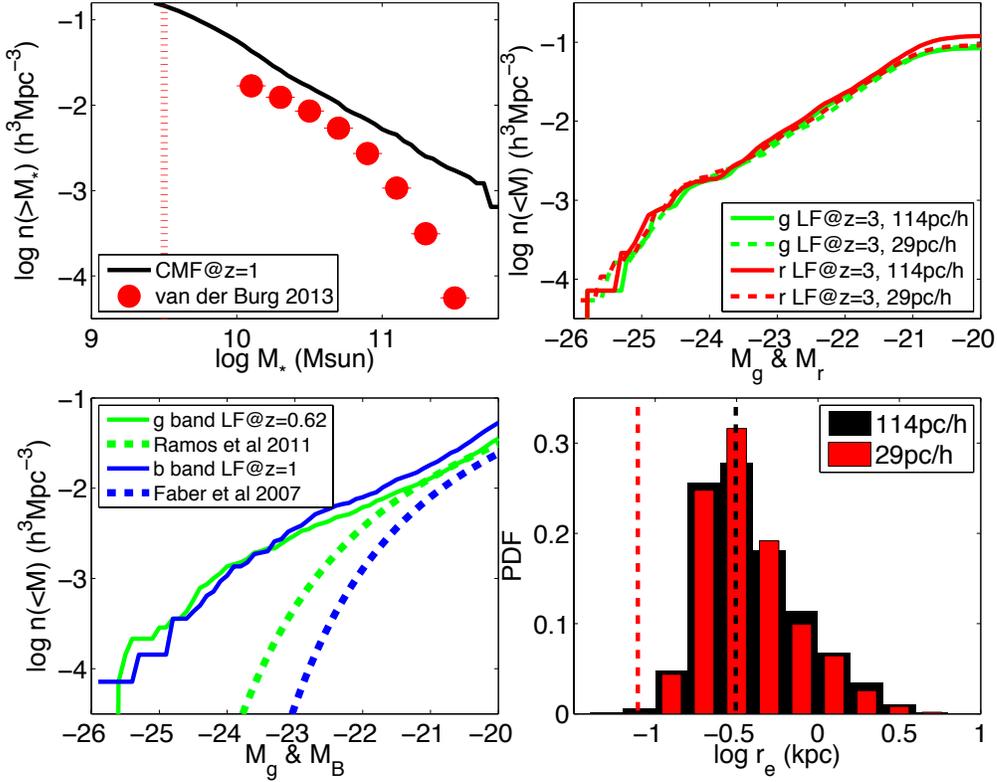}}   
\vskip -0.5cm
\caption{
Top-left panel: galaxy stellar mass function (SMF) at $z=1$ from the fiducial run with spatial 
resolution of $114$pc/h (solid curve) and observations by \citet{2013vanderBurg} at $z=1$;  
the observational points have been shifted upward by $0.4$dex.
The vertical dotted line indicates the stellar mass cutoff of $\ge 10^{9.5}\msun$ that we apply to simulated galaxies.
Top-right panel: galaxy luminosity function (LF) in SDSS g band (green) and 
r band (red) at $z=3$ from the fiducial run (solid curves) and the higher resolution run (dashed curves)
for galaxies with $M_*\ge 10^{9.5}\msun$.
Bottom-left panel: galaxy luminosity function (LF) in restframe g band (green) at $z=0.62$ and 
B band (blue) at $z=1$ for galaxies with $M_*\ge 10^{9.5}\msun$,
to compared from the corresponding observations of the same color dashed curves
at $z=0.62$\citep{2011Ramos} 
and $z=1$\citep{2007Faber}, respectively.
The observational points have been shifted upward by $0.4$dex.
No dust effects have been included in the simulated luminosity functions.
Bottom-right panel: the distribution of the recovered true stellar radii of all galaxies 
with $M_*\ge 10^{9.5}\msun$ at $z=3$ from the fiducial (black) and the higher resolution run (red).
The vertical lines indicate the effective resolutions of the two simulations,
which are found to be $1.84$ times the maximum refined resolution. 
}
\label{fig:mfr50test}
\end{figure}

In order to determine what galaxies in our simulations to use in our subsequent analysis,
we make an empirical numerical convergence test.
Top-right panel in Figure~\ref{fig:mfr50test} shows comparisons between 
galaxies of two simulations at $z=3$ with different resolutions for the luminosity functions in
rest-frame $g$ and $r$ bands.
The fiducial simulation has a resolution of $114$pc/h and 
an identical comparison run has four times better resolution of $29$pc/h.
We are not able to make comparisons at redshift substantially lower than $z=3$ at this time. 
In any case, we expect that the comparison at $z=3$ is a more stringent test,
because the resolution effect is likely more severe at higher redshift than at lower redshift
in a hierarchical growth model where galaxies become increasingly larger with time.
The comparisons are best done statistically, because not all individual 
galaxies can be identified at a one-to-one basis due to resolution-dependent star formation and merging histories.
Comparisons with respect to other measures, such as stellar mass function, SFR, etc, give comparable convergence.
Based on results shown, we decide to place a lower stellar mass limit of $10^{9.5}\msun$, 
which is more than $75\%$ complete for almost all relevant quantities, 
to the extent that we are able to make statistical comparisons between these two runs
with respect to the global properties of galaxies (stellar mass, luminosity, SFR, sSFR, etc).

In terms of checking the validity and applicability of the simulations,
we also make comparisons for the galaxy cumulative mass function at $z=1$ with observations in the top-left panel of  
Figure~\ref{fig:mfr50test}.
We see that the simulated galaxies have a higher abundance than observed by a factor of $4-5$ in the low mass end
and the difference increases towards higher mass end.
This difference is expected, because the simulation volume is an overdense region that has
a higher galaxy density overall and progressively higher densities for more rare, higher mass galaxies.
This difference is also borne out in the comparisons 
between simulated and observed rest-frame $g$ band galaxy luminosity function at $z=0.62$ 
and $B$ band luminosity function at $z=1$ (lower-left panel of Figure~\ref{fig:mfr50test}).
The difference between simulation and observations is also seen to increase with decreasing redshift, as expected.
We note that at the high luminosity end dust effects may become very important, causing the apparent 
discrepancies between simulations and observations larger than they actually are.
Overall, our simulation serves our purposes well for two reasons.
First, it contains massive group/cluster systems as well as field regions that allow us to probe environment effects with enough leverage/range.
Second, the faint end slope of the galaxy population matches observations well, albeit with a larger overall amplitude;
as a result, relative, comparative statements that we will make across the stellar mass range $10^{9.5}$ and $10^{11}\msun$ is 
approximately valid even though the absolute number of galaxies may be off.

As will become clear later, during quenching by ram-pressure stripping and gas starvation,
star formation often continues in the central $10$kpc regions of galaxies.
Therefore, it is useful to have an estimate of the effective resolutions of simulations empirically as follows.
Let us denote the actual resolution, given the maximum refinement resolution of $\Delta l$,
as $\Delta x=C\Delta l$, where $C$ is a contant that is expected to be larger than but of order unity
for a mesh code.
We then assume that the computed effective stellar radius (enclosing $50\%$ of stellar mass) of any galaxy is
\begin{equation}
\label{eq:re}
r_{\rm c}^2 = r_{\rm t}^2 + (\Delta x)^2 = r_{\rm t}^2 + (C \Delta l)^2, 
\end{equation}
\noindent
where $r_{\rm c}$ is the actual effective radius from simulation
and $r_{\rm t}$ is the true effective radius if one had infinite resolution.
We have two equations of Eq~\ref{eq:re} for the two simulations with different resolutions 
($\Delta l=114$pc/h and $\Delta l=29$pc/h)
and two unknowns (constant $C$ and vector $r_{\rm t}$).
We solve for $C$ by requiring the distributions of $r_{\rm t}$ derived 
from the two simulations have the closest agreement;
$C=1.84$ is found.
The bottom-right panel of Figure~\ref{fig:mfr50test} shows the distributions of derived $r_{\rm t}$ for all galaxies with stellar mass
$\ge 10^{9.5}\msun$ from the two simulations with $C=1.84$.
The effective resolutions ($C\Delta l$) for the two runs, $312$pc and $78$pc, 
are indicated by the vertical dashed lines corresponding to the histograms with the same colors.
As we will show later, the central gas within $10$kpc is largely retained by the galaxy
that is being subject to ram-pressure stripping.
Clearly, our resolution is adequate to resolve gas distribution, where
it is demanded to address ram-pressure stripping physics.
On the other hand, consistent with other panels of Figure~\ref{fig:mfr50test}, we see that
at the stellar mass cut of $10^{9.5}\msun$,
while the majority of galaxies are effectively resolved with respect to their stellar distribution,
a significant fraction of galaxies in our selected sample with stellar mass close to $10^{9.5}\msun$
is under-resolved in the central regions.
We expect that, although our simulation may overestimate the ram-pressure stripping effects for the most inner regions
for some of the smaller galaxies,
the numerical effect on overall ram-pressure stripping process due to this limitation is likely small.
This is because the amount of gas within our resolution limit ($313$pc) is small compared to the total mass of gas within (say) $10$kpc,
within which, we will show later, gas is little affected by ram-pressure stripping. 
In other words, an under-resolution should not materially impact the overall 
the amount of ram-pressure stripped gas, since the gas in the central region much larger than our resolution
is not stripped anyway.

\begin{figure}[!hb]
\centering
\vskip -0.0cm
\resizebox{5.0in}{!}{\includegraphics[angle=0]{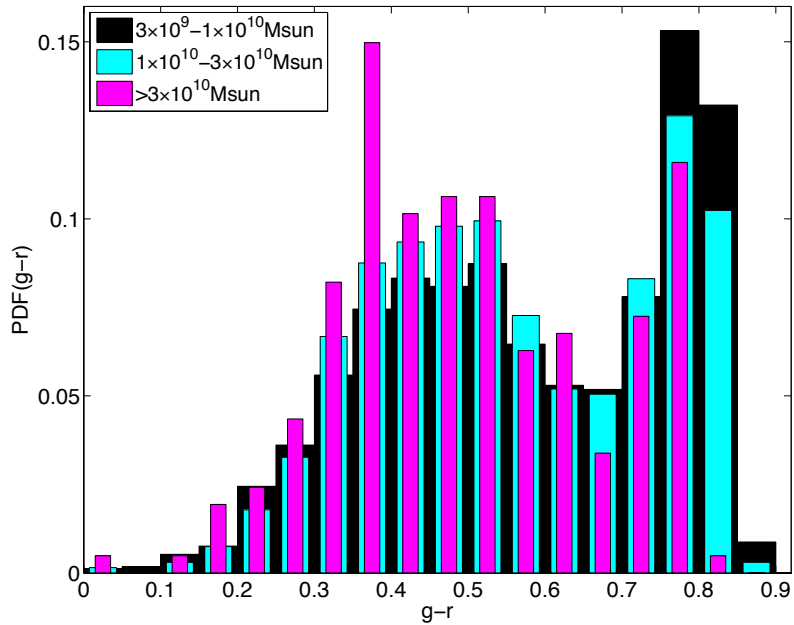}}    
\vskip -0.8cm
\caption{
shows $g-r$ color distributions of simulated galaxies in three stellar mass ranges,
$3\times 10^9-1\times 10^{10}\msun$ (black),
$1\times 10^{10}-3\times 10^{10}\msun$ (cyan) and
$>3\times 10^{10}\msun$ (magenta), at $z=0.62$.
}
\label{fig:bimodal}
\end{figure}

\subsection{Bimodal Distribution of Galaxy Colors at $z=0.62$}

The main goal of this paper is to identify and understand the physical processes that produce
the observed galaxy color bimodality. 
We first examine if our simulations actually reproduce the observed color bimodality.
Figure~\ref{fig:bimodal} shows $g-r$ color distributions of simulated galaxies in three stellar mass ranges,
$3\times 10^9-1\times 10^{10}\msun$ (black),
$1\times 10^{10}-3\times 10^{10}\msun$ (cyan) and
$>3\times 10^{10}\msun$ (magenta), at $z=0.62$.
The g-r color distributions show clear bimodalities for all three subsets of galaxies,
with the red peak becoming more prominent for less luminous galaxies at $z=0.62$,
consistent with recent observations \citep[e.g.,][]{2004Bell, 2006Willmer, 2006Bundy, 2007Faber}.
We also caution that one should not overstate the success in this regard for two reasons.
First, on the simulation side, since our simulation volume does not necessarily represent
an ``average" volume of the universe, a direct comparison to observations would be difficult.
Second, observations at high redshift (i.e., $z\sim 0.62$) are perhaps less complete than 
at low redshift, and identification of low mass (and especially low surface brightness) galaxies,
in particular those that are satellite galaxies and red, may be challenged at present \citep[e.g.,][]{2013Knobel}.
Our main purpose is to make a comparative study of galaxies of different types in the simulation
and to understand how blue galaxies turn red. 

It is intriguing to note that there is no lack of red dwarf galaxies.
While a direct comparison to observations with respect to abundant red dwarf galaxies
can not be made at $z=0.62$, future observations may be able to check this.
Since our simulation does not include AGN mechanical feedback,
this suggests that the bimodal nature of galaxy colors 
does not necessarily require AGN feedback for galaxies in the mass ranges examined.
This finding is in agreement with \citet[][]{2011Feldmann}, who find that 
AGN feedback is not an essential ingredient for producing quiescent, red elliptical galaxies in galaxy groups.
While SF feedback is included in our simulation,
our subsequent analysis shows that environmental effects
play the dominant role in driving galaxy color evolution and consequently color bimodality.
Our results do not, however, exclude the possibility 
that AGN feedback may play 
an important role in regulating larger, central galaxies, such as cD galaxies
at the centers of rich clusters of galaxies,
for which we do not have a sufficient sample to make a statistical statement.
Our earlier comparison between simulated luminosity functions of galaxies at $z=0$
and SDSS observations indicates that some additional feedback, likely in the way 
of AGN, may be required to suppress star formation in the most massive galaxies \citep[][]{2011bCen}.

\section{Results}

Most of our results shown are presented through a variety of comparisons
of the dependencies of galaxies of different types on a set of environmental variables,
to learn how galaxies change color.
We organize our analysis in an approximately chronological order.
In \S 3.1 we focus on processes around the ``quenching" time, $t_{\rm q}$, 
followed by the ensuing period of gas starvation in hot environment in \S 3.2.
In \S 3.3 we discuss stellar mass growth,
evolution of stellar mass function of red galaxies
and present galaxy color migration tracks.
\S 3.4 gives an example of consequences of the color migration picture - galaxy age-mass relation.
We present observable environmental dependence of galaxy makeup at $z=0.62$ in \S 3.5.

\subsection{Ram-Pressure Stripping: Onset of Star Formation Quenching}

\begin{figure}[!ht]
\centering
\vskip -0.0cm
\resizebox{6.0in}{!}{\includegraphics[angle=0]{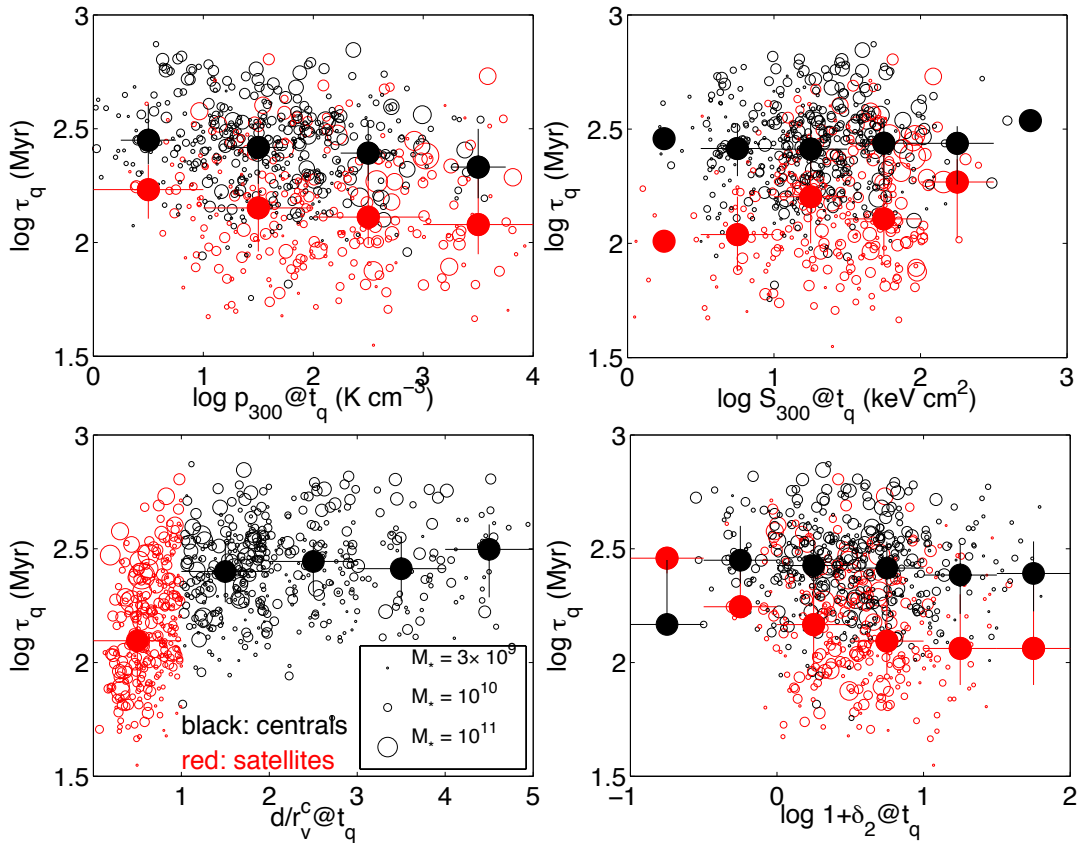}}   
\vskip -0.5cm
\caption{\footnotesize 
shows $\tau_{\rm q}$, the exponential decay time of SFR, against four environmental variables at $t_{\rm q}$:
ram-pressure $p_{300}$ on 300kpc proper scale, environmental entropy $S_{300}$ on 300kpc proper scale, 
distance to primary galaxy $d/r_v^c$ in units of the primary galaxy's virial radius and environmental overdensity $\delta_2$ on $2h^{-1}$Mpc comoving scale.
The magenta solid dots with dispersions are the means.
}
\label{fig:tauminSFQ}
\end{figure}

Figure~\ref{fig:tauminSFQ} shows the quenching time scale $\tau_{\rm q}$ (star formation rate exponential decay time)
against four environmental variables at the quenching time $t_{\rm q}$:
ram-pressure $p_{300}$, environmental entropy $S_{300}$, distance to primary galaxy $d/r_v^c$ and environmental overdensity $\delta_2$.
It is useful to make clear some nomenclature here.
We have used the distance to the primary galaxy, $d/r_v^c$, as an environment variable,
which runs from zero to values significant above unity.
This is merely saying that any galaxy (except the most massive galaxy in the simulation) 
can find a larger galaxy at some distance, not necessarily at $d/r_v^c\le 1$.
The definition of ``satellite galaxies" is reserved only for those galaxies with $d/r_v^c\le 1$,
shown clearly as red circles in the low-left panel of Figure~\ref{fig:tauminSFQ}.
The black circles, labeled as ``centrals" are galaxies with $d/r_v^c>1$,
i.e., those that are not not ``satellite galaxies".

The observation that galaxies are being quenched at all radii - $d/r_v^c>1$ as well as $d/r_v^c<1$ - indicates that the most likey physical mechanism for the onset of quenching
is ram-pressure. 
Tidal stripping is not expected to be effective at removing gas (or stars) at $d/r_v^c>1$ (see \S 2.3 for a discussion).
The fact that $\tau_{\rm q}$ decreases with increasing $p_{300}$ is self-consistent with ram-pressure
being responsible for the onset of quenching.
The outcome that $\tau_{\rm q}$ only very weakly anti-correlates with $p_{300}$ 
indicates that the onset of quenching is some ``threshold" event, which presumably 
occurs when the ram-pressure exceeds gravitational restoring force (i.e., the threshold), thus strongly re-enforcing
the observation that ram-pressure is largely responsible for the onset of quenching.
A ``threshold" type mechanism fits nicely with the fact that the dispersion of 
$\tau_{\rm q}$ at a given $p_{300}$ is substantially larger than the correlation trend,
because galaxies that cross the ``threshold" are expected to depend on very inhomogeneous internal properties
among galaxies (see Figure~\ref{fig:tauminSIGMA} below). 
The weak anti-correlation between $\tau_{\rm q}$ and $\delta_2$ stems from
a broad positive correlation between $p_{300}$ and $\delta_2$.
The fact that there is no discernible correlation between 
$\tau_{\rm q}$ and $S_{300}$ indicates that the onset of quenching is {\it not} initiated by gas starvation.

The most noticeable contrast to the weak trends noted above is the difference between 
satellite galaxies (at $d/r_v^c<1$) and central galaxies (at $d/r_v^c>1$),
in that $\tau_{\rm q}$ of the former is lower than that of the latter by a factor of $\sim 2$.
This is naturally explained as follows.
First, at $d/r_v^c<1$ ram-pressure stripping and tidal stripping operate in tandem to 
accelerate the gas removal process,
whereas at $d/r_v^c>1$ ram-pressure stripping operates ``alone" to remove gas on somewhat longer time scales.
Second, at $d/r_v^c>1$ ram-pressure stripping is, on average, less strong than
at $d/r_v^c<1$.


Possible internal variables that affect the effectiveness of ram-pressure stripping include the relative orientation 
of the normal of the gas disk and the motion vector, rotation velocity of the gas disk,
whether gas disk spiral arms are trailing or not at time of ram-pressure stripping,
gas surface density amplitude and profile, dark matter halo density profile.
As an obvious example, galaxies that have their motion vector and disk normal aligned 
are likely to have maximum ram-pressure stripping effect, everything else being equal.
In the other extreme when the two vectors are perpendicular to each other, the ram-pressure stripping effect may be minimized.
Needless to say, given many factors involved, 
the onset of ram-pressure stripping effect will be multi-variant.
We elaborate on the multi-variant nature of ram-pressure stripping with one example.
The top-panel of Figure~\ref{fig:tauminSIGMA} shows $\tau_{\rm q}$ as a function of 
the stellar surface density $\Sigma_{\rm e}$ within the effective stellar radius $r_{\rm e}$.
We see a significant positive correlation between 
$\tau_{\rm q}$ and $\Sigma_{\rm e}$ in the sense that
it takes longer to ram-pressure-remove cold gas with higher central surface density (hence higher gravitational restoring force) galaxies.
While this positive correlation between $\tau_{\rm q}$ and $\Sigma_{\rm e}$ is consistent with observational indications
\citep[e.g.,][]{2012Cheung},
the underlying physical origin is in a sense subtle.
Since ram-pressure stripping is a ``threshold" event, as noted earlier,
when ram-pressure force just exceeds the internal gravitational restoring force,
one would have expected that a high surface density would yield a shorter dynamic time
hence a shorter $\tau_{\rm q}$.
This is in fact an incorrect interpretation.
Rather, the gas in the central regions where 
$\Sigma_{\rm e}$ is measured is immune to ram-pressure stripping in the vast majority of cases (see Figure (\ref{fig:rat}) below).
Instead, a higher $\Sigma_{\rm e}$ translates, on average, to a larger scale where gas is removed, which has a longer dynamic time hence 
a longer $\tau_{\rm q}$.

\begin{figure}[!ht]
\centering
\vskip -0.0cm
\resizebox{5.0in}{!}{\includegraphics[angle=0]{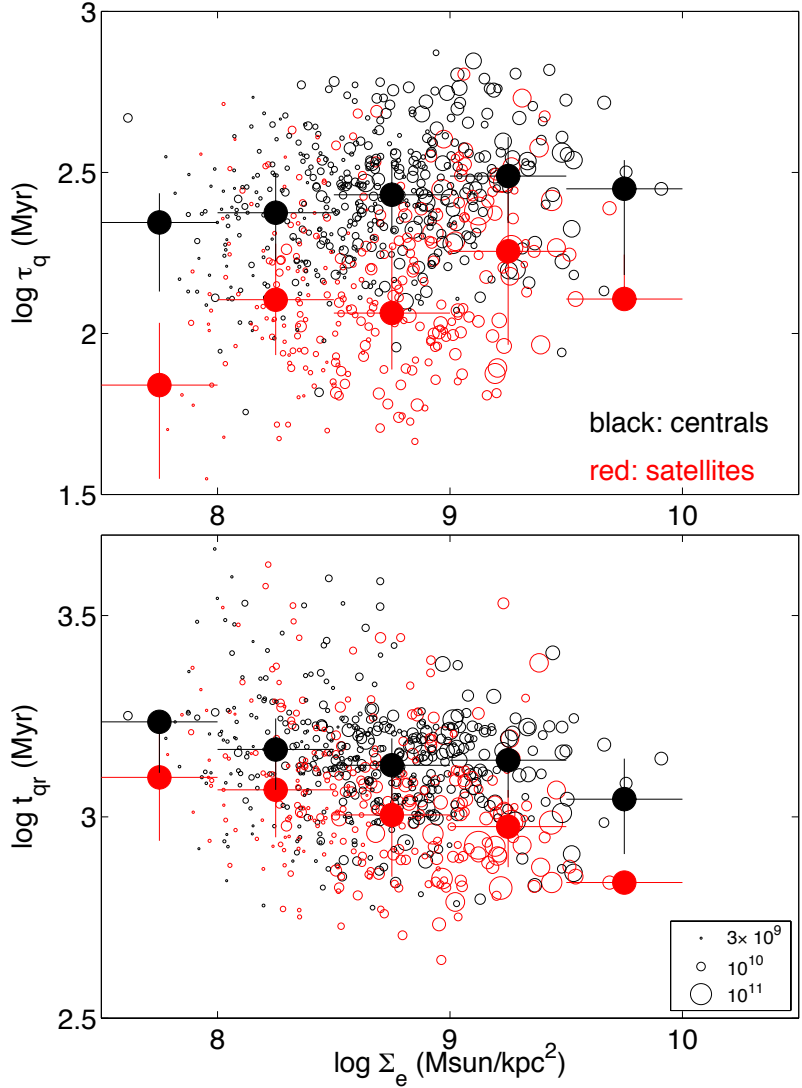}}    
\vskip -0.5cm
\caption{\footnotesize 
Top panel: the exponential decay time scale of SFR, $\tau_{\rm q}$,
as a function of the stellar surface density $\Sigma_{\rm e}$ within the stellar effective radius $r_{\rm e}$.
Bottom panel: the time interval between onset of quenching and the time the galaxy turns red, $t_{\rm qr}$
as a function of $\Sigma_{\rm e}$.
The magenta dots are the averages at a given x-axis value.
}
\label{fig:tauminSIGMA}
\end{figure}

Taken together, we conclude that, while a high ram-pressure provides the conditions for ram-pressure stripping to take effect,
the effectiveness or timescale for gas removal by ram-pressure stripping 
also depend on the internal structure of galaxies.
It is very interesting to note that, unlike between $t_{\rm q}$ 
and $\Sigma_{\rm e}$, 
$t_{\rm qr}$ (the time interval between the onset of quenching $t_{\rm q}$ and the time when the galaxy turns red) 
and $\Sigma_{\rm e}$ shown in the bottom-panel of Figure~\ref{fig:tauminSIGMA},
if anything, is weakly anti-correlated.
We attribute this outcome to the phenomenon that galaxies with higher central surface density have a shorter 
time scale for consuming the existing cold gas hence, once the overall cold gas reservoir is removed.
This explanation will be elaborated more later.

\begin{figure}[!h]
\centering
\vskip -0.0cm
\resizebox{5.0in}{!}{\includegraphics[angle=0]{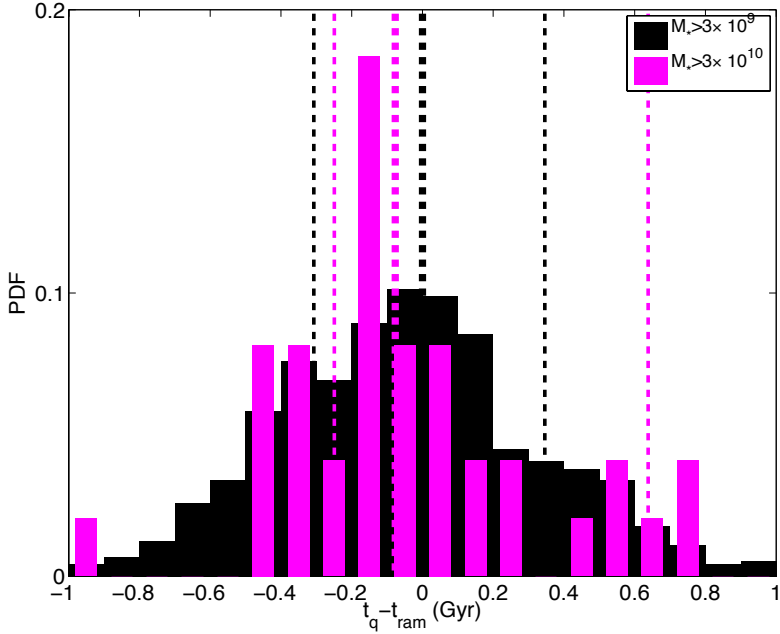}}   
\vskip -0.5cm
\caption{\footnotesize 
the histograms of $t_{\rm q}-t_{\rm ram}$ for red galaxies at $z=0.62$,
where $t_{\rm ram}$ is the point in time when the derivative of $p_{300}$ with respect to time is maximum,
and $t_{\rm q}$ is the onset of quenching time for SFR.
The vertical thick lines show the medians of the corresponding histograms of the same colors,
and the vertical thin lines for each color are for $25\%$ and $75\%$ percentiles.
}
\label{fig:tqtram}
\end{figure}

\begin{figure}[!h]
\centering
\vskip -0.0cm
\resizebox{6.0in}{!}{\includegraphics[angle=0]{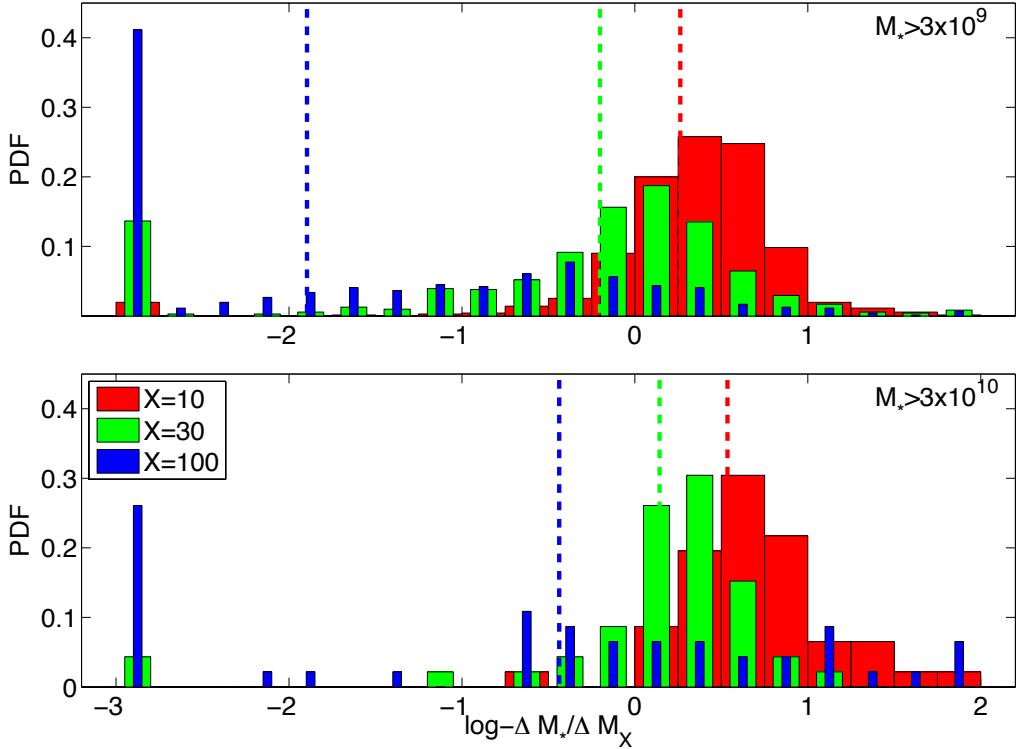}}    
\vskip -0.0cm
\caption{\footnotesize 
shows the distribution of $-\Delta M_*/\Delta M_{\rm X}$ at three radial ranges, $X=(10, 30, 100)$.
$\Delta M_*$ is the amount of stars formed during the time interval from the onset of quenching $t_{\rm q}$ to the time the galaxy turns red $t_{\rm r}$,
and $\Delta M_{\rm X}$ 
is the difference of the amount of cold ($T<10^5$K) gas within a radius $X~$kpc between $t_{\rm r}$ and $t_{\rm q}$.
The vertical dashed lines show the medians of the corresponding histograms of the same colors.
}
\label{fig:rat}
\end{figure}

We have made the case above that ram-pressure stripping is primarily responsible for the onset of quenching process
based on evidence on the dependence of exponential decay time of SFR at the onset of quenching on
environment variables. We now make a direct comparison between $t_{\rm q}$ and $t_{\rm ram}$.
Figure (\ref{fig:tqtram}) shows the histograms of $t_{\rm q}-t_{\rm ram}$.
We see that the time difference between the two is centered around zero,
indicating a casual connection between the onset of SFR quenching and the rapid rise of ram-pressure.
The width of the distribution of a few hundred Myrs reflects the fact that 
the exact strength of ram-pressure stripping required to dislodge the gas 
varies greatly, depending on many variables as discussed above.
This is as yet the strongest supporting evidence for ram-pressure stripping being 
responsible for the onset of quenching, 
especially considering the conjunctional evidence that the onset of quenching could occur 
outside the virial radius of a larger neighboring galaxy where tidal stripping is expected to be less effective
and yet ram-pressure is expected to become important.

\begin{figure}[!ht]
\centering
\vskip -0.0cm
\hskip -1.7cm
\resizebox{4.20in}{!}{\includegraphics[angle=0]{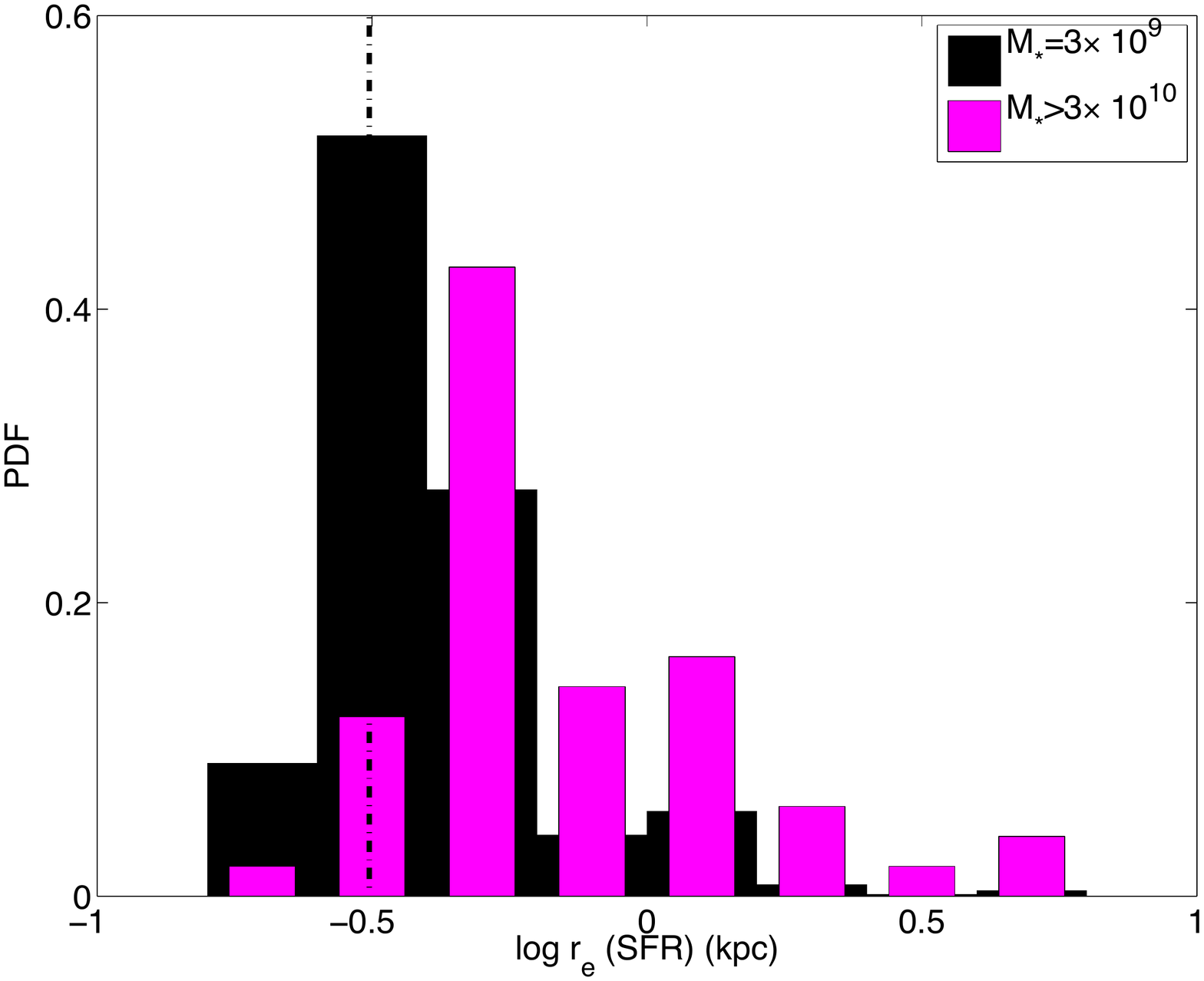}}    
\hskip -2.15cm
\resizebox{4.20in}{!}{\includegraphics[angle=0]{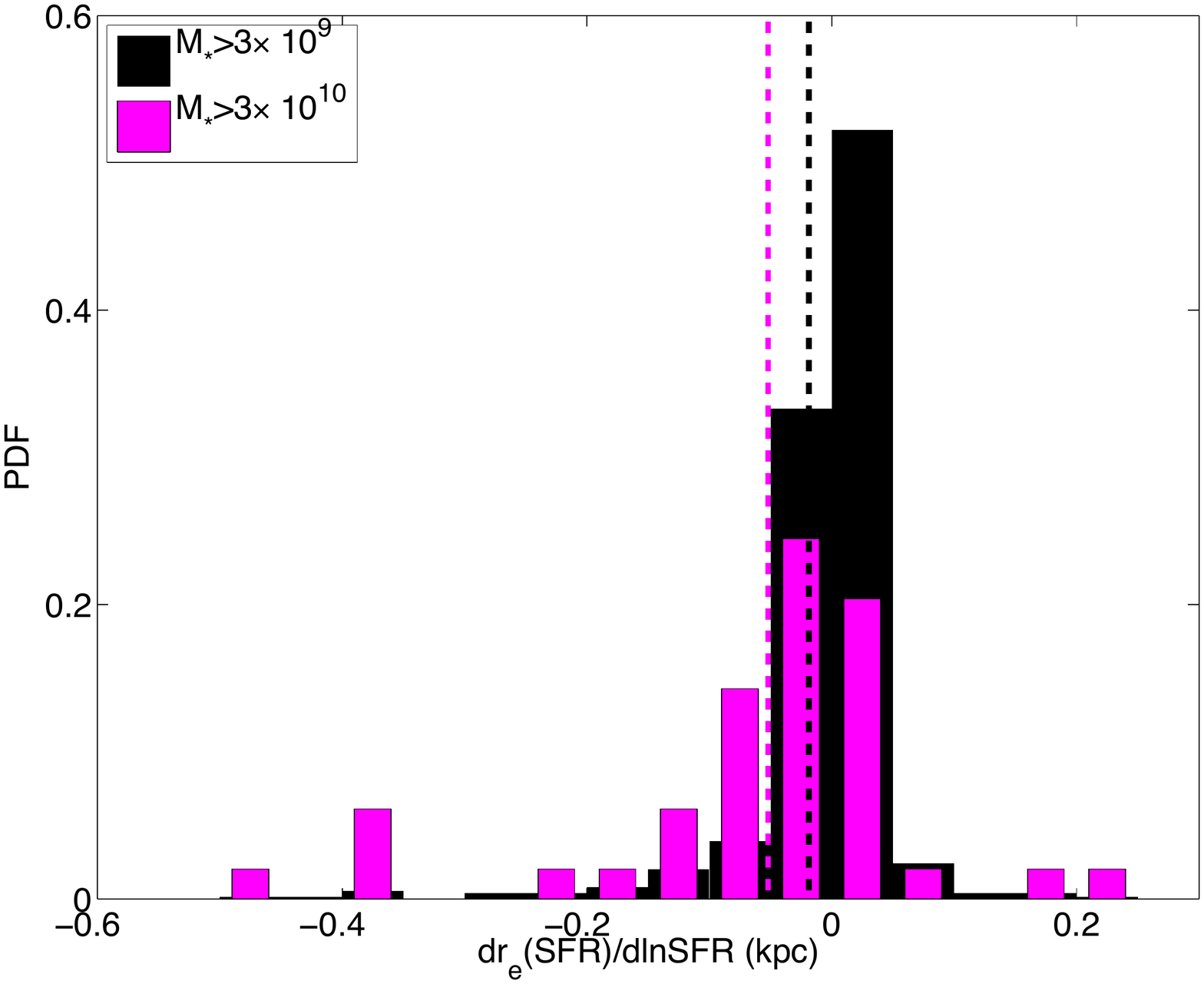}}   
\vskip -0.5cm
\caption{\footnotesize 
Left panel shows the distribution of the effective radius of stars formed in the last $100$Myrs, $r_e(\rm SFR)$,
at $t_{\rm q}$ for galaxies in two stellar mass ranges.   
The vertical dot-dashed line indicates the effective resolution of the simulation, taken from the bottom-right panel in Figure~\ref{fig:mfr50test}.
Right panel shows the distribution of the ratio of the decline of $r_e(\rm SFR)$ with respect to the decline of SFR,
$d~r_e({\rm SFR})/d~\ln~{\rm SFR}$ at $t_{\rm q}$.
In each panel, we differentiate between galaxies in three separate stellar mass ranges.
The vertical lines show the medians of the corresponding histograms of the same colors.
}
\label{fig:r50his}
\end{figure}

Evidence so far supports the notion that ram-pressure stripping 
is the initial driver for the decline of SFR in galaxies that are en route to the red sequence.
The immediate question is then:
What region in galaxies does ram-pressure stripping affect?
To answer this question, we need to compare 
the amount of cold gas available at $t_{\rm q}$ 
with the amount of star formation that ocurrs subsequently.
We compute the following ratios:
the ratios of the amount of stars formed during the time interval from $t_{\rm q}$ to the time the galaxy turns red ($t_{\rm r}$) 
to the difference between the amount of cold ($T<10^5$K) gas at $t_{\rm q}$ and $t_{\rm r}$,
denoted as ($-\Delta M_*/\Delta M_{\rm 10}$, $-\Delta M_*/\Delta M_{\rm 30}$, $-\Delta M_*/\Delta M_{\rm 100}$) within three radii $(10,30,100)$kpc.
The minus signs are intended to make the ratios positive, since
stellar mass typically increase with time, whereas the cold gas mass for galaxies being quenched decreases.
Note that $\Delta M_*$ could be negative.
We set a floor value to the above ratios at $10^{-3}$.
Figure (\ref{fig:rat}) shows the distributions of $-\Delta M_*/\Delta M_{\rm \rm X}$,
where $X=(10,30,100)$. 
We see that for $r\le 10$kpc the peak of the distribution (red histograms) for all stellar masses 
is at $-\Delta M_*/\Delta M_{\rm 10} > 1$, typically in the range of $2-10$, with the vast majority of cases at $>1$.
For a larger radius $r\le 30$kpc the distribution (green histograms) for all stellar masses
is now peaked at $-\Delta M_*/\Delta M_{\rm 30} \sim 1$ with about 50\% at $<1$, 
while for a still larger radius $r\le 100$kpc the distribution (green histograms) 
it is now shifted to $-\Delta M_*/\Delta M_{\rm 100} \le 1$ and more than 50\% at less than 
$(0.01,0.4)$ for galaxies of stellar mass $>(3\times 10^9, 3\times 10^{10})\msun$, respectively. 
This is unambiguous evidence that ram-pressure stripping removes 
the majority of cold gas on scales $\ge 30$kpc, while
the cold gas within $10$kpc is unaffected by ram-pressure stripping and consumed
by subsequent in situ star formation.
It is noted that the lack of effect on the cold gas at $r<10$kpc by ram-pressure stripping seems universal,
the gas removal by ram-pressure stripping at larger radii $r>30$kpc varies substantially from galaxy to galaxy,
which we argue is consistent with the large variations of $\tau_{\rm q}$ seen in Figure~\ref{fig:tauminSFQ}. 
We thus conclude that there is continued nuclear SF in the quenching phase.
\citet[][]{2011Feldmann}, based on a smaller sample of simulated galaxies that form a group of galaxies with 
a spatial resolution of $300$pc (compared to $160$pc here), 
find that in situ star formation is responsible for consuming a substantial fraction of the residual gas on small scales
after gas accretion is stopped subsequent to the infall, consistent with our results.
This outside-in ram-pressure stripping picture and continuous SF in the inner region that emerges from the above analysis 
has important implications and observable consequences,
consistent with the latest observations \citep[e.g.,][]{2013Gavazzi}.

We quantify how centrally concentrated the star formation is at the outset of SF quenching
in Figure (\ref{fig:r50his}), in part to assess our ability to resolve SF during the quenching phase.
The left panel shows the distribution of the effective radius of stars formed 
in the last $100$Myrs prior to $t_{\rm q}$, denoted as $r_e^{\rm SFR}$,  for galaxies in two stellar mass ranges. 
The right panel shows the distribution of the ratio of the decline of $r_e^{\rm SFR}$ with respect to the decline of SFR,
$dr_e^{\rm SFR}/d\ln{\rm SFR}$ at $t_{\rm q}$.
It is evident from Figure (\ref{fig:r50his}) 
that more massive galaxies tend to have larger $r_e^{\rm SFR}$, as expected.
It is also evident that the recent formation for the vast majority of galaxies occurs within a radius of a few kiloparsecs.
It is noted that ongoing SF in a significant fraction of galaxies with stellar masses $\le 3\times 10^{9}\msun$ 
is under-resolved, as indicated by the vertical dot-dashed line in the left panel.
However, none of our subsequent conclusions would be much altered by this numerical effect, 
because (1) all of our conclusions appear to be universal across the stellar mass ranges
and (2) the inner region of $10$kpc is not much affected by ram-pressure stripping anyway (thus under-resolving 
a small central fraction within $10$kpc does not affect the overall ram-pressure stripping effects).
What is interesting is that more than 50\% of galaxies in both stellar mass ranges
have negative values of $dr_e^{\rm SFR}/d\ln{\rm SFR}$ at $t_{\rm q}$,
indicating that, when the SFR decreases in the quenching phase,
star formation proceeds at progressively larger radii in the central region. 
This result, while maybe somewhat counter-intuitive, 
is physically understandable.
We attribute this inside-out star formation picture 
to the star formation rate surface density being a superlinear function of gas surface density 
in the Kennicutt-Schmidt \citep[][]{1959Schmidt, 1998Kennicutt} law.
The picture goes as follows: when gas supply from large scales ($\sim 100$kpc) is cut off and under the assumption that
gas in the central region does not re-distribute radially,
the SFR diminishes faster with decreasing radius in the central region where SF occurs,
causing the effective SF radius to increase with time and star formation rate to decline faster than cold gas content,
while the overall SFR is declining.

In summary, ram-pressure stripping is ineffective in removing cold gas that is already present on scales of $\le 10$kpc but 
most effective in removing less dense gas on larger scales of $\ge 30$kpc.
The chief role played by ram-pressure stripping appears to disconnect galaxies from their cold reservoir
on scales that are much larger the typical stellar radii.
The time scale in question is then on the order of the dynamical time of galaxies at
close to the virial radius. 

\subsection{Starving Galaxies to the Red Sequence and Environmental Sphere of Influence}

\begin{figure}[!h]
\centering
\resizebox{6.5in}{!}{\includegraphics[angle=0]{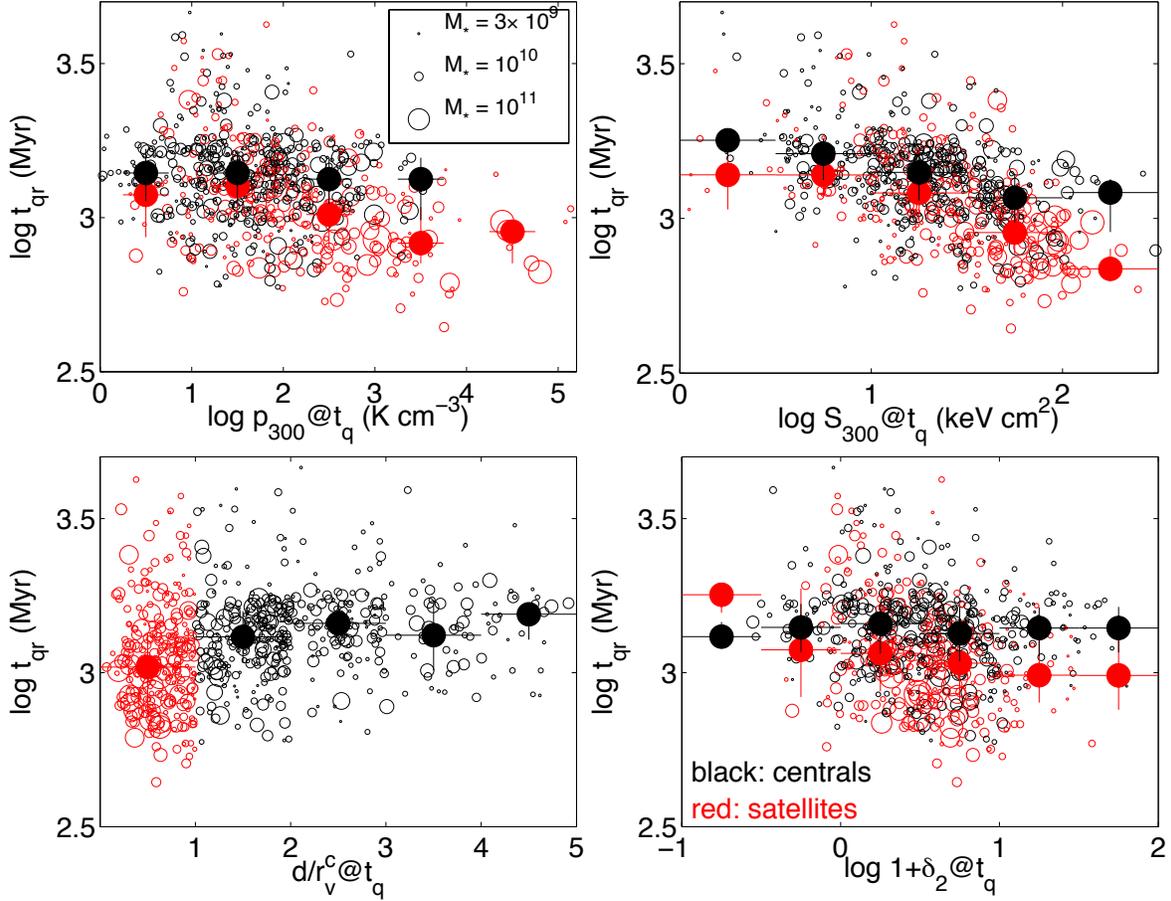}}   
\caption{\footnotesize 
shows $t_{\rm qr}$ (time interval from the onset of quenching to the time the galaxy turns red) against four environmental variables at $t_{\rm q}$:
ram-pressure $p_{300}$ on 300kpc proper scale, environmental entropy $S_{300}$ on 300kpc proper scale, 
distance to primary galaxy $d/r_v^c$ in units of the primary galaxy's virial radius and environmental overdensity $\delta_2$ on $2h^{-1}$Mpc comoving scale.
The red dash line in the upper-right panel is intended to indicate a visually noticeable trend.
Red circles are satellite galaxies at $t_{\rm q}$, i.e., within the virial radius of a larger galaxy,
and black circles are for non-satellite galaxies.
The size of each circle indicates the stellar mass 
of a galaxy, as shown in the legend in the lower-left panel.
}
\label{fig:taubrSFQ}
\end{figure}

The previous subsection details some of the effects on galaxies being quenched 
due to gas removal by ram-pressure stripping (in conjunction with other hydrodynamical processes)
along with consumption by concurrent SF.  Our attention is now turned to the subsequent evolution.
Figure~\ref{fig:taubrSFQ} plots $t_{\rm qr}$ against four environmental variables at $t_{\rm q}$.
From all panels we consistently see the expected trends:
the time interval $t_{\rm qr}$ from onset of quenching $t_{\rm q}$ to turning red $t_{\rm r}$,
on average, decreases with increasing environmental pressure, increasing environmental entropy, increasing
environmental overdensity and decreasing distance to the primary galaxy.
While there is a discernible difference in $t_{\rm qr}$ between satellite galaxies and central galaxies,
the difference is substantially smaller than that in the initial exponential decay time scale of SFR
$\tau_{\rm q}$ (see Figure~\ref{fig:tauminSFQ}).
This observation makes it clear that 
the onset of quenching initiated by ram-pressure stripping does not determine 
the overall duration of quenching.
Since all the environment variables used tend to broadly correlate with one another - 
higher density regions tend to have higher temperatures, higher gas entropy and higher pressure -
it is not surprising that we see $t_{\rm qr}$ are correlated with all of them in the expected sense.
Earlier we have shown that
$t_{\rm qr}$ is weakly anti-correlated with the stellar surface density at $r_e$, $\Sigma_{\rm e}$ (see bottom-panel of Figure~\ref{fig:tauminSIGMA}).
This suggests that the overall duration from onset of quenching to turning red
is not a matter of a galaxy's ability to hold on to its existing cold gas
but rather the extent of the external gas supply condition, i.e., environment.
This hypothesis is significantly affirmed 
by noticing that
the strongest anti-correlation is found between $t_{\rm qr}$ and $S_{300}$,
among all environment variables examined.
Thus, we conclude, given available evidence,
that the eventual ``push" of galaxies into the red sequence is not as a spectacular event as the initial
onset of quenching that is triggered by a cutoff of large-scale gas supply due to ram-pressure stripping,
and is essentially the process of gas starvation,
when the galaxy has entered a low cold gas density and/or high temperature and/or high velocity dispersion environment.

\begin{figure}[!h]
\centering
\vskip -0.5cm
\hskip -2.55cm
\resizebox{8.0in}{!}{\includegraphics[angle=0]{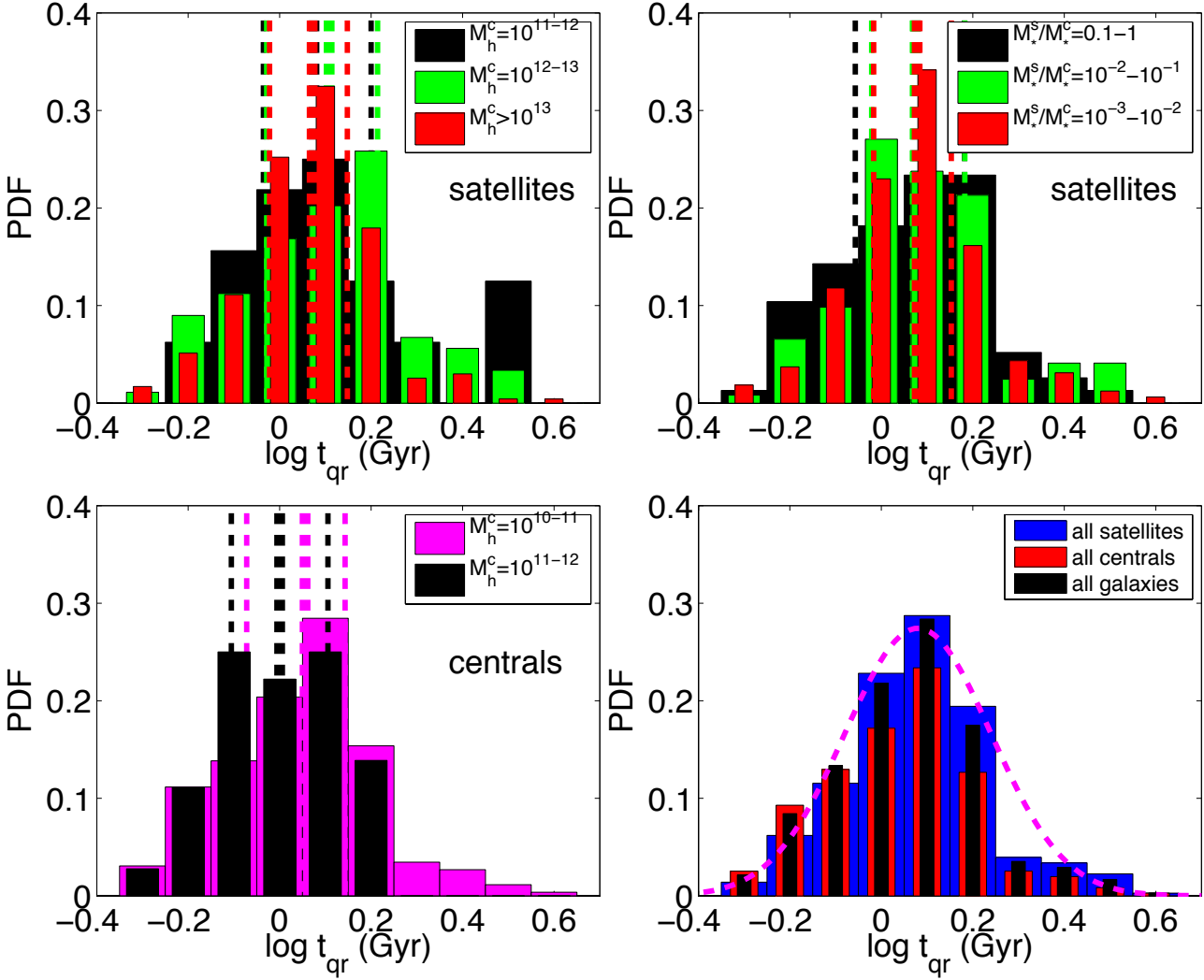}}   
\vskip -0.5cm
\caption{\footnotesize 
Top-left panel: shows the distribution of $t_{\rm qr}$ for satellite galaxies at $z=0.62$,
separated into three primary halo mass ranges:
$M_h^c=10^{11}-10^{12}\msun$ (black),
$M_h^c=10^{12}-10^{13}\msun$ (green),
$M_h^c>10^{13}\msun$ (red).
Top-right panel:
shows the distribution of $t_{\rm qr}$ for satellite galaxies at $z=0.62$,
separated into three ranges of satellite stellar mass to primary stellar mass ratio:
$M_*^s/M_*^c=0.1-1$ (black),
$M_*^s/M_*^c=0.01-0.1$ (green),
$M_*^s/M_*^c=0.001-0.01$ (red).
Bottom left panel: shows the distribution of $t_{\rm qr}$ for primary galaxies at $z=0.62$,
separated into three primary halo mass ranges:
$M_h^c=10^{10}-10^{11}\msun$ (black),
$M_h^c=10^{11}-10^{12}\msun$ (green),
$M_h^c>10^{12}\msun$ (red).
The three vertical dashed lines of order (thin, thick, thin) 
are the (25\%, 50\%, 75\%) percentiles for the histograms of the same color.
Bottom right panel: shows the distribution of $t_{\rm qr}$ for all satellite galaxies (blue),
all primary galaxies (red) and all galaxies (black) at $z=0.62$.
An eye-balling lognormal fit is shown as the magenta line (see Eq~\ref{eq:tqr}).
}
\label{fig:tbrhis}
\end{figure}

We present distributions of $t_{\rm qr}$ in Figure~\ref{fig:tbrhis}. 
The top-left panel shows the distribution of $t_{\rm qr}$ for satellite galaxies (those with $d/r_h^c\le 1$),
grouped into three primary halo mass ranges:
$M_h^c=10^{11}-10^{12}\msun$ (black),
$M_h^c=10^{12}-10^{13}\msun$ (green),
$M_h^c>10^{13}\msun$ (red); the medians of the distributions are (1.2,1.3,1.2)Gyr, respectively.
The top-right panel shows the distribution of $t_{\rm qr}$ for satellite galaxies 
grouped into three ranges of the ratio of satellite to cental stellar mass:
$M_*^s/M_*^c=0.1-1$ (black),
$M_*^s/M_*^c=0.01-0.1$ (green),
$M_*^s/M_*^c=0.001-0.01$ (red); the medians of the distributions of the three groups are nearly identical at $\sim 1.3$Gyr.
The bottom-left panel shows
the distribution of $t_{\rm qr}$ for primary galaxies (those with $d/r_h^c>1$),
grouped into two halo mass ranges:
$M_h^c=10^{10}-10^{11}\msun$ (black),
and $M_h^c=10^{11}-10^{12}\msun$ (green).  
We see that the medians of the distributions are $1.2$Gyr for both mass ranges.
The bottom-right panel plots the distribution of all satellite galaxies and all central galaxies,
along with a simple gaussian fit to the combined set.
A look of the bottom-right panel of Figure~\ref{fig:tbrhis}
suggests that there is practically no difference between the two distributions.
At first sight, this may seem incomprehensible.
A closer examination reveals the underlying physics.

\begin{figure}[!ht]
\centering
\vskip -1.0cm
\resizebox{6.5in}{!}{\includegraphics[angle=0]{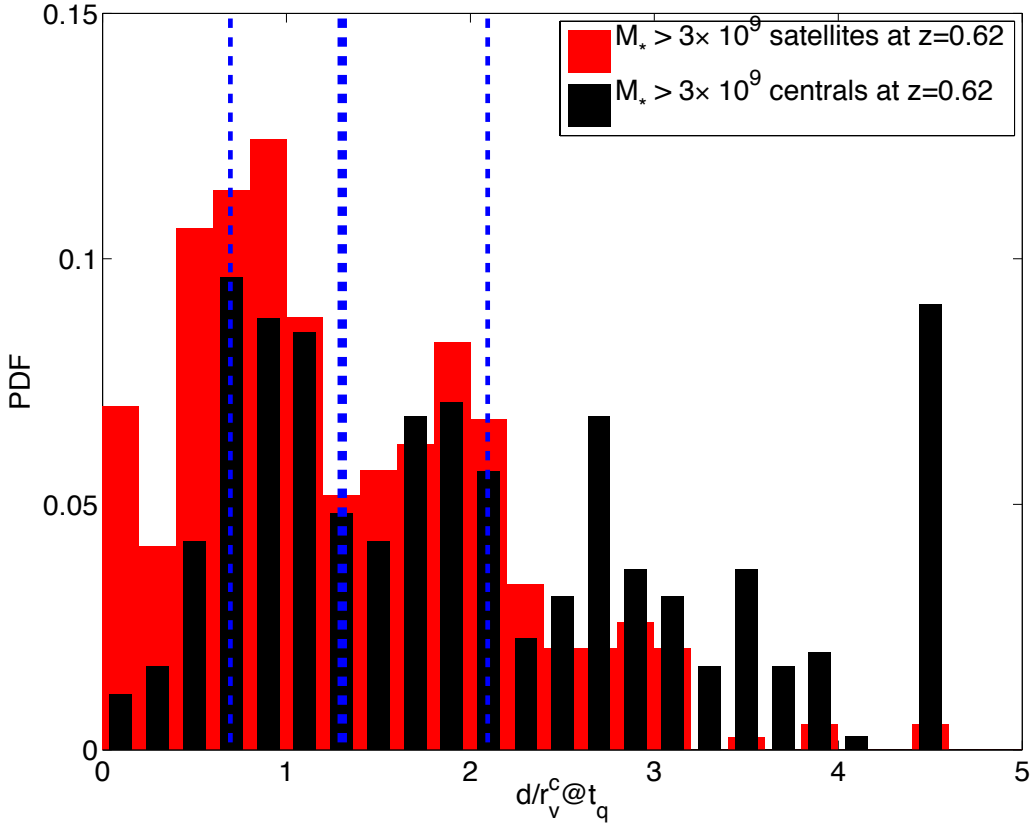}}    
\vskip -0.5cm
\caption{\footnotesize 
shows the distribution of the relative distance $d/r_v^c$ of progenitors at $t_{\rm q}$ of red galaxies at $z=0.62$ 
for two subsets of galaxies: 
the red histogram for those that are within the virial radius of a larger galaxy (i.e., satellite galaxies at $z=0.62$) 
and the black histogram for those that are not within the virial radius of a larger galaxy at $z=0.62$.
The thick blue vertical dashed lines are 50\% percentiles for all galaxies being quenched
and the thin blue vertical dashed lines are 25\% and 75\% percentiles.
}
\label{fig:disSFQ}
\end{figure}

Figure~\ref{fig:disSFQ} shows the distributions of $d/r_v^c$ at $t_{\rm q}$
for satellite (red) and central (black) red galaxies at $z=0.62$.
While it is not a surprise that the vast majority of the satellite galaxies at $z=0.62$ have their onset of quenching taking place
at $d/r_v^c\le 3$ at $t_{\rm q}$,
it is evident that the same appears to be true for the central galaxies at $z=0.62$.
This observation supports the picture that
both satellite and central red galaxies at $z=0.62$ 
have been subject to similar environment effects that turn them red.
It is noted again that this statement that 
red central galaxies have been subject to similar processes as the red satellite galaxies
has been quantitatively confirmed in Figure~\ref{fig:tbrhis}.
The suggestion by \citet[][]{2013bWetzel} that some central galaxies are ejected satellite galaxies
is consistent with our findings here.
Our study thus clearly indicates that one should not confuse red central galaxies with
their being quenched by processes other than environment.
In fact, all available evidence suggests that it is 
environment quenching that plays the dominant role
for the vast majority of galaxies that turn red, whether they 
become satellite galaxies at $z=0.62$ or not.
\citet[][]{2011Feldmann}, using a much smaller sample of simulated galaxies that form a group of galaxies, 
find that quenching of gas accretion starts at a few virial radii from the group center,
in good agreement with our results.
It is seen in Figure~\ref{fig:disSFQ} that
only about 20\% of the onset of galaxy quenching 
occurs as satellites, i.e., within the virial radius of a larger galaxy,
consistent with conclusion derived by others 
\citep[e.g.,][]{2008vandenBosch}.

In the bottom-right panel of Figure~\ref{fig:tbrhis}
we provide an approximate fit to 
the distribution of $t_{\rm qr}$ for all quenched galaxies normalized to galaxies at $z=0.62$ as
\begin{equation}
\label{eq:tqr}
f(\log t_{\rm qr}) = {1\over 2\log t_{\rm med}\sqrt{2\pi}} \exp{\left[-(\log t_{\rm qr}/\log t_{\rm med}-1)^2/8\right]},
\end{equation}
\noindent
where $t_{\rm qr}$ and $t_{\rm med}$ are in Gyr and $\log t_{\rm med}=0.08-1.5\times \log ((1+z)/1.62)$.
The adopted $\log t_{\rm med}=0.08-1.5\times \log (1+z)/1.62$ dependence on $z$
is merely an estimate of the time scale, had it scaled with redshift 
proportional to the dynamical time of the universe. 
One is cautioned not to apply this literally.
Nevertheless, it is likely that the median quenching time at lower redshift
is longer than $\sim 1.2$Gyr at $z=0.62$, perhaps in the range of $2-3$Gyr. 
Incidentally, this estimated quenching time, if extrapolated to $z=0$, 
is consistent with theoretical interpretation of observational data in semi-analytic modeling
or N-Body simulations \citep[e.g.,][]{2012Taranu, 2013aWetzel}. 

In semi-analytic modeling \citep[e.g.,][]{2009Kimm}, 
the quenching time is often taken to be a delta function.
In other words, the satellite quenching process is assumed to be uniform, independent of 
the internal and external properties of the satellites.
Our simulation results (see Eq~\ref{eq:tqr}) 
indicate that such a simplistic 
approach is not well motivated physically.
We suggest that, if a spread in quenching time is introduced in the
semi-analytic modeling, 
an improvement on the agreement between predictions based on semi-analytic modeling
and observations may result in.

In summary, we find that, within the environmental sphere of influence, 
galaxies are disconnected with their large-scale cold gas supply by ram-pressure stripping,  
and subsequently lack of gas cooling and/or accretion
in high velocity environment ensures a prolonged period of gas starvation that
ultimately turns galaxies red.
This applies to satellite galaxies as well as the vast majority of ``apparent" central red galaxies.
The dominance of environment quenching that is found in ab initio cosmological simulations here
is in accord with observations \citep[e.g.,][]{2008vandenBosch, 2012Peng, 2013Kovac}.

\subsection{Color Migration Tracks}

On its way to the red sequence, a galaxy has to pass through the green valley.
Do all galaxies in the green valley migrate to the red sequence?
We examine the entire population of green galaxies in the redshift range $z=1-1.5$. 
Tracing these green galaxies to $z=0.62$, we find that for galaxies with stellar masses greater than
($10^{9.5}$, $10^{10}$, $10^{10.5})\msun$, respectively, 
(40\%, 40\%, 48\%) of galaxies in the green valley at $z=1-1.5$, do not become red galaxies by $z=0.62$. 
While this is an important prediction of our simulations,
we do not provide more information on how one might tell apart these two different population of galaxies in the green valley,
except to point out that attempts to identify galaxies in the green valley as progenitors of red galaxies may generate some confusion.
We examine the distributions (not shown) of the time that red galaxies spent in the green valley, $t_{\rm green}$, en route to the red sequence.
The trends with respect to $M_h$ and $M_*^s/M_*^c$ seen 
are similar to those seen in Figure~\ref{fig:tbrhis}.
No significant differentiation among halo masses of central galaxies is visible,
once again supportive of environment quenching. 
Overall, one may summarize the results in three points.
First, $t_{\rm green}$ is almost universal, independent of being satellites or not, the mass, or the ratio of masses.
Second, the range $t_{\rm green}=0.30\pm 0.15$Gyr appears to enclose most of the galaxies,
although there is a significant tail towards the high end for satellites in low mass central halos.
Third, comparing $t_{\rm green}\sim 0.3$Gyr to the interval from onset of quenching to the time galaxy turning red of $t_{\rm qr}=1.2-1.3$Gyr,
it indicates that, from the onset of quenching to turning red, typical galaxies spend about 25\% of the time in the green valley.

\begin{figure}[ht]
\centering
\vskip -0.0cm
\resizebox{5.0in}{!}{\includegraphics[angle=0]{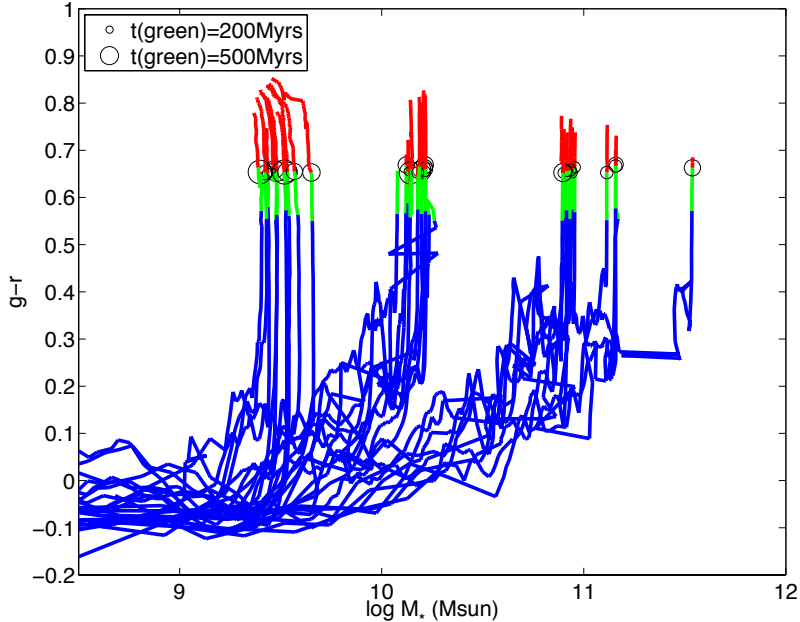}}  
\vskip -0.5cm
\caption{\footnotesize 
shows the evolutionary tracks of 30 semi-randomly selected galaxies on the stellar mass $M_*$-g-r color plane.
The 30 galaxies are selected to be clustered around three masses, $M_*=(10^{9.5},10^{10.1}, 10^{11})\msun$.
Each track has a circle attached at the end of the green period to indicate
the time spent in the green valley.
We shall call this diagram ``skyrockets" diagram of galaxy color migration.
}
\label{fig:migrate}
\end{figure}

Let us now examine the migration tracks of galaxies that eventually enter the red sequence.
Figure~\ref{fig:migrate}  
shows the color-stellar mass diagram for 30 semi-randomly selected red galaxies.
It is striking that the color evolution in the green valley and red sequence is mostly vertical,
i.e., not accompanied by significant change in stellar mass.
This means that the stellar mass growth of most galaxies must occur in the blue cloud.
One can see easily that the blue tracks are mostly moving from lower left to upper right with time
for $g-r\le 0.3$, indicating that galaxies grow when in the blue cloud.
In the blue cloud it is seen that there are occasional horizontal tracks, representing 
mergers that maintain overall color. 
These are mergers that do not result in red galaxies. 
The examples of these include the two most massive
galaxies in the plot with final stellar masses of $\sim 10^{11.6}\msun$, where
there is a major binary merger of $(10^{11.25} + 10^{11.25})\msun$ at $g-r=0.26$.
There are also cases where the tracks temporarily go from north-west to south-east,
indicating significant/major mergers that trigger starbursts that render the remnant galaxies bluer.
This anecdotal evidence that galaxies do not significantly grow mass in the red sequence 
will be confirmed below quantitatively.
\citet[][]{2011Feldmann}, using a small sample of simulated galaxies that form a group of galaxies, 
find that mergers and significant mass growth in galaxies occur, prior to their entering
groupd environment, consistent with the findings here. 
Thus, this ``skyrockets" diagram of color-stellar mass evolution in Figure~\ref{fig:migrate} 
turns out to be a fair representation of typical tracks of galaxies that become red galaxies.

\begin{figure}[!ht]
\centering
\vskip -0.1cm
\hskip -2.1cm
\resizebox{4.20in}{!}{\includegraphics[angle=0]{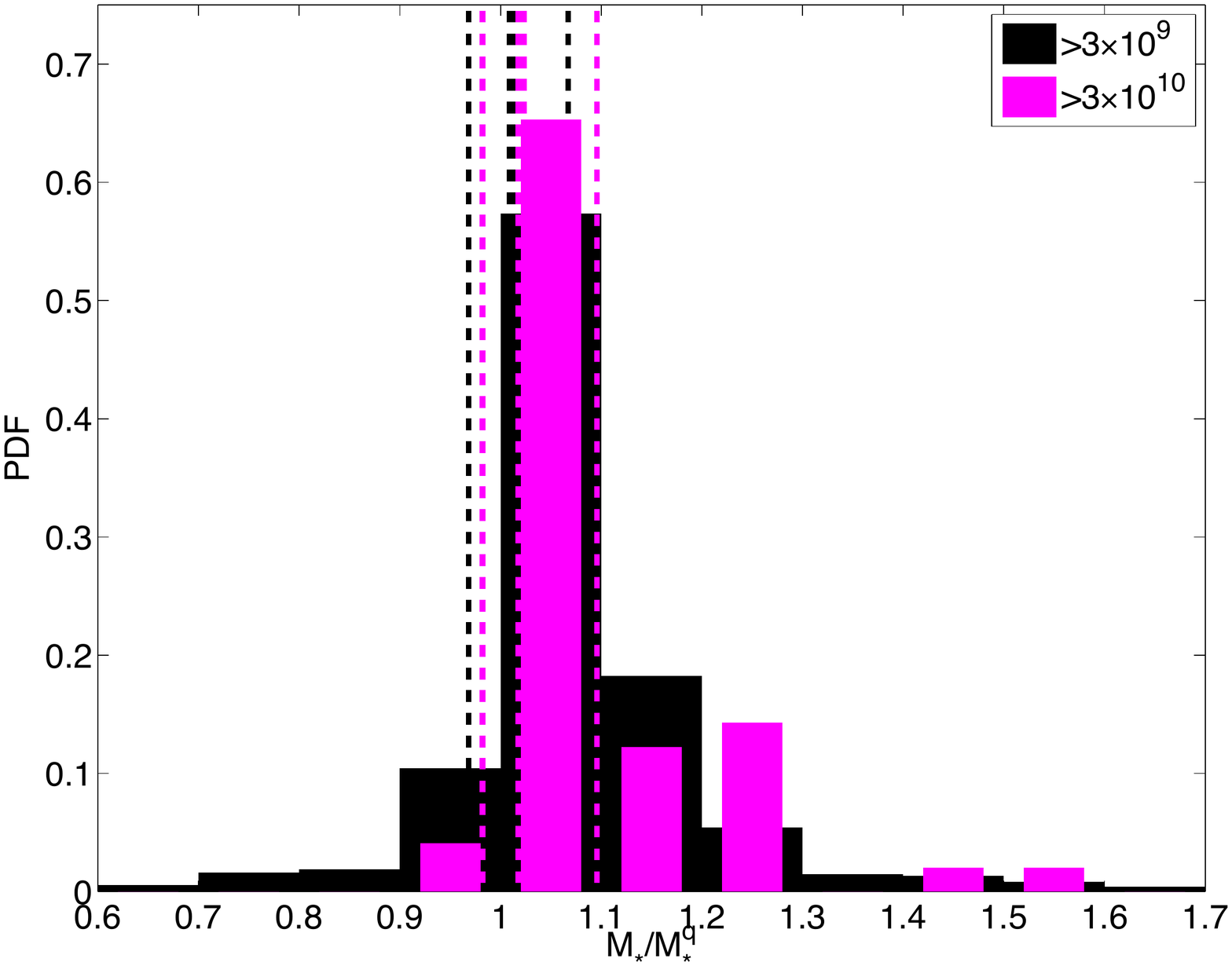}}    
\hskip -2.0cm
\resizebox{4.20in}{!}{\includegraphics[angle=0]{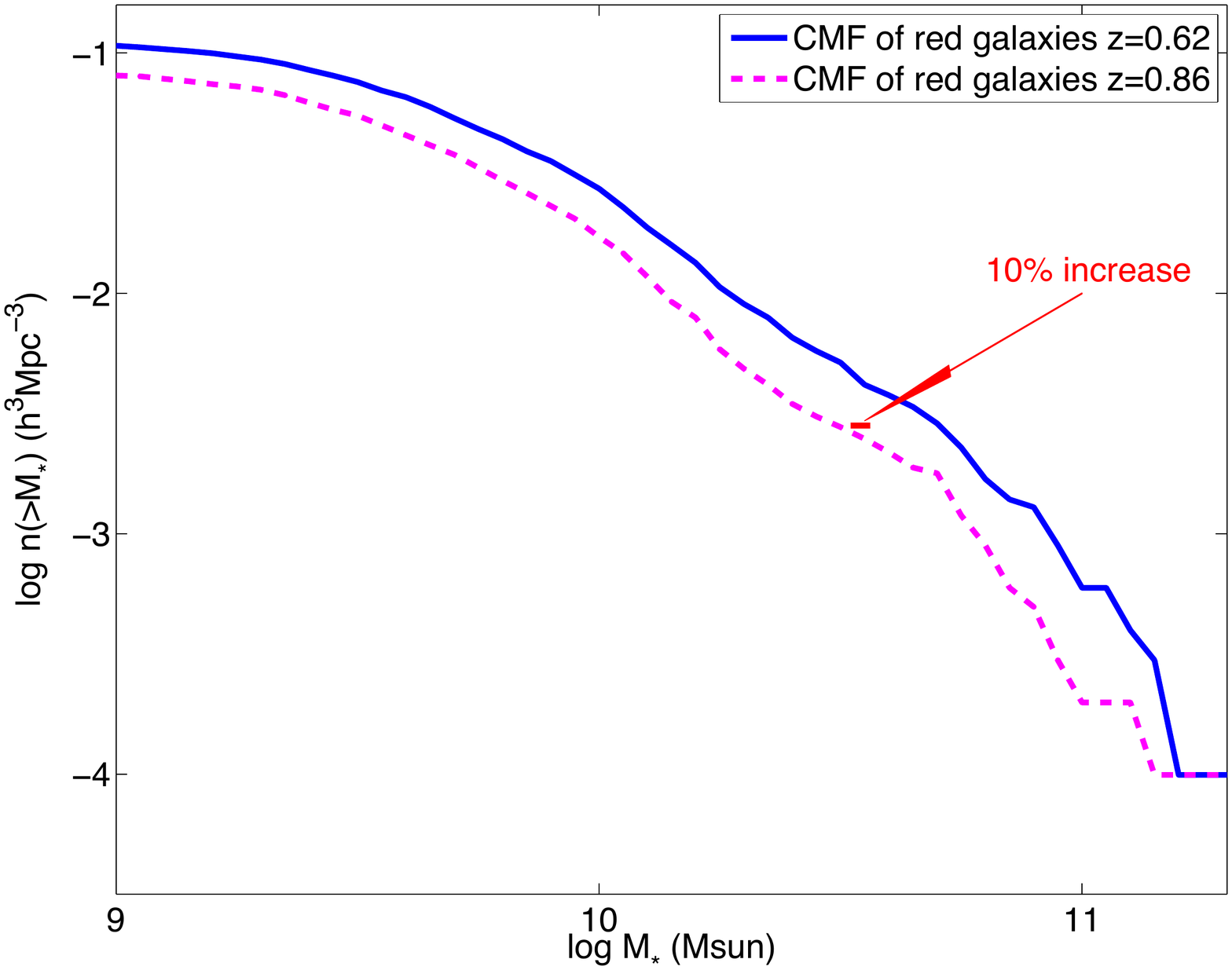}}      
\vskip -0.5cm
\caption{\footnotesize 
Left panel: the histogram of the ratio of stellar mass of red galaxies at $z=0.62$ to their progenitor's stellar mass at the onset of quenching $t_{\rm q}$,
for two stellar mass ranges of the red galaxies at $z=0.62$.
Right panel: cumulative stellar mass functions of red galaxies at $z=0.62$ (blue) and $z=1$ (magenta). 
For the red galaxies at $z=0.62$ we find that the median value of $t_{\rm q}$ corresponds to redshift $z\sim 0.86$,
thus the choice of $z=0.86$.
}
\label{fig:redmass}
\end{figure}

We address the stellar mass growth of red galaxies quantitatively in two different ways.
The left panel of Figure~\ref{fig:redmass} shows the histogram of the ratio of stellar mass 
of red galaxies at $z=0.62$ to their progenitor's stellar mass at the onset of quenching $t_{\rm q}$.
We see that the overall stellar mass growth of red galaxies since the onset of quenching 
is relatively moderate, with the vast majority of galaxies gaining less than 30\% of their 
stellar mass during this period,
consistent with observations \citep[e.g.,][]{2010Peng, 2012Peng}.
There is a non-negligible fraction of galaxies that experience a decline of stellar mass, due to tidal interactions and collisions.
There is $5-10\%$ of red galaxies that gain more than $\ge 40\%$ of their stellar mass during this period,
possibly due to mergers and accretion of satellite galaxies.
We do not address red galaxies more massive than $10^{12}\msun$ because of lack of a statistically significant red sample.
Since these larger galaxies tend to reside at the centers of groups and clusters,
there is a larger probability that AGN feedback may play a significant role in them.
Empirical evidence suggests that radio jets get extinguished in the near vicinity 
of the central galaxies in groups/clusters \citep[e.g.,][]{2007McNamara}, 
in sharp contrast to AGNs in isolated galaxies where jets, seen as large radio lobes,
 appear to deposit most of their energy on scales much larger than the star formation regions.
Thus, AGN feedback in the central massive galaxies in clusters/groups may be energetically important 
to have a major effect on gas cooling and star formation in them \citep[e.g.,][]{2004Omma}.
Thus, our neglect of AGN feedback in the simulation cautions us
to not draw any definitive conclusion with respect to this special class of galaxies at this time.

The stellar mass growth of individual red galaxies shown in the left panel of Figure~\ref{fig:redmass} 
contains very useful information.
However, it does not address a related but separate question:
How does the stellar mass function of red galaxies evolve with redshift?
We address this question here.
We compute the cumulative stellar mass function of red galaxies at $z=0.62$ and $z=0.86$, separately,
and show them in the right panel of Figure~\ref{fig:redmass}.
We see that for red galaxies with stellar masses greater than $\sim 3\times 10^{10}\msun$, when matched in abundance,
the stellar masses grow a factor of $\sim 1.6$ from $z=0.86$ to $z=0.62$, much larger than $10\%$ 
(for about 75\% of galaxies) seen in the left panel of Figure~\ref{fig:redmass}.
We refrain from making a direct comparison to observations in this case, 
because our limited simulation volume is highly biased with respect to the massive end of the mass function.

We strict ourselves to a comparative analysis of galaxies in our simulation volume
and ask the question of how red galaxies in our simulation volume grow with time.
The most important point to note is that this apparent growth of stellar mass of red galaxies based on abundance matching 
could not be due to growth of individual red galaxies in the red sequence,
since the actual stellar mass increase since the onset of quenching is moderate, $\le 10\%$ typically, 
seen in the left panel of Figure~\ref{fig:redmass}.
Physically, this suggests that dry mergers do not play a major role in the ``apparent" 
stellar mass growth of red galaxies, 
consistent with observations \citep[e.g.,][]{2007Pozzetti}.
Rather, galaxies grow their stellar mass when they are still in the blue cloud,
illustrated in Figure~\ref{fig:migrate}.

A physical picture of galaxy color migration emerges based on our results.
{\it The migration from the blue cloud to the red sequence proceeds in a staggered fashion: 
stellar masses of individual galaxies continuously grow, predominantly in the blue cloud,
and blue galaxies over the entire mass range continuously migrate into the red sequence over time.}
Galaxies migrate from the blue cloud to the red sequence almost vertically
in the usual color-magnitude diagram 
(see Figure~\ref{fig:migrate}).
For simplicity we will call this type of color migration ``Vertical Tracks",
which correspond most closely to ``B tracks" proposed by \citet[][]{2007Faber}, 
with the growth since the onset of quenching being moderate ($\le 30\%$).

\subsection{Galaxy Age-Mass and Age-Environment Relations}

\begin{figure}[ht]
\centering
\vskip -0.cm
\resizebox{6.5in}{!}{\includegraphics[angle=0]{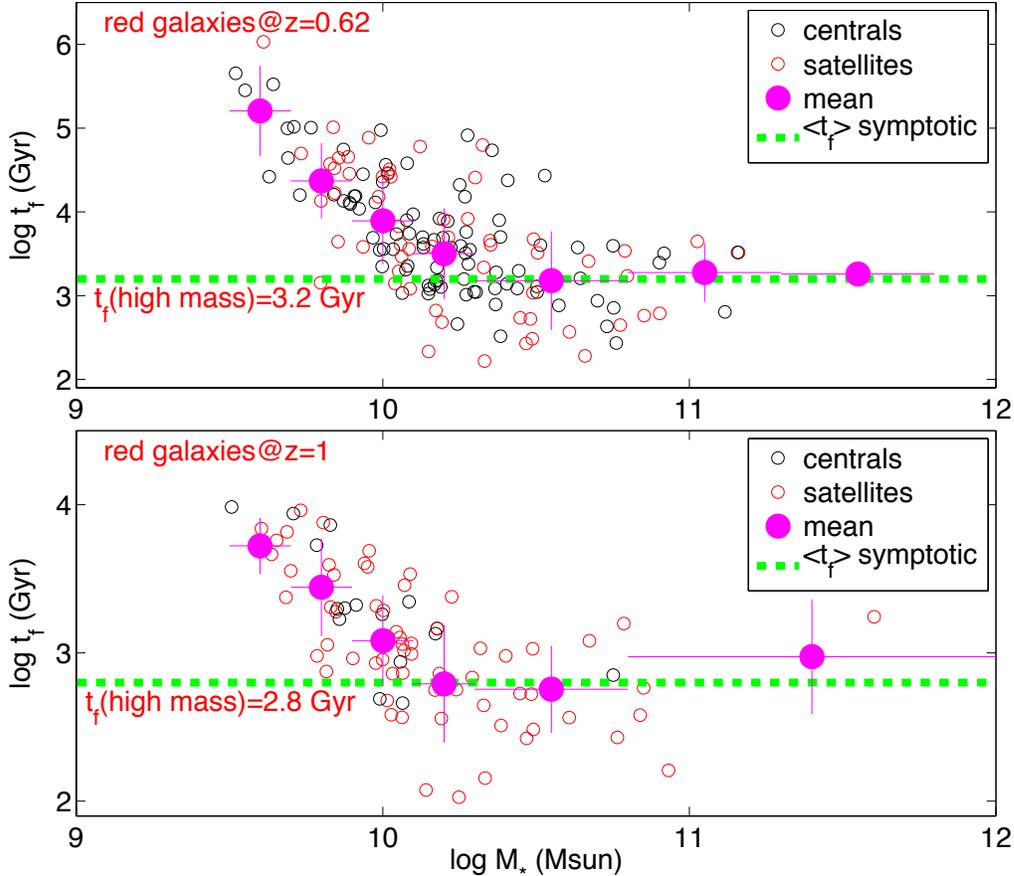}}   
\vskip -0.5cm
\caption{\footnotesize 
shows a scatter plot of red galaxies in the stellar mass $M_*$-mean galaxy formation time $t_f$ plane at $z=0.62$ (top)
and $z=1$ (bottom), where $t_f$ is stellar formation time, not lookback time.
The red galaxies are subdivided into two groups: centrals (black circles) and satellites (red circles).
For the purpose of comparison to observations, we only show
galaxies with high surface brightness of $<23$B~mag~arcsec$^{-2}$ \citep[e.g.,][]{1997Impey}.
The green horizontal dashed lines indicate the mean formation redshift of the most luminous red galaxies 
at the two redshifts.
The magenta dots are the averages of $t_{\rm f}$ at the stellar mass bins.
}
\label{fig:RedMstarage2panel}
\end{figure}

The vertical tracks found have many implications on observables.
The first question one asks is this:
if galaxies follow the vertical tracks, is the galaxy age-mass relation consistent with observations?
We address this question in this subsection.
Figure~\ref{fig:RedMstarage2panel} 
shows a scatter plot of red galaxies in the stellar mass $M_*$-mean galaxy formation time $t_f$ plane at $z=0.62$ (top)
and $z=1$ (bottom), where $t_f$ is stellar formation time, not lookback time.
The red galaxies are subdivided into two groups: centrals (black circles) and satellites (red circles).
For the purpose of comparison to observations, we only show
galaxies with high surface brightness of $\mu_B<23$~mag~arcsec$^{-2}$ \citep[e.g.,][]{1997Impey}.
Several interesting results can be learned.

First, no systematic difference between satellite and central galaxies is visible,
supporting earlier findings that there is no appreciable differences between satellites and centrals
with respect to duration from quenching to turning red $t_{\rm qr}$ (Figure~\ref{fig:tbrhis}).
Second, at any given redshift, the brightest red galaxies are relatively ``old" (but not necessarily the oldest), 
of ages of several billion years (age = $t_{\rm H}-t_{\rm f}$ and $t_{\rm H}=(7.85,5.94)$Gyr for $z=(0.62,1)$),
consistent with observations.
Third, at stellar masses greater than $10^{10.2-10.7}\msun$ red galaxies have a nearly uniform mean age;
the age spread at a given stellar mass of $\sim 1$Gyrs is consistent with observations \citep[e.g.,][]{2010Demarco}. 
Fourth, fainter red galaxies are younger than brighter red galaxies in the mass range $10^{9.5-10.5}\msun$;
we see that the age difference between the two ends of the mass range is $\sim 2.5$Gyr and $1.3$Gyr, respectively, at $z=0.62$ and $z=1$,
suggesting a steepening with decreasing redshift of the age difference between galaxies of different masses in the red sequence.
\citet[][]{2010Demarco} find an age difference between the faint and bright ends of red sequence galaxies
of $\sim 2$Gyr at $z=0.84$, in excellent agreement with our results.
The physical origin for the steepening with decreasing redshift of the age difference between galaxies of different masses in the red sequence
is traceable to the steepening of specific SFR with stellar mass with decreasing redshift
that is, in a fundamental way, related to the cosmic downsizing phenomenon \citep{2011bCen}.

\begin{figure}[ht]
\centering
\vskip -0.cm
\resizebox{6.0in}{!}{\includegraphics[angle=0]{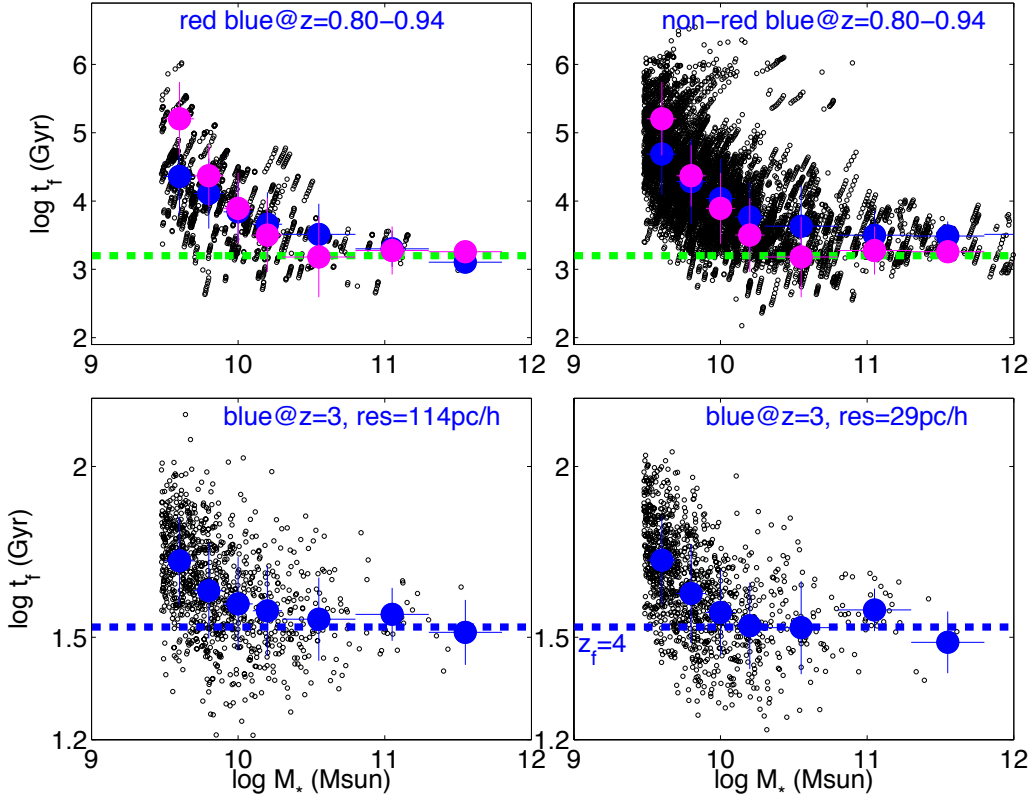}}   
\vskip -0.5cm
\caption{\footnotesize 
shows the stellar mass $M_*$-mean galaxy formation time $t_f$ scatter plot for 
blue galaxies at $z=0.80-0.94$ that become red galaxies (top left)
and that do not become red galaxies (top right).
Each small group of mostly linearly aligned circles is one galaxy that appears multiple 
times (maximum is 8).
The blue dots indicate average values.
The green horizontal dashed lines and the magenta dots are the same in the top panel of Figure (\ref{fig:RedMstarage2panel}),
indicating the mean formation time of the most luminous red galaxies and the average formation time of red galaxies at $z=0.62$. 
Bottom left panel: 
shows the stellar mass $M_*$-mean galaxy formation time $t_f$ scatter plot for 
blue galaxies at $z=3$ with the fiducial resolution $114$pc/h.
Bottom right panel: 
shows the stellar mass $M_*$-mean galaxy formation time $t_f$ scatter plot for 
blue galaxies at $z=3$ with the four times better resolution of $29$pc/h.
The blue dots indicate average values.
}
\label{fig:RBtf}
\end{figure}

It is interesting to note that, in Figure~\ref{fig:RedMstarage2panel},
scatters notwithstanding, there appears to be a critical stellar mass of $\sim 10^{10.2-10.7}\msun$, above which 
the age (or formation time) of red galaxies flattens out to a constant value.
At least for the redshift range that we have examined, $z=0.62-1$, 
this critical stellar mass appears to be redshift independent.
At still higher redshift we do not have enough statistics to see if this critical mass remains the same.
This critical mass is tantalizingly close to 
the division mass of $\sim 10^{10.5}\msun$ discovered by \citet[][]{2003Kauffmann} at low redshift,
which appears to demarcate a number of interesting trends in galaxy properties.
This physical origin of this mass is unclear and deferred to a future study.

Given the ``vertical tracks", i.e., lack of significant stellar mass growth subsequently to quenching,
one may ask this: is the age-mass relation of red galaxies inherited from their blue progenitors?
We will now address this question.
To select progenitors of red galaxies at $z=0.62$,
we note that the majority of galaxies that turn red by $z=0.62$ have $t_{\rm qr}=1-1.7$Gyr.
Thus, we choose galaxies in the redshift range $z=0.80-0.94$ (8 snapshots with $z=(0.80,0.82,0.84,0.86,0.88,0.90,0.92,0.94)$),
where the Hubble time differences between $z=0.62$ and $z=0.80$ and $z=0.94$ are $(1.0,1.7)$Gyr, respectively,
enclosing the vast majority of blue progenitors of red galaxies at $z=0.62$ near the onset of quenching.
We separate the blue galaxies into two groups: one group contains the blue progenitors of $z=0.62$ red galaxies
and the other group other blue galaxies that have not turned into red galaxies by $z=0.62$. 
Figure~\ref{fig:RBtf} shows 
the stellar mass $M_*$-mean galaxy formation time $t_f$ scatter plot for 
blue galaxies at $z=0.80-0.94$ that are progenitors of red galaxies at $z=0.62$ (top left)
and those that do not become red galaxies (top right).
Each small group of mostly linearly aligned circles is one galaxy that appears multiple times (maximum is 8).
Within the scatters we see that the green dashed line, borrowed from Figure~\ref{fig:RedMstarage2panel},
provides a good match to the near constant age at the high mass end for the progenitors of red galaxies.
The magenta dots, borrowed from Figure~\ref{fig:RedMstarage2panel},
match well the trend for the blue dots in the mass range $10^{9.5-10.5}\msun$.
These results are fully consistent with our initial expectation based on the observation (of our simulation) 
of two physical processes:
(1) that stellar mass growth is moderate during $t_{\rm qr}$ hence evolution during $t_{\rm qr}$ does not significantly alter 
the mean star formation time of each galaxy,
(2) less massive forming galaxies have higher sSFR than massive galaxies, 
causing a steepening of the age-mass relation at the low mass end.
This explains the physical origin of the age-mass relation seen in Figure~\ref{fig:RedMstarage2panel}.
It is prudent to make sure that these important general trends seen in the simulation are robust.
In the bottom two panels of Figure~\ref{fig:RBtf} we make a comparison between 
blue galaxies of two simulations with different resolutions, at $z=3$.
The bottom-left panel is from the fiducial simulation with a resolution of $114$pc/h and 
the bottom-right panel is from an identical simulation with four times better resolution of $29$pc/h.
We see that both the age-mass trend at low mass end and the near constancy of stellar age at the high mass end
are shared by the two simulations, suggesting that results from our fiducial simulation
are sufficiently converged for the general trends presented at the level of concerned accuracies.

A comparison between the top left and top right panels in 
Figure~\ref{fig:RBtf} makes it clear that
the age-mass relation of the blue progenitors
of red galaxies at quenching is, to a large degree, 
shared by blue galaxies that do not become red galaxies by $z=0.62$.
One subtle difference is that the most massive non-progenitor blue galaxies 
are slightly younger than the most massive progenitors of red galaxies,
suggesting that the blue progenitors of red galaxies, on ``their way" to become red galaxies,
have started to ``foreshadow" quenching effects mildly.

\subsection{Environmental Dependencies of Various Galaxy Populations}  

\begin{figure}[ht]
\centering
\vskip -0.5cm
\resizebox{4.0in}{!}{\includegraphics[angle=0]{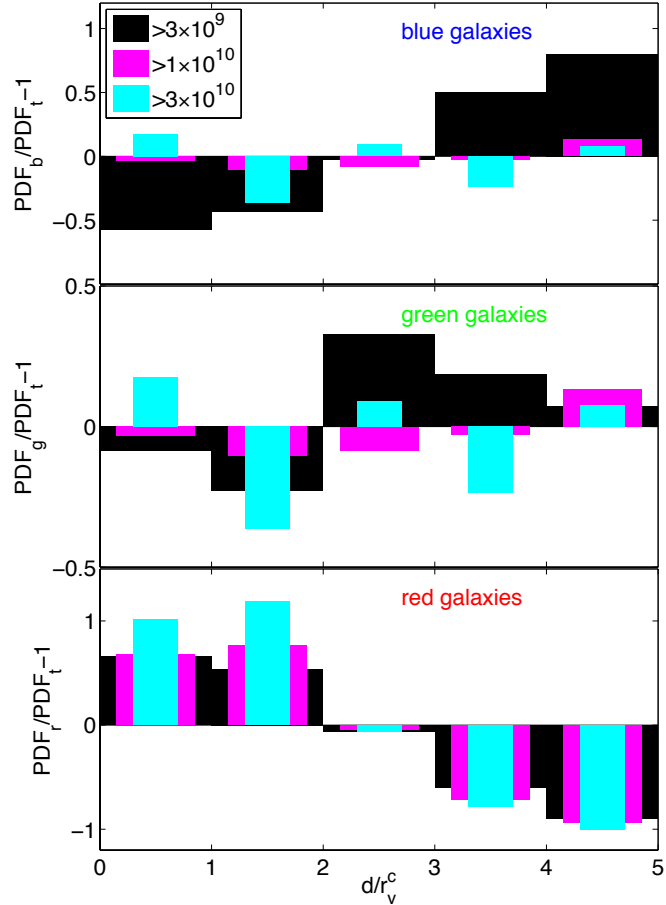}}   
\vskip -0.5cm
\caption{\footnotesize 
Top panel shows distribution of the distance to the nearest primary galaxy
for blue galaxies at $z=0.62$, ${\rm PDF}_{\rm b}/{\rm PDF}_{\rm t}-1$.
The middle panel shows the normalized distribution
of green galaxies, ${\rm PDF}_{\rm g}/{\rm PDF}_{\rm t}-1$.
The bottom panel shows the normalized distribution
of red galaxies, ${\rm PDF}_{\rm r}/{\rm PDF}_{\rm t}-1$.
All galaxies with distance larger than 4 virial radii of the primary galaxy are added to the bin with $d/r_v^c=4-5$.
}
\label{fig:fracdis}
\end{figure}

\begin{figure}[ht]
\centering
\vskip -0,5cm
\resizebox{4.0in}{!}{\includegraphics[angle=0]{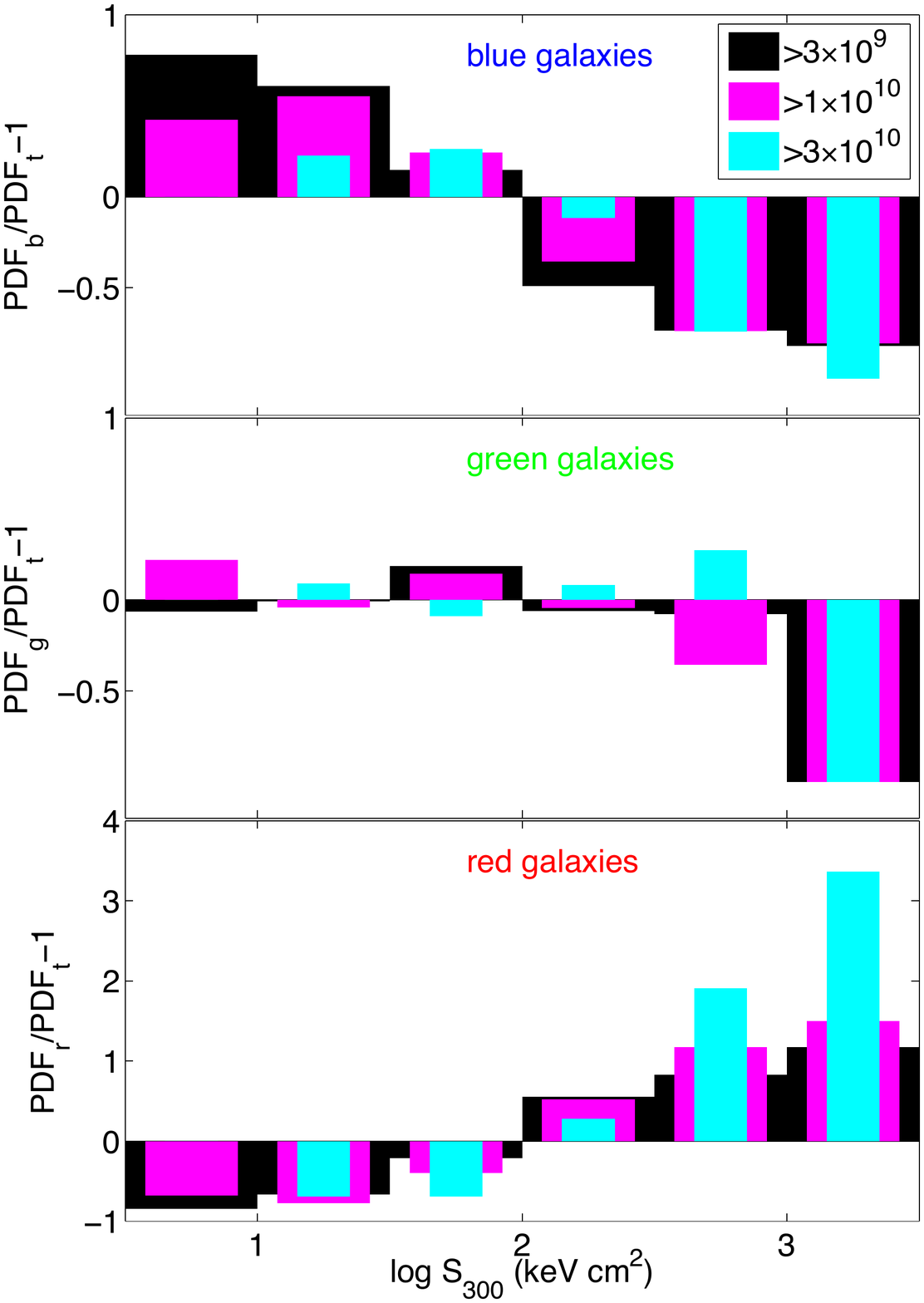}}   
\vskip -0,5cm
\caption{\footnotesize 
Top panel shows the normalized environmental entropy distribution of 
blue galaxies at $z=0.62$, ${\rm PDF}_{\rm b}/{\rm PDF}_{\rm t}-1$.
The middle panel shows the normalized difference distribution
of green and blue galaxies, ${\rm PDF}_{\rm g}/{\rm PDF}_{\rm b}-1$.
The bottom panel shows the normalized difference distribution
of red and blue galaxies, ${\rm PDF}_{\rm r}/{\rm PDF}_{\rm b}-1$.
}
\label{fig:fracent}
\end{figure}

At a given redshift the cumulative environmental effects are imprinted 
on the relative distribution of galaxies of different color types and
possibly on the properties of galaxies within each type.
We now present predictions of our simulations with respect to these aspects.
Figure~\ref{fig:fracdis} shows distributions of three types of galaxies as a function of distance to the primary galaxy 
in units of the virial radius of the primary galaxy at $z=0.62$.
All galaxies with distance larger than 4 virial radii of the primary galaxy are added to the bin with $d/r_v^c=4-5$.
We use the total galaxy population above the respective stellar mass threshold as a reference sample 
and distributions in the top (blue galaxies), 
middle (green galaxies) and bottom (red galaxies) panels are normalized relative to reference sample.
Comparing the top (blue galaxies), middle (green galaxies) and bottom (red galaxies) panels, 
we see clear differences of environmental dependencies of the three types of galaxies.
For blue galaxies, there is a deficit at $d/r_v^c\le 2$, which is compensated 
by a comparable excess at $d/r_v^c\ge 3$. 
The range $d/r_v^c=2-3$ seems to mark the region where an excess of green galaxies,
about one half of which will become red galaxies during the next $1-1.7$Gyr.
It is useful to recall that 
not all galaxies in the green valley will turn into red galaxies,
which perhaps has contributed in part to some of the ``irregularities" of
the distribution of the green galaxies (middle panel).
For red galaxies, we see a mirror image of blue galaxies: 
there is an excess at $d/r_v^c\le 2$ and a deficit $d/r_v^c\ge 3$.
This trend is in agreement with observational indications \citep[e.g.,][]{2013Woo}. 
The emerging picture found here that satellite quenching plays a dominant role in quenching galaxies
is in accord with observations \citep[e.g.,][]{2008vandenBosch, 2010Peng, 2012Peng}.

Figure~\ref{fig:fracent} shows distributions of three types of galaxies as a function of environmental entropy $S_{300}$.
We see that the excess of red galaxies starts at $S_{300}=100~{\rm keV~cm}^{2}$ and rises toward higher entropy regions for red galaxies.
The trend for blue galaxies is almost an inverted version of that for red galaxies.
The trend for green galaxies lie in-between those for blue and red galaxies, as expected.
In \citet{2011bCen} we put forth the notion 
that a critical entropy 
$S_{c}=100~{\rm keV~cm}^{2}$ (at $z=0.62$ and weakly dependent on redshift), 
marks a transition to a regime of inefficient gas cooling hence cold gas starvation,
because above this entropy the gas cooling time exceeds the Hubble time.
This is borne out with our more detailed analysis here.
We also plot (not shown here) 
distributions of three types of galaxies as a function of the environmental pressure $p_{300}$ and environmental overdensity $\delta_{2}$, respectively,
and find that the trend is broadly similar to that see in Figure~\ref{fig:fracent}.
Overall, our results are in accord with the observed density-morphology relation 
\citep[e.g.,][]{1974Oemler, 1980Dressler,1984Postman, 2006Cooper, 2007Tanaka, 2006Bundy, 2012Quadri, 2012Muzzin}, 
and with the general observed trend of galaxy population becoming bluer or mean/median specific star formation rate becoming higher 
towards underdense regions in the local universe \citep[e.g.,][]{2002Lewis, 2003Goto,2003Gomez,2004Tanaka,2004Rojas}.

\begin{figure}[ht]
\centering
\hskip -1.5cm
\resizebox{7.5in}{!}{\includegraphics[angle=0]{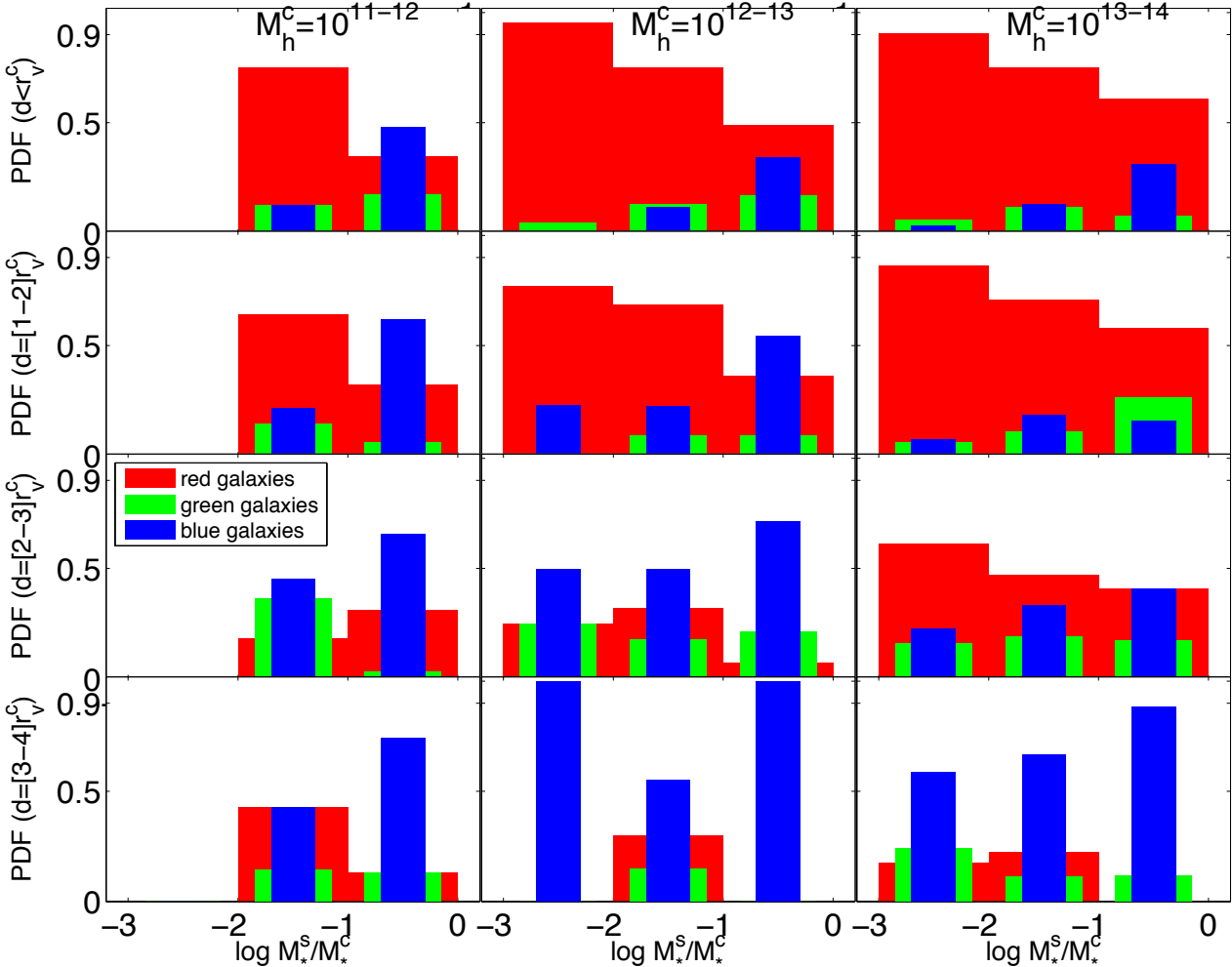}}    
\vskip -1.0cm
\caption{\footnotesize 
shows fractions of three populations of galaxies in terms of color, (red, green, blue), as a function of satellite to primary galaxy stellar mass ratio.
The (left, middle, right) columns are for primary halo mass of ($10^{11-12}\msun$,$10^{12-13}\msun$,$10^{13-14}\msun$).
The four rows from top to bottom are for satellites within 
four different radial shells centered on the primary galaxy ($\le r_v^c$, $[1-2]r_v^c$, $[2-3]r_v^c$, $[3-4]r_v^c$).
}
\label{fig:Msat}
\end{figure}

Having examined the dependencies of three types of galaxies on environmental variables, 
we now explore the dependencies on two additional variables: 
the mass of the halo of the primary galaxy and the secondary to primary galaxy stellar mass ratio.
Figure~\ref{fig:Msat} 
shows fractions of three populations of galaxies in terms of color (red, green, blue) as a function of secondary to primary galaxy stellar mass ratio.
The (left, middle, right) columns are for primary galaxies of halo masses in three ranges ($10^{11-12}\msun$,$10^{12-13}\msun$,$10^{13-14}\msun$) respectively.
The four rows from top to bottom are for secondaries within 
four different radial shells centered on the primary galaxy ($\le r_v^c$, $[1-2]r_v^c$, $[2-3]r_v^c$, $[3-4]r_v^c$).
We adopt the following language to make comparative statements:
the environment quenching is important if the fraction of blue galaxies is less than the fraction of red galaxies and vice versa.
We see two separate trends in Figure~\ref{fig:Msat}. 
First, more massive environments are more able to quench star formation;
for primary galaxies with halo masses in the range of $10^{13}-10^{14}\msun$
the quenching appears to extend at least to $[2-3]r_v^c$,
whereas for primary galaxies with lower halo masses 
the quenching effect is no longer significant at $[2-3]r_v^c$.
Second, for any given primary galaxy halo mass,
the quenching is more effective when the ratio of secondary to primary stellar mass is smaller.
That the impact of environmental effects increases with decreasing secondary to primary galaxy stellar mass ratio
is perhaps not surprising and consistent with observations \citep[e.g.,][]{2009Poggianti,2010Thomas, 2012Wetzel}, 
but at odds with the assumption of mass-independent environment quenching in empirical modeling \citep[e.g.,][]{2010Peng}. 
Our finding that quenching is still effective down to at least halo mass range 
$10^{11}-10^{12}\msun$ for the primary
is in agreement with observations \citep[e.g.,][]{2012Wetzel}. 

\citet[][]{2013Mok}, using deep GMOS-S spectroscopy for 11 galaxy groups at $z=0.8-1$,
show that the strongest environmental dependence is observed in the fraction of passive galaxies, 
which make up only $\sim 20$ percent of the field in the mass range $M_{\rm star}=10^{10.3}-10^{11.0}\msun$ 
but are the dominant component of groups. 
If we take the radial range $3-4r_v^c$ as the ``field", we see that the fraction of red galaxies 
is in the range $20-30\%$ (Figure~\ref{fig:Msat}),
reaching $65-90\%$ within $2 r_v^c$, in agreement with \citet[][]{2013Mok}.
In a large galaxy sample study of a $z=0.83$ cluster, \citet[][]{2009Patel} find that 
red galaxy fraction is $93\pm 3\%$ in the central region of the cluster and declines to a level of 
$64\pm 3\%$ at a projected cluster-centric radius of $\sim 3$Mpc.
In the right column of Figure~\ref{fig:Msat}, for halos in the mass range of $10^{13-14}\msun$,
we see that within the virial radius the red fractions are in the range of $60-90\%$, depending on the satellite to primary galaxy mass ratio,
with the overall mean of red fraction within the virial radius in the range $80-90\%$.
The more central region has a red fraction higher than $80-90\%$, consistent with their observation.
\citet[][]{2009Patel} also find that the environmental effect extends even beyond 3Mpc,
which is about 3 virial radii for their cluster.
From the right column in Figure~\ref{fig:Msat} we see that red galaxies dominate
up to 3 virial radii, in reasonable accord with observations.

Some of our results are not necessarily the most straightforward to imagine physically in the absence of simulations and detailed analysis.
For example, the interpretation by \citet[][]{2012Presotto} that the rate of SF quenching is faster in groups than in the field
may be interpreted, in a different way, in the context of our results.
The excess of quenched galaxies in group environments is larger than in
less dense environments as seen in Figures~(\ref{fig:fracdis},\ref{fig:fracent}),
consistent with observations \citep[][]{2012Presotto}.
However, we show that the quenching time depends weakly on environment as shown in Figure~\ref{fig:tbrhis}. 
In other words, our simulations suggest the following physical picture:
environment quenching of star formation in galaxies is {\it more efficient} in more hostile environment
(closer to a larger galaxy, higher density, high pressure, high entropy, etc), but {\it not faster, for the most part}.

\begin{figure}[ht]
\centering
\vskip -0.0cm
\resizebox{6.5in}{!}{\includegraphics[angle=0]{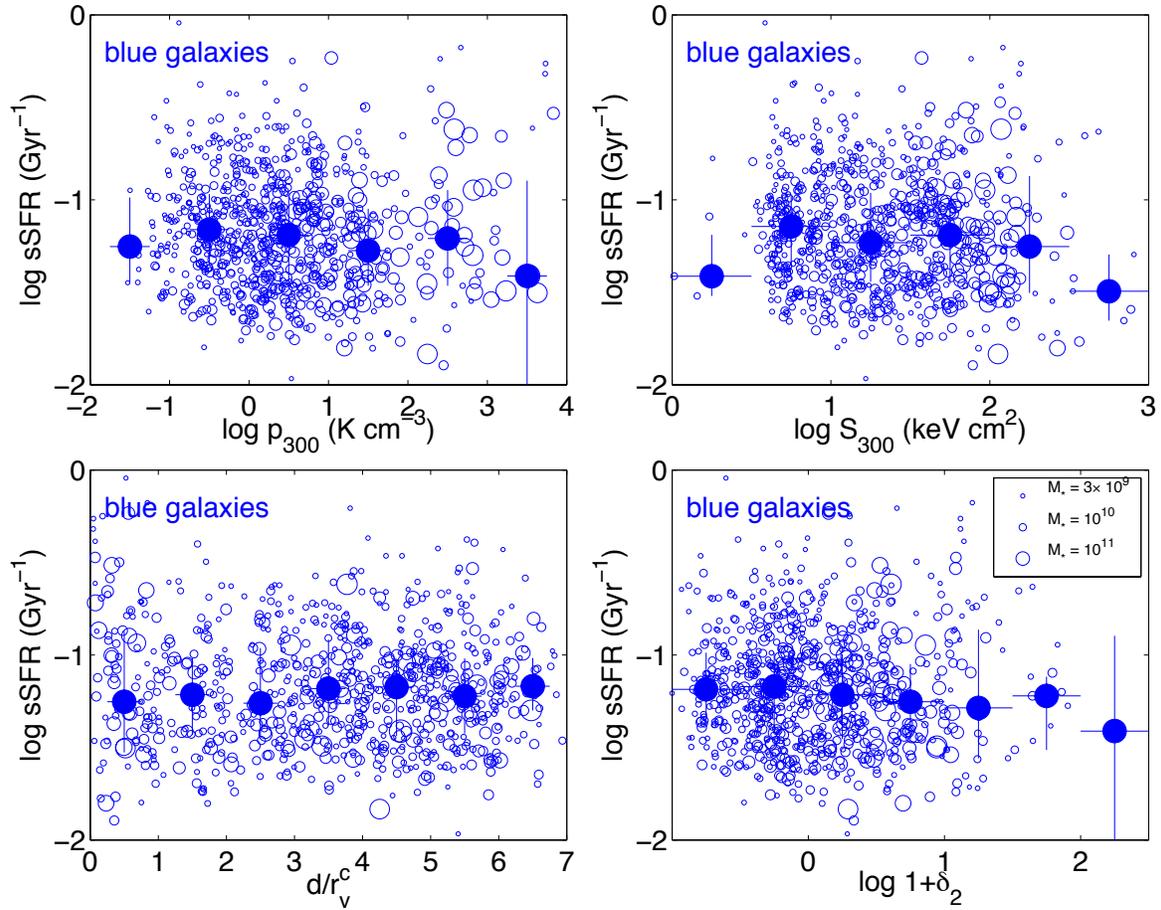}}   
\vskip -0.5cm
\caption{\footnotesize 
shows the specific SFR of blue galaxies at $z=0.62$ as a function of each of the four environmental variables.
Black circles are for central galaxies and red for satellites.
The size of each circle indicates the stellar mass 
of a galaxy, as shown in the legend.
The blue dots indicate median values.
}
\label{fig:SFR}
\end{figure}

Given the unambiguous environmental effects on the relative distributions of galaxy colors,
we ask the following reverse question:
Can we decipher environmental effects on a set of blue galaxies alone across different environments at any given epoch?
Figure (\ref{fig:SFR}) shows the specific SFR (sSFR) distributions of blue galaxies
as a function of the four environmental variables at $z=0.62$.
While one sees trends of the overall number of galaxies with respect to each
of the environmental variables, as already shown in Figures~(\ref{fig:fracdis},\ref{fig:fracent}),
there is no strong visible trend of the mean value of sSFR with respect to environment for blue galaxies.
To put it slightly differently, 
the variations of sSFR among blue galaxies at any given mass are sufficiently large 
and the quenching effects on galaxies at a given sSFR and stellar mass are sufficiently varied that 
at any environment the scatter in sSFR at a given stellar mass is larger than
the difference in average sSFR.
We attribute this result physically to the combined effects of substantial non-uniformity of the properties
of blue galaxies prior to entering the 
environmental ``spheres" of influence and 
substantial non-uniformity of the influence of the environment on galaxies illustrated in Figures~(\ref{fig:tauminSFQ} and \ref{fig:tauminSIGMA}).
This result - that the intrinsic properties of star-forming galaxies 
appear independent of environment - is, however, in accord with observations \citep[e.g.,][]{2012Ideue, 2012Wijesinghe}.

\section{Conclusions}

Utilizing {\it ab initio}
{\color{red}\bf L}arge-scale {\color{red}\bf A}daptive-mesh-refinement 
{\color{red}\bf O}mniscient {\color{red}\bf Z}oom-{\color{red}\bf I}n cosmological hydrodynamic simulations 
({\color{red}\bf LAOZI Simulations}) of the standard cold dark matter model,
we perform a chronological and statistical study of formation and evolution 
of ({\color{blue}1664}, {\color{green}367}, {\color{red}1296})
({\color{blue}blue}, {\color{green}green}, {\color{red} red}) galaxies of stellar mass $10^{9.5}-10^{12.5}\msun$ at $z=0.62$. 
The simulations have an ultra-high resolution of $\le 114$pc/h with an additional calibration run of resolution $29$pc/h.
The soundness of treatment of relevant physical processes,
especially the uncertain feedback processes from star formation, is checked and confidence built by 
comparisons with a large set of independent observations over a range of scales and redshift.
Here we examine some global trends of formation and evolution of galaxies,
focusing on galaxy star formation quenching and color migration.
We do not include AGN feedback, in part because of its large uncertainties and 
primarily because of our intention to focus on external effects.
Our main findings may be summarized in several points.

(1) Environment quenching is shown to be responsible for making the vast majority, if not all, of red galaxies.
Two environmental conditions appear to be necessary in the making of a typical red galaxy:
ram-pressure stripping and cold gas starvation.
The ram-pressure stripping disconnects a galaxy from its large-scale cold gas reservoir,
while existing cold gas in the central ($\sim 10$~kpc) region is unaffected by ram-pressure 
but consumed by diminishing SF with an exponential time of several hundred Myr.
Subsequently, gas starvation in the high-velocity environment guards against further significant SF,
pushing it through the green valley to the red sequence.
The total duration from the initial sharp drop of SFR to the time of entry into the red sequence 
is $\sim 0.6-3$Gyr (see Eq~\ref{eq:tqr} for a fitting formula) with the median of $1.2$Gyr for red galaxies at $z=0.62$.
We suggest that adopting a spread in quenching time in the semi-analytic modeling
may result in an improved treatment in that approach.

(2) The radius of a galaxy's quenching sphere depends on properties of the infalling galaxy being quenched,
causing large variations in the quenching time scales at a given radius
and large variations in the quenching radius at a given mass of infalling galaxy.
Nevertheless, the vast majority of red galaxies are found to be within 
three virial radii of a larger galaxy, at the onset of quenching when the specific star formation rate experiences
the sharpest decline to fall below $\sim (10^{-2}-10^{-1})$Gyr$^{-1}$ (depending on the redshift when it occurs).
The exponential decay time of SFR at the onset of quenching is, on average, about a factor of 
two shorter for events occurring within the virial radius of the quenching galaxy than for those
outside the virial radius of the quenching galaxy,
which may be evidence of enhanced quenching with the additional aid of tidal stripping when
the onset of quenching takes place within the virial radius.
However, it is stressed that the overall duration from the onset of quenching to the time entering the sequence
does not depend much on environment.

(3) Not all galaxies in the green valley migrate to the red sequence;
(40\%,40\%,48\%) of green valley galaxies of stellar mass $>(3\times 10^9,10^{10},3\times 10^{10})\msun$
do not proceed to become red galaxies at $z=0.62$ after having turned green for $\ge 2$Gyrs.
For those galaxies that are en route to the red sequence,
the time spent in the green valley is brief, typically of $300$Myr.

(4) Throughout the quenching period and the ensuing period in the red sequence
galaxies follow nearly vertical tracks in the color-stellar-mass diagram,
which correspond most closely to ``C tracks" proposed by \citet[][]{2007Faber}. 
In contrast, individual galaxies of all masses grow most of their stellar masses in the blue cloud, prior to the onset of quenching,  
and progressively more massive blue galaxies with already relatively old mean stellar ages continue to enter the red sequence.
Consequently, correlations among observables of red galaxies - such as the age-mass relation - are largely inherited from their blue progenitors.  
The age-mass relation of simulated red galaxies is found to be in good agreement with observations.

(5) While environmental effects are responsible for producing the environmental dependence 
of the color makeup of the galaxy population,
the average properties (e.g., SFR) of blue galaxies as a sub-population display little dependence on environment,
which is in agreement with observations.
Overall, the excess (deficit) of red (blue) galaxies occurs within about three virial radii, in good agreement
with a wide range of observations.
Our detailed examination suggests that the excess of quenched galaxies in progressively denser environment (groups, clusters, etc)
is, for the most part, a result of quenching {\it being more efficient},
{\it not faster}, on average, in denser environment.
Physically, this comes about because most of the time it takes to drive a blue star-forming galaxy to the red sequence
is spent during the starvation phase, not the initial gas removal phase by ram-pressure stripping that displays
a stronger dependence on environment.

\vskip 1cm

I would like to thank Claire Lackner for providing the SQL based merger tree construction software,
and Drs. John Wise, Matthew Turk and Sam Skillmans for help with analysis and visualization program yt \citep[][]{2011Turk}.
I would like to thank Michael Vogeley for a very careful reading of the manuscript, many colleagues for useful discussions,
and an anonymous referee for a critical and constructive report.
Computing resources were in part provided by the NASA High-
End Computing (HEC) Program through the NASA Advanced
Supercomputing (NAS) Division at Ames Research Center.
This work is supported in part by grant NASA NNX11AI23G.
The simulation data are available from the author upon request.


\begin{thebibliography}{117}
\expandafter\ifx\csname natexlab\endcsname\relax\def\natexlab#1{#1}\fi

\bibitem[{{Abel} {et~al.}(1997){Abel}, {Anninos}, {Zhang}, \&
  {Norman}}]{1997Abel}
{Abel}, T., {Anninos}, P., {Zhang}, Y., \& {Norman}, M.~L. 1997, New Astronomy,
  2, 181

\bibitem[{{Aird} {et~al.}(2012){Aird}, {Coil}, {Moustakas}, {Blanton},
  {Burles}, {Cool}, {Eisenstein}, {Smith}, {Wong}, \& {Zhu}}]{2012Aird}
{Aird}, J., {Coil}, A.~L., {Moustakas}, J., {Blanton}, M.~R., {Burles}, S.~M.,
  {Cool}, R.~J., {Eisenstein}, D.~J., {Smith}, M.~S.~M., {Wong}, K.~C., \&
  {Zhu}, G. 2012, \apj, 746, 90

\bibitem[{{Baldry} {et~al.}(2004){Baldry}, {Glazebrook}, {Brinkmann},
  {Ivezi{\'c}}, {Lupton}, {Nichol}, \& {Szalay}}]{2004Baldry}
{Baldry}, I.~K., {Glazebrook}, K., {Brinkmann}, J., {Ivezi{\'c}}, {\v Z}.,
  {Lupton}, R.~H., {Nichol}, R.~C., \& {Szalay}, A.~S. 2004, \apj, 600, 681

\bibitem[{{Balogh} {et~al.}(2004){Balogh}, {Baldry}, {Nichol}, {Miller},
  {Bower}, \& {Glazebrook}}]{2004Balogh}
{Balogh}, M.~L., {Baldry}, I.~K., {Nichol}, R., {Miller}, C., {Bower}, R., \&
  {Glazebrook}, K. 2004, \apjl, 615, L101

\bibitem[{{Balogh} {et~al.}(2000){Balogh}, {Navarro}, \& {Morris}}]{2000Balogh}
{Balogh}, M.~L., {Navarro}, J.~F., \& {Morris}, S.~L. 2000, \apj, 540, 113

\bibitem[{{Bekki}(2009)}]{2009Bekki}
{Bekki}, K. 2009, \mnras, 399, 2221

\bibitem[{{Bell} {et~al.}(2004){Bell}, {Wolf}, {Meisenheimer}, {Rix}, {Borch},
  {Dye}, {Kleinheinrich}, {Wisotzki}, \& {McIntosh}}]{2004Bell}
{Bell}, E.~F., {Wolf}, C., {Meisenheimer}, K., {Rix}, H., {Borch}, A., {Dye},
  S., {Kleinheinrich}, M., {Wisotzki}, L., \& {McIntosh}, D.~H. 2004, \apj,
  608, 752

\bibitem[{{Binney}(1977)}]{1977Binney}
{Binney}, J. 1977, \apj, 215, 483

\bibitem[{{Blanton} {et~al.}(2003{\natexlab{a}}){Blanton}, {Hogg}, {Bahcall},
  {Baldry}, {Brinkmann}, {Csabai}, {Eisenstein}, {Fukugita}, {Gunn},
  {Ivezi{\'c}}, {Lamb}, {Lupton}, {Loveday}, {Munn}, {Nichol}, {Okamura},
  {Schlegel}, {Shimasaku}, {Strauss}, {Vogeley}, \& {Weinberg}}]{2003bBlanton}
{Blanton}, M.~R., {Hogg}, D.~W., {Bahcall}, N.~A., {Baldry}, I.~K.,
  {Brinkmann}, J., {Csabai}, I., {Eisenstein}, D., {Fukugita}, M., {Gunn},
  J.~E., {Ivezi{\'c}}, {\v Z}., {Lamb}, D.~Q., {Lupton}, R.~H., {Loveday}, J.,
  {Munn}, J.~A., {Nichol}, R.~C., {Okamura}, S., {Schlegel}, D.~J.,
  {Shimasaku}, K., {Strauss}, M.~A., {Vogeley}, M.~S., \& {Weinberg}, D.~H.
  2003{\natexlab{a}}, \apj, 594, 186

\bibitem[{{Blanton} {et~al.}(2003{\natexlab{b}}){Blanton}, {Hogg}, {Bahcall},
  {Brinkmann}, {Britton}, {Connolly}, {Csabai}, {Fukugita}, {Loveday},
  {Meiksin}, {Munn}, {Nichol}, {Okamura}, {Quinn}, {Schneider}, {Shimasaku},
  {Strauss}, {Tegmark}, {Vogeley}, \& {Weinberg}}]{2003Blanton}
{Blanton}, M.~R., {Hogg}, D.~W., {Bahcall}, N.~A., {Brinkmann}, J., {Britton},
  M., {Connolly}, A.~J., {Csabai}, I., {Fukugita}, M., {Loveday}, J.,
  {Meiksin}, A., {Munn}, J.~A., {Nichol}, R.~C., {Okamura}, S., {Quinn}, T.,
  {Schneider}, D.~P., {Shimasaku}, K., {Strauss}, M.~A., {Tegmark}, M.,
  {Vogeley}, M.~S., \& {Weinberg}, D.~H. 2003{\natexlab{b}}, \apj, 592, 819

\bibitem[{{Bongiorno} {et~al.}(2012){Bongiorno}, {Merloni}, {Brusa},
  {Magnelli}, {Salvato}, {Mignoli}, {Zamorani}, {Fiore}, {Rosario}, {Mainieri},
  {Hao}, {Comastri}, {Vignali}, {Balestra}, {Bardelli}, {Berta}, {Civano},
  {Kampczyk}, {Le Floc'h}, {Lusso}, {Lutz}, {Pozzetti}, {Pozzi}, {Riguccini},
  {Shankar}, \& {Silverman}}]{2012Bongiorno}
{Bongiorno}, A., {Merloni}, A., {Brusa}, M., {Magnelli}, B., {Salvato}, M.,
  {Mignoli}, M., {Zamorani}, G., {Fiore}, F., {Rosario}, D., {Mainieri}, V.,
  {Hao}, H., {Comastri}, A., {Vignali}, C., {Balestra}, I., {Bardelli}, S.,
  {Berta}, S., {Civano}, F., {Kampczyk}, P., {Le Floc'h}, E., {Lusso}, E.,
  {Lutz}, D., {Pozzetti}, L., {Pozzi}, F., {Riguccini}, L., {Shankar}, F., \&
  {Silverman}, J. 2012, \mnras, 427, 3103

\bibitem[{{Bruzual} \& {Charlot}(2003)}]{Bruzual03}
{Bruzual}, G., \& {Charlot}, S. 2003, \mnras, 344, 1000

\bibitem[{Bryan \& Norman(1999)}]{1999bBryan}
Bryan, G.~L., \& Norman, M.~L. 1999, in Structured Adaptive Mesh Refinement
  Grid Methods, ed. N.~P.~C. S.~B.~Baden (IMA Volumes on Structured Adaptive
  Mesh Refinement Methods, No. 117), 165

\bibitem[{{Bundy} {et~al.}(2006){Bundy}, {Ellis}, {Conselice}, {Taylor},
  {Cooper}, {Willmer}, {Weiner}, {Coil}, {Noeske}, \& {Eisenhardt}}]{2006Bundy}
{Bundy}, K., {Ellis}, R.~S., {Conselice}, C.~J., {Taylor}, J.~E., {Cooper},
  M.~C., {Willmer}, C.~N.~A., {Weiner}, B.~J., {Coil}, A.~L., {Noeske}, K.~G.,
  \& {Eisenhardt}, P.~R.~M. 2006, \apj, 651, 120

\bibitem[{{Bundy} {et~al.}(2008){Bundy}, {Georgakakis}, {Nandra}, {Ellis},
  {Conselice}, {Laird}, {Coil}, {Cooper}, {Faber}, {Newman}, {Pierce},
  {Primack}, \& {Yan}}]{2008Bundy}
{Bundy}, K., {Georgakakis}, A., {Nandra}, K., {Ellis}, R.~S., {Conselice},
  C.~J., {Laird}, E., {Coil}, A., {Cooper}, M.~C., {Faber}, S.~M., {Newman},
  J.~A., {Pierce}, C.~M., {Primack}, J.~R., \& {Yan}, R. 2008, \apj, 681, 931

\bibitem[{{Cen}(2011{\natexlab{a}})}]{2011bCen}
{Cen}, R. 2011{\natexlab{a}}, \apj, 741, 99

\bibitem[{{Cen}(2011{\natexlab{b}})}]{2011cCen}
---. 2011{\natexlab{b}}, \apjl, 742, L33

\bibitem[{{Cen}(2012{\natexlab{a}})}]{2012bCen}
---. 2012{\natexlab{a}}, \apj, 753, 17

\bibitem[{{Cen}(2012{\natexlab{b}})}]{2012Cen}
---. 2012{\natexlab{b}}, \apj, 748, 121

\bibitem[{{Cen}(2013)}]{2013Cen}
---. 2013, ArXiv e-prints

\bibitem[{{Cen} {et~al.}(1995){Cen}, {Kang}, {Ostriker}, \& {Ryu}}]{1995Cen}
{Cen}, R., {Kang}, H., {Ostriker}, J.~P., \& {Ryu}, D. 1995, \apj, 451, 436

\bibitem[{{Cen} {et~al.}(2005){Cen}, {Nagamine}, \& {Ostriker}}]{2005Cen}
{Cen}, R., {Nagamine}, K., \& {Ostriker}, J.~P. 2005, \apj, 635, 86

\bibitem[{{Cen} \& {Ostriker}(1992)}]{1992CenOstriker}
{Cen}, R., \& {Ostriker}, J.~P. 1992, \apjl, 399, L113

\bibitem[{{Cen} \& {Riquelme}(2008)}]{2008Cen}
{Cen}, R., \& {Riquelme}, M.~A. 2008, \apj, 674, 644

\bibitem[{{Cheung} {et~al.}(2012){Cheung}, {Faber}, {Koo}, {Dutton}, {Simard},
  {McGrath}, {Huang}, {Bell}, {Dekel}, {Fang}, {Salim}, {Barro}, {Bundy},
  {Coil}, {Cooper}, {Conselice}, {Davis}, {Dom{\'{\i}}nguez}, {Kassin},
  {Kocevski}, {Koekemoer}, {Lin}, {Lotz}, {Newman}, {Phillips}, {Rosario},
  {Weiner}, \& {Willmer}}]{2012Cheung}
{Cheung}, E., {Faber}, S.~M., {Koo}, D.~C., {Dutton}, A.~A., {Simard}, L.,
  {McGrath}, E.~J., {Huang}, J.-S., {Bell}, E.~F., {Dekel}, A., {Fang}, J.~J.,
  {Salim}, S., {Barro}, G., {Bundy}, K., {Coil}, A.~L., {Cooper}, M.~C.,
  {Conselice}, C.~J., {Davis}, M., {Dom{\'{\i}}nguez}, A., {Kassin}, S.~A.,
  {Kocevski}, D.~D., {Koekemoer}, A.~M., {Lin}, L., {Lotz}, J.~M., {Newman},
  J.~A., {Phillips}, A.~C., {Rosario}, D.~J., {Weiner}, B.~J., \& {Willmer},
  C.~N.~A. 2012, \apj, 760, 131

\bibitem[{{Coil} {et~al.}(2008){Coil}, {Newman}, {Croton}, {Cooper}, {Davis},
  {Faber}, {Gerke}, {Koo}, {Padmanabhan}, {Wechsler}, \& {Weiner}}]{2008Coil}
{Coil}, A.~L., {Newman}, J.~A., {Croton}, D., {Cooper}, M.~C., {Davis}, M.,
  {Faber}, S.~M., {Gerke}, B.~F., {Koo}, D.~C., {Padmanabhan}, N., {Wechsler},
  R.~H., \& {Weiner}, B.~J. 2008, \apj, 672, 153

\bibitem[{{Cooper} {et~al.}(2006){Cooper}, {Newman}, {Croton}, {Weiner},
  {Willmer}, {Gerke}, {Madgwick}, {Faber}, {Davis}, {Coil}, {Finkbeiner},
  {Guhathakurta}, \& {Koo}}]{2006Cooper}
{Cooper}, M.~C., {Newman}, J.~A., {Croton}, D.~J., {Weiner}, B.~J., {Willmer},
  C.~N.~A., {Gerke}, B.~F., {Madgwick}, D.~S., {Faber}, S.~M., {Davis}, M.,
  {Coil}, A.~L., {Finkbeiner}, D.~P., {Guhathakurta}, P., \& {Koo}, D.~C. 2006,
  \mnras, 370, 198

\bibitem[{{Dalgarno} \& {McCray}(1972)}]{1972Dalgarno}
{Dalgarno}, A., \& {McCray}, R.~A. 1972, \araa, 10, 375

\bibitem[{{Danforth} \& {Shull}(2008)}]{2008Danforth}
{Danforth}, C.~W., \& {Shull}, J.~M. 2008, \apj, 679, 194

\bibitem[{{Davis} \& {Geller}(1976)}]{1976Davis}
{Davis}, M., \& {Geller}, M.~J. 1976, \apj, 208, 13

\bibitem[{{Dekel} \& {Birnboim}(2006)}]{2006Dekel}
{Dekel}, A., \& {Birnboim}, Y. 2006, \mnras, 368, 2

\bibitem[{{Demarco} {et~al.}(2010){Demarco}, {Gobat}, {Rosati}, {Lidman},
  {Rettura}, {Nonino}, {van der Wel}, {Jee}, {Blakeslee}, {Ford}, \&
  {Postman}}]{2010Demarco}
{Demarco}, R., {Gobat}, R., {Rosati}, P., {Lidman}, C., {Rettura}, A.,
  {Nonino}, M., {van der Wel}, A., {Jee}, M.~J., {Blakeslee}, J.~P., {Ford},
  H.~C., \& {Postman}, M. 2010, \apj, 725, 1252

\bibitem[{{Dressler}(1980)}]{1980Dressler}
{Dressler}, A. 1980, \apj, 236, 351

\bibitem[{Eisenstein \& Hu(1999)}]{1999Eisenstein}
Eisenstein, D., \& Hu, P. 1999, ApJ, 511, 5

\bibitem[{{Faber} {et~al.}(2007){Faber}, {Willmer}, {Wolf}, {Koo}, {Weiner},
  {Newman}, {Im}, {Coil}, {Conroy}, {Cooper}, {Davis}, {Finkbeiner}, {Gerke},
  {Gebhardt}, {Groth}, {Guhathakurta}, {Harker}, {Kaiser}, {Kassin},
  {Kleinheinrich}, {Konidaris}, {Kron}, {Lin}, {Luppino}, {Madgwick},
  {Meisenheimer}, {Noeske}, {Phillips}, {Sarajedini}, {Schiavon}, {Simard},
  {Szalay}, {Vogt}, \& {Yan}}]{2007Faber}
{Faber}, S.~M., {Willmer}, C.~N.~A., {Wolf}, C., {Koo}, D.~C., {Weiner}, B.~J.,
  {Newman}, J.~A., {Im}, M., {Coil}, A.~L., {Conroy}, C., {Cooper}, M.~C.,
  {Davis}, M., {Finkbeiner}, D.~P., {Gerke}, B.~F., {Gebhardt}, K., {Groth},
  E.~J., {Guhathakurta}, P., {Harker}, J., {Kaiser}, N., {Kassin}, S.,
  {Kleinheinrich}, M., {Konidaris}, N.~P., {Kron}, R.~G., {Lin}, L., {Luppino},
  G., {Madgwick}, D.~S., {Meisenheimer}, K., {Noeske}, K.~G., {Phillips},
  A.~C., {Sarajedini}, V.~L., {Schiavon}, R.~P., {Simard}, L., {Szalay}, A.~S.,
  {Vogt}, N.~P., \& {Yan}, R. 2007, \apj, 665, 265

\bibitem[{{Feldmann} {et~al.}(2011){Feldmann}, {Carollo}, \&
  {Mayer}}]{2011Feldmann}
{Feldmann}, R., {Carollo}, C.~M., \& {Mayer}, L. 2011, \apj, 736, 88

\bibitem[{{Gavazzi} {et~al.}(2013){Gavazzi}, {Savorgnan}, {Fossati}, {Dotti},
  {Fumagalli}, {Boselli}, {Guti{\'e}rrez}, {Hern{\'a}ndez Toledo},
  {Giovanelli}, \& {Haynes}}]{2013Gavazzi}
{Gavazzi}, G., {Savorgnan}, G., {Fossati}, M., {Dotti}, M., {Fumagalli}, M.,
  {Boselli}, A., {Guti{\'e}rrez}, L., {Hern{\'a}ndez Toledo}, H., {Giovanelli},
  R., \& {Haynes}, M.~P. 2013, \aap, 553, A90

\bibitem[{{G{\'o}mez} {et~al.}(2003){G{\'o}mez}, {Nichol}, {Miller}, {Balogh},
  {Goto}, {Zabludoff}, {Romer}, {Bernardi}, {Sheth}, {Hopkins}, {Castander},
  {Connolly}, {Schneider}, {Brinkmann}, {Lamb}, {SubbaRao}, \&
  {York}}]{2003Gomez}
{G{\'o}mez}, P.~L., {Nichol}, R.~C., {Miller}, C.~J., {Balogh}, M.~L., {Goto},
  T., {Zabludoff}, A.~I., {Romer}, A.~K., {Bernardi}, M., {Sheth}, R.,
  {Hopkins}, A.~M., {Castander}, F.~J., {Connolly}, A.~J., {Schneider}, D.~P.,
  {Brinkmann}, J., {Lamb}, D.~Q., {SubbaRao}, M., \& {York}, D.~G. 2003, \apj,
  584, 210

\bibitem[{{Goto} {et~al.}(2003){Goto}, {Yamauchi}, {Fujita}, {Okamura},
  {Sekiguchi}, {Smail}, {Bernardi}, \& {Gomez}}]{2003Goto}
{Goto}, T., {Yamauchi}, C., {Fujita}, Y., {Okamura}, S., {Sekiguchi}, M.,
  {Smail}, I., {Bernardi}, M., \& {Gomez}, P.~L. 2003, \mnras, 346, 601

\bibitem[{{Gunn} \& {Gott}(1972)}]{1972Gunn}
{Gunn}, J.~E., \& {Gott}, J.~R.~I. 1972, \apj, 176, 1

\bibitem[{{Haardt} \& {Madau}(1996)}]{1996Haardt}
{Haardt}, F., \& {Madau}, P. 1996, \apj, 461, 20

\bibitem[{{Harrison} {et~al.}(2012){Harrison}, {Alexander}, {Mullaney},
  {Altieri}, {Coia}, {Charmandaris}, {Daddi}, {Dannerbauer}, {Dasyra}, {Del
  Moro}, {Dickinson}, {Hickox}, {Ivison}, {Kartaltepe}, {Le Floc'h}, {Leiton},
  {Magnelli}, {Popesso}, {Rovilos}, {Rosario}, \& {Swinbank}}]{2012Harrison}
{Harrison}, C.~M., {Alexander}, D.~M., {Mullaney}, J.~R., {Altieri}, B.,
  {Coia}, D., {Charmandaris}, V., {Daddi}, E., {Dannerbauer}, H., {Dasyra}, K.,
  {Del Moro}, A., {Dickinson}, M., {Hickox}, R.~C., {Ivison}, R.~J.,
  {Kartaltepe}, J., {Le Floc'h}, E., {Leiton}, R., {Magnelli}, B., {Popesso},
  P., {Rovilos}, E., {Rosario}, D., \& {Swinbank}, A.~M. 2012, \apjl, 760, L15

\bibitem[{{Heckman}(2001)}]{2001Heckman}
{Heckman}, T.~M. 2001, in Astronomical Society of the Pacific Conference
  Series, Vol. 240, Gas and Galaxy Evolution, ed. J.~E. {Hibbard}, M.~{Rupen},
  \& J.~H. {van Gorkom}, 345

\bibitem[{{Hogg} {et~al.}(2003){Hogg}, {Blanton}, {Eisenstein}, {Gunn},
  {Schlegel}, {Zehavi}, {Bahcall}, {Brinkmann}, {Csabai}, {Schneider},
  {Weinberg}, \& {York}}]{2003Hogg}
{Hogg}, D.~W., {Blanton}, M.~R., {Eisenstein}, D.~J., {Gunn}, J.~E.,
  {Schlegel}, D.~J., {Zehavi}, I., {Bahcall}, N.~A., {Brinkmann}, J., {Csabai},
  I., {Schneider}, D.~P., {Weinberg}, D.~H., \& {York}, D.~G. 2003, \apjl, 585,
  L5

\bibitem[{{Ideue} {et~al.}(2012){Ideue}, {Taniguchi}, {Nagao}, {Shioya},
  {Kajisawa}, {Trump}, {Vergani}, {Iovino}, {Koekemoer}, {Le F{\`e}vre},
  {Ilbert}, \& {Scoville}}]{2012Ideue}
{Ideue}, Y., {Taniguchi}, Y., {Nagao}, T., {Shioya}, Y., {Kajisawa}, M.,
  {Trump}, J.~R., {Vergani}, D., {Iovino}, A., {Koekemoer}, A.~M., {Le
  F{\`e}vre}, O., {Ilbert}, O., \& {Scoville}, N.~Z. 2012, \apj, 747, 42

\bibitem[{{Impey} \& {Bothun}(1997)}]{1997Impey}
{Impey}, C., \& {Bothun}, G. 1997, \araa, 35, 267

\bibitem[{{Joung} {et~al.}(2009){Joung}, {Cen}, \& {Bryan}}]{2009Joung}
{Joung}, M.~R., {Cen}, R., \& {Bryan}, G.~L. 2009, \apjl, 692, L1

\bibitem[{{Kauffmann} {et~al.}(2003){Kauffmann}, {Heckman}, {White}, {Charlot},
  {Tremonti}, {Peng}, {Seibert}, {Brinkmann}, {Nichol}, {SubbaRao}, \&
  {York}}]{2003Kauffmann}
{Kauffmann}, G., {Heckman}, T.~M., {White}, S.~D.~M., {Charlot}, S.,
  {Tremonti}, C., {Peng}, E.~W., {Seibert}, M., {Brinkmann}, J., {Nichol},
  R.~C., {SubbaRao}, M., \& {York}, D. 2003, \mnras, 341, 54

\bibitem[{{Kauffmann} {et~al.}(2004){Kauffmann}, {White}, {Heckman},
  {M{\'e}nard}, {Brinchmann}, {Charlot}, {Tremonti}, \&
  {Brinkmann}}]{2004Kauffmann}
{Kauffmann}, G., {White}, S.~D.~M., {Heckman}, T.~M., {M{\'e}nard}, B.,
  {Brinchmann}, J., {Charlot}, S., {Tremonti}, C., \& {Brinkmann}, J. 2004,
  \mnras, 353, 713

\bibitem[{{Kennicutt}(1998)}]{1998Kennicutt}
{Kennicutt}, Jr., R.~C. 1998, \apj, 498, 541

\bibitem[{{Kimm} {et~al.}(2009){Kimm}, {Somerville}, {Yi}, {van den Bosch},
  {Salim}, {Fontanot}, {Monaco}, {Mo}, {Pasquali}, {Rich}, \&
  {Yang}}]{2009Kimm}
{Kimm}, T., {Somerville}, R.~S., {Yi}, S.~K., {van den Bosch}, F.~C., {Salim},
  S., {Fontanot}, F., {Monaco}, P., {Mo}, H., {Pasquali}, A., {Rich}, R.~M., \&
  {Yang}, X. 2009, \mnras, 394, 1131

\bibitem[{{Knobel} {et~al.}(2013){Knobel}, {Lilly}, {Kova{\v c}}, {Peng},
  {Bschorr}, {Carollo}, {Contini}, {Kneib}, {Le Fevre}, {Mainieri}, {Renzini},
  {Scodeggio}, {Zamorani}, {Bardelli}, {Bolzonella}, {Bongiorno}, {Caputi},
  {Cucciati}, {de la Torre}, {de Ravel}, {Franzetti}, {Garilli}, {Iovino},
  {Kampczyk}, {Lamareille}, {Le Borgne}, {Le Brun}, {Maier}, {Mignoli},
  {Pello}, {Perez Montero}, {Presotto}, {Silverman}, {Tanaka}, {Tasca},
  {Tresse}, {Vergani}, {Zucca}, {Barnes}, {Bordoloi}, {Cappi}, {Cimatti},
  {Coppa}, {Koekemoer}, {L{\'o}pez-Sanjuan}, {McCracken}, {Moresco}, {Nair},
  {Pozzetti}, \& {Welikala}}]{2013Knobel}
{Knobel}, C., {Lilly}, S.~J., {Kova{\v c}}, K., {Peng}, Y., {Bschorr}, T.~J.,
  {Carollo}, C.~M., {Contini}, T., {Kneib}, J.-P., {Le Fevre}, O., {Mainieri},
  V., {Renzini}, A., {Scodeggio}, M., {Zamorani}, G., {Bardelli}, S.,
  {Bolzonella}, M., {Bongiorno}, A., {Caputi}, K., {Cucciati}, O., {de la
  Torre}, S., {de Ravel}, L., {Franzetti}, P., {Garilli}, B., {Iovino}, A.,
  {Kampczyk}, P., {Lamareille}, F., {Le Borgne}, J.-F., {Le Brun}, V., {Maier},
  C., {Mignoli}, M., {Pello}, R., {Perez Montero}, E., {Presotto}, V.,
  {Silverman}, J., {Tanaka}, M., {Tasca}, L., {Tresse}, L., {Vergani}, D.,
  {Zucca}, E., {Barnes}, L., {Bordoloi}, R., {Cappi}, A., {Cimatti}, A.,
  {Coppa}, G., {Koekemoer}, A.~M., {L{\'o}pez-Sanjuan}, C., {McCracken}, H.~J.,
  {Moresco}, M., {Nair}, P., {Pozzetti}, L., \& {Welikala}, N. 2013, \apj, 769,
  24

\bibitem[{{Komatsu} {et~al.}(2010){Komatsu}, {Smith}, {Dunkley}, {Bennett},
  {Gold}, {Hinshaw}, {Jarosik}, {Larson}, {Nolta}, {Page}, {Spergel},
  {Halpern}, {Hill}, {Kogut}, {Limon}, {Meyer}, {Odegard}, {Tucker}, {Weiland},
  {Wollack}, \& {Wright}}]{2010Komatsu}
{Komatsu}, E., {Smith}, K.~M., {Dunkley}, J., {Bennett}, C.~L., {Gold}, B.,
  {Hinshaw}, G., {Jarosik}, N., {Larson}, D., {Nolta}, M.~R., {Page}, L.,
  {Spergel}, D.~N., {Halpern}, M., {Hill}, R.~S., {Kogut}, A., {Limon}, M.,
  {Meyer}, S.~S., {Odegard}, N., {Tucker}, G.~S., {Weiland}, J.~L., {Wollack},
  E., \& {Wright}, E.~L. 2010, ArXiv e-prints

\bibitem[{{Kormendy} {et~al.}(2009){Kormendy}, {Fisher}, {Cornell}, \&
  {Bender}}]{2009Kormendy}
{Kormendy}, J., {Fisher}, D.~B., {Cornell}, M.~E., \& {Bender}, R. 2009, \apjs,
  182, 216

\bibitem[{{Kovac} {et~al.}(2013){Kovac}, {Lilly}, {Knobel}, {Bschorr}, {Peng},
  {Carollo}, {Contini}, {Kneib}, {Le Fevre}, {Mainieri}, {Renzini},
  {Scodeggio}, {Zamorani}, {Bardelli}, {Bolzonella}, {Bongiorno}, {Caputi},
  {Cucciati}, {de la Torre}, {de Ravel}, {Franzetti}, {Garilli}, {Iovino},
  {Kampczyk}, {Lamareille}, {Le Borgne}, {Le Brun}, {Maier}, {Mignoli},
  {Oesch}, {Pello}, {Perez Montero}, {Presotto}, {Silverman}, {Tanaka},
  {Tasca}, {Tresse}, {Vergani}, {Zucca}, {Aussel}, {Koekemoer}, {Le Floch},
  {Moresco}, \& {Pozzetti}}]{2013Kovac}
{Kovac}, K., {Lilly}, S.~J., {Knobel}, C., {Bschorr}, T.~J., {Peng}, Y.,
  {Carollo}, C.~M., {Contini}, T., {Kneib}, J.-P., {Le Fevre}, O., {Mainieri},
  V., {Renzini}, A., {Scodeggio}, M., {Zamorani}, G., {Bardelli}, S.,
  {Bolzonella}, M., {Bongiorno}, A., {Caputi}, K., {Cucciati}, O., {de la
  Torre}, S., {de Ravel}, L., {Franzetti}, P., {Garilli}, B., {Iovino}, A.,
  {Kampczyk}, P., {Lamareille}, F., {Le Borgne}, J.-F., {Le Brun}, V., {Maier},
  C., {Mignoli}, M., {Oesch}, P., {Pello}, R., {Perez Montero}, E., {Presotto},
  V., {Silverman}, J., {Tanaka}, M., {Tasca}, L., {Tresse}, L., {Vergani}, D.,
  {Zucca}, E., {Aussel}, H., {Koekemoer}, A.~M., {Le Floch}, E., {Moresco}, M.,
  \& {Pozzetti}, L. 2013, ArXiv e-prints

\bibitem[{{Kronberger} {et~al.}(2008){Kronberger}, {Kapferer}, {Ferrari},
  {Unterguggenberger}, \& {Schindler}}]{2008Kronberger}
{Kronberger}, T., {Kapferer}, W., {Ferrari}, C., {Unterguggenberger}, S., \&
  {Schindler}, S. 2008, \aap, 481, 337

\bibitem[{{Larson} {et~al.}(1980){Larson}, {Tinsley}, \&
  {Caldwell}}]{1980Larson}
{Larson}, R.~B., {Tinsley}, B.~M., \& {Caldwell}, C.~N. 1980, \apj, 237, 692

\bibitem[{{Lewis} {et~al.}(2002){Lewis}, {Balogh}, {De Propris}, {Couch},
  {Bower}, {Offer}, {Bland-Hawthorn}, {Baldry}, {Baugh}, {Bridges}, {Cannon},
  {Cole}, {Colless}, {Collins}, {Cross}, {Dalton}, {Driver}, {Efstathiou},
  {Ellis}, {Frenk}, {Glazebrook}, {Hawkins}, {Jackson}, {Lahav}, {Lumsden},
  {Maddox}, {Madgwick}, {Norberg}, {Peacock}, {Percival}, {Peterson},
  {Sutherland}, \& {Taylor}}]{2002Lewis}
{Lewis}, I., {Balogh}, M., {De Propris}, R., {Couch}, W., {Bower}, R., {Offer},
  A., {Bland-Hawthorn}, J., {Baldry}, I.~K., {Baugh}, C., {Bridges}, T.,
  {Cannon}, R., {Cole}, S., {Colless}, M., {Collins}, C., {Cross}, N.,
  {Dalton}, G., {Driver}, S.~P., {Efstathiou}, G., {Ellis}, R.~S., {Frenk},
  C.~S., {Glazebrook}, K., {Hawkins}, E., {Jackson}, C., {Lahav}, O.,
  {Lumsden}, S., {Maddox}, S., {Madgwick}, D., {Norberg}, P., {Peacock}, J.~A.,
  {Percival}, W., {Peterson}, B.~A., {Sutherland}, W., \& {Taylor}, K. 2002,
  \mnras, 334, 673

\bibitem[{{McGee} {et~al.}(2011){McGee}, {Balogh}, {Wilman}, {Bower},
  {Mulchaey}, {Parker}, \& {Oemler}}]{2011McGee}
{McGee}, S.~L., {Balogh}, M.~L., {Wilman}, D.~J., {Bower}, R.~G., {Mulchaey},
  J.~S., {Parker}, L.~C., \& {Oemler}, A. 2011, \mnras, 413, 996

\bibitem[{{McNamara} \& {Nulsen}(2007)}]{2007McNamara}
{McNamara}, B.~R., \& {Nulsen}, P.~E.~J. 2007, \araa, 45, 117

\bibitem[{{Mendel} {et~al.}(2013){Mendel}, {Simard}, {Ellison}, \&
  {Patton}}]{2013Mendel}
{Mendel}, J.~T., {Simard}, L., {Ellison}, S.~L., \& {Patton}, D.~R. 2013,
  \mnras, 429, 2212

\bibitem[{{Mendez} {et~al.}(2011){Mendez}, {Coil}, {Lotz}, {Salim},
  {Moustakas}, \& {Simard}}]{2011Mendez}
{Mendez}, A.~J., {Coil}, A.~L., {Lotz}, J., {Salim}, S., {Moustakas}, J., \&
  {Simard}, L. 2011, ArXiv e-prints

\bibitem[{{Mok} {et~al.}(2013){Mok}, {Balogh}, {McGee}, {Wilman}, {Finoguenov},
  {Tanaka}, {Giodini}, {Bower}, {Connelly}, {Hou}, {Mulchaey}, \&
  {Parker}}]{2013Mok}
{Mok}, A., {Balogh}, M.~L., {McGee}, S.~L., {Wilman}, D.~J., {Finoguenov}, A.,
  {Tanaka}, M., {Giodini}, S., {Bower}, R.~G., {Connelly}, J.~L., {Hou}, A.,
  {Mulchaey}, J.~S., \& {Parker}, L.~C. 2013, ArXiv e-prints

\bibitem[{{Moore} {et~al.}(1996){Moore}, {Katz}, {Lake}, {Dressler}, \&
  {Oemler}}]{1996Moore}
{Moore}, B., {Katz}, N., {Lake}, G., {Dressler}, A., \& {Oemler}, A. 1996,
  \nat, 379, 613

\bibitem[{{Mori} \& {Burkert}(2000)}]{2000Mori}
{Mori}, M., \& {Burkert}, A. 2000, \apj, 538, 559

\bibitem[{{Murray} {et~al.}(1993){Murray}, {White}, {Blondin}, \&
  {Lin}}]{1993Murray}
{Murray}, S.~D., {White}, S.~D.~M., {Blondin}, J.~M., \& {Lin}, D.~N.~C. 1993,
  \apj, 407, 588

\bibitem[{{Muzzin} {et~al.}(2012){Muzzin}, {Wilson}, {Yee}, {Gilbank},
  {Hoekstra}, {Demarco}, {Balogh}, {van Dokkum}, {Franx}, {Ellingson}, {Hicks},
  {Nantais}, {Noble}, {Lacy}, {Lidman}, {Rettura}, {Surace}, \&
  {Webb}}]{2012Muzzin}
{Muzzin}, A., {Wilson}, G., {Yee}, H.~K.~C., {Gilbank}, D., {Hoekstra}, H.,
  {Demarco}, R., {Balogh}, M., {van Dokkum}, P., {Franx}, M., {Ellingson}, E.,
  {Hicks}, A., {Nantais}, J., {Noble}, A., {Lacy}, M., {Lidman}, C., {Rettura},
  A., {Surace}, J., \& {Webb}, T. 2012, \apj, 746, 188

\bibitem[{{Oemler}(1974)}]{1974Oemler}
{Oemler}, Jr., A. 1974, \apj, 194, 1

\bibitem[{{Omma} \& {Binney}(2004)}]{2004Omma}
{Omma}, H., \& {Binney}, J. 2004, \mnras, 350, L13

\bibitem[{{Park} {et~al.}(2007){Park}, {Choi}, {Vogeley}, {Gott}, {Blanton}, \&
  {SDSS Collaboration}}]{2007Park}
{Park}, C., {Choi}, Y.-Y., {Vogeley}, M.~S., {Gott}, III, J.~R., {Blanton},
  M.~R., \& {SDSS Collaboration}. 2007, \apj, 658, 898

\bibitem[{{Patel} {et~al.}(2009){Patel}, {Kelson}, {Holden}, {Illingworth},
  {Franx}, {van der Wel}, \& {Ford}}]{2009Patel}
{Patel}, S.~G., {Kelson}, D.~D., {Holden}, B.~P., {Illingworth}, G.~D.,
  {Franx}, M., {van der Wel}, A., \& {Ford}, H. 2009, \apj, 694, 1349

\bibitem[{{Peng} {et~al.}(2010){Peng}, {Lilly}, {Kova{\v c}}, {Bolzonella},
  {Pozzetti}, {Renzini}, {Zamorani}, {Ilbert}, {Knobel}, {Iovino}, {Maier},
  {Cucciati}, {Tasca}, {Carollo}, {Silverman}, {Kampczyk}, {de Ravel},
  {Sanders}, {Scoville}, {Contini}, {Mainieri}, {Scodeggio}, {Kneib}, {Le
  F{\`e}vre}, {Bardelli}, {Bongiorno}, {Caputi}, {Coppa}, {de la Torre},
  {Franzetti}, {Garilli}, {Lamareille}, {Le Borgne}, {Le Brun}, {Mignoli},
  {Perez Montero}, {Pello}, {Ricciardelli}, {Tanaka}, {Tresse}, {Vergani},
  {Welikala}, {Zucca}, {Oesch}, {Abbas}, {Barnes}, {Bordoloi}, {Bottini},
  {Cappi}, {Cassata}, {Cimatti}, {Fumana}, {Hasinger}, {Koekemoer},
  {Leauthaud}, {Maccagni}, {Marinoni}, {McCracken}, {Memeo}, {Meneux}, {Nair},
  {Porciani}, {Presotto}, \& {Scaramella}}]{2010Peng}
{Peng}, Y.-j., {Lilly}, S.~J., {Kova{\v c}}, K., {Bolzonella}, M., {Pozzetti},
  L., {Renzini}, A., {Zamorani}, G., {Ilbert}, O., {Knobel}, C., {Iovino}, A.,
  {Maier}, C., {Cucciati}, O., {Tasca}, L., {Carollo}, C.~M., {Silverman}, J.,
  {Kampczyk}, P., {de Ravel}, L., {Sanders}, D., {Scoville}, N., {Contini}, T.,
  {Mainieri}, V., {Scodeggio}, M., {Kneib}, J.-P., {Le F{\`e}vre}, O.,
  {Bardelli}, S., {Bongiorno}, A., {Caputi}, K., {Coppa}, G., {de la Torre},
  S., {Franzetti}, P., {Garilli}, B., {Lamareille}, F., {Le Borgne}, J.-F., {Le
  Brun}, V., {Mignoli}, M., {Perez Montero}, E., {Pello}, R., {Ricciardelli},
  E., {Tanaka}, M., {Tresse}, L., {Vergani}, D., {Welikala}, N., {Zucca}, E.,
  {Oesch}, P., {Abbas}, U., {Barnes}, L., {Bordoloi}, R., {Bottini}, D.,
  {Cappi}, A., {Cassata}, P., {Cimatti}, A., {Fumana}, M., {Hasinger}, G.,
  {Koekemoer}, A., {Leauthaud}, A., {Maccagni}, D., {Marinoni}, C.,
  {McCracken}, H., {Memeo}, P., {Meneux}, B., {Nair}, P., {Porciani}, C.,
  {Presotto}, V., \& {Scaramella}, R. 2010, \apj, 721, 193

\bibitem[{{Peng} {et~al.}(2012){Peng}, {Lilly}, {Renzini}, \&
  {Carollo}}]{2012Peng}
{Peng}, Y.-j., {Lilly}, S.~J., {Renzini}, A., \& {Carollo}, M. 2012, \apj, 757,
  4

\bibitem[{{Planck Collaboration} {et~al.}(2013){Planck Collaboration}, {Ade},
  {Aghanim}, {Armitage-Caplan}, {Arnaud}, {Ashdown}, {Atrio-Barandela},
  {Aumont}, {Baccigalupi}, {Banday}, \& et~al.}]{2013Planck}
{Planck Collaboration}, {Ade}, P.~A.~R., {Aghanim}, N., {Armitage-Caplan}, C.,
  {Arnaud}, M., {Ashdown}, M., {Atrio-Barandela}, F., {Aumont}, J.,
  {Baccigalupi}, C., {Banday}, A.~J., \& et~al. 2013, ArXiv e-prints

\bibitem[{{Poggianti} {et~al.}(2009){Poggianti}, {Arag{\'o}n-Salamanca},
  {Zaritsky}, {De Lucia}, {Milvang-Jensen}, {Desai}, {Jablonka}, {Halliday},
  {Rudnick}, {Varela}, {Bamford}, {Best}, {Clowe}, {Noll}, {Saglia},
  {Pell{\'o}}, {Simard}, {von der Linden}, \& {White}}]{2009Poggianti}
{Poggianti}, B.~M., {Arag{\'o}n-Salamanca}, A., {Zaritsky}, D., {De Lucia}, G.,
  {Milvang-Jensen}, B., {Desai}, V., {Jablonka}, P., {Halliday}, C., {Rudnick},
  G., {Varela}, J., {Bamford}, S., {Best}, P., {Clowe}, D., {Noll}, S.,
  {Saglia}, R., {Pell{\'o}}, R., {Simard}, L., {von der Linden}, A., \&
  {White}, S. 2009, \apj, 693, 112

\bibitem[{{Postman} \& {Geller}(1984)}]{1984Postman}
{Postman}, M., \& {Geller}, M.~J. 1984, \apj, 281, 95

\bibitem[{{Pozzetti} {et~al.}(2007){Pozzetti}, {Bolzonella}, {Lamareille},
  {Zamorani}, {Franzetti}, {Le F{\`e}vre}, {Iovino}, {Temporin}, {Ilbert},
  {Arnouts}, {Charlot}, {Brinchmann}, {Zucca}, {Tresse}, {Scodeggio}, {Guzzo},
  {Bottini}, {Garilli}, {Le Brun}, {Maccagni}, {Picat}, {Scaramella},
  {Vettolani}, {Zanichelli}, {Adami}, {Bardelli}, {Cappi}, {Ciliegi},
  {Contini}, {Foucaud}, {Gavignaud}, {McCracken}, {Marano}, {Marinoni},
  {Mazure}, {Meneux}, {Merighi}, {Paltani}, {Pell{\`o}}, {Pollo}, {Radovich},
  {Bondi}, {Bongiorno}, {Cucciati}, {de la Torre}, {Gregorini}, {Mellier},
  {Merluzzi}, {Vergani}, \& {Walcher}}]{2007Pozzetti}
{Pozzetti}, L., {Bolzonella}, M., {Lamareille}, F., {Zamorani}, G.,
  {Franzetti}, P., {Le F{\`e}vre}, O., {Iovino}, A., {Temporin}, S., {Ilbert},
  O., {Arnouts}, S., {Charlot}, S., {Brinchmann}, J., {Zucca}, E., {Tresse},
  L., {Scodeggio}, M., {Guzzo}, L., {Bottini}, D., {Garilli}, B., {Le Brun},
  V., {Maccagni}, D., {Picat}, J.~P., {Scaramella}, R., {Vettolani}, G.,
  {Zanichelli}, A., {Adami}, C., {Bardelli}, S., {Cappi}, A., {Ciliegi}, P.,
  {Contini}, T., {Foucaud}, S., {Gavignaud}, I., {McCracken}, H.~J., {Marano},
  B., {Marinoni}, C., {Mazure}, A., {Meneux}, B., {Merighi}, R., {Paltani}, S.,
  {Pell{\`o}}, R., {Pollo}, A., {Radovich}, M., {Bondi}, M., {Bongiorno}, A.,
  {Cucciati}, O., {de la Torre}, S., {Gregorini}, L., {Mellier}, Y.,
  {Merluzzi}, P., {Vergani}, D., \& {Walcher}, C.~J. 2007, \aap, 474, 443

\bibitem[{{Presotto} {et~al.}(2012){Presotto}, {Iovino}, {Scodeggio},
  {Cucciati}, {Knobel}, {Bolzonella}, {Oesch}, {Finoguenov}, {Tanaka}, {Kova{\v
  c}}, {Peng}, {Zamorani}, {Bardelli}, {Pozzetti}, {Kampczyk},
  {L{\'o}pez-Sanjuan}, {Vergani}, {Zucca}, {Tasca}, {Carollo}, {Contini},
  {Kneib}, {Le F{\`e}vre}, {Lilly}, {Mainieri}, {Renzini}, {Bongiorno},
  {Caputi}, {de la Torre}, {de Ravel}, {Franzetti}, {Garilli}, {Lamareille},
  {Le Borgne}, {Le Brun}, {Maier}, {Mignoli}, {Pell{\`o}}, {Perez-Montero},
  {Ricciardelli}, {Silverman}, {Tresse}, {Barnes}, {Bordoloi}, {Cappi},
  {Cimatti}, {Coppa}, {Koekemoer}, {McCracken}, {Moresco}, {Nair}, \&
  {Welikala}}]{2012Presotto}
{Presotto}, V., {Iovino}, A., {Scodeggio}, M., {Cucciati}, O., {Knobel}, C.,
  {Bolzonella}, M., {Oesch}, P., {Finoguenov}, A., {Tanaka}, M., {Kova{\v c}},
  K., {Peng}, Y., {Zamorani}, G., {Bardelli}, S., {Pozzetti}, L., {Kampczyk},
  P., {L{\'o}pez-Sanjuan}, C., {Vergani}, D., {Zucca}, E., {Tasca}, L.~A.~M.,
  {Carollo}, C.~M., {Contini}, T., {Kneib}, J.-P., {Le F{\`e}vre}, O., {Lilly},
  S., {Mainieri}, V., {Renzini}, A., {Bongiorno}, A., {Caputi}, K., {de la
  Torre}, S., {de Ravel}, L., {Franzetti}, P., {Garilli}, B., {Lamareille}, F.,
  {Le Borgne}, J.-F., {Le Brun}, V., {Maier}, C., {Mignoli}, M., {Pell{\`o}},
  R., {Perez-Montero}, E., {Ricciardelli}, E., {Silverman}, J.~D., {Tresse},
  L., {Barnes}, L., {Bordoloi}, R., {Cappi}, A., {Cimatti}, A., {Coppa}, G.,
  {Koekemoer}, A.~M., {McCracken}, H.~J., {Moresco}, M., {Nair}, P., \&
  {Welikala}, N. 2012, \aap, 539, A55

\bibitem[{{Quadri} {et~al.}(2012){Quadri}, {Williams}, {Franx}, \&
  {Hildebrandt}}]{2012Quadri}
{Quadri}, R.~F., {Williams}, R.~J., {Franx}, M., \& {Hildebrandt}, H. 2012,
  \apj, 744, 88

\bibitem[{{Quilis} {et~al.}(2000){Quilis}, {Moore}, \& {Bower}}]{2000Quilis}
{Quilis}, V., {Moore}, B., \& {Bower}, R. 2000, Science, 288, 1617

\bibitem[{{Rafelski} {et~al.}(2012){Rafelski}, {Wolfe}, {Prochaska},
  {Neeleman}, \& {Mendez}}]{2012Rafelski}
{Rafelski}, M., {Wolfe}, A.~M., {Prochaska}, J.~X., {Neeleman}, M., \&
  {Mendez}, A.~J. 2012, ArXiv e-prints

\bibitem[{{Ramos} {et~al.}(2011){Ramos}, {Pellegrini}, {Benoist}, {da Costa},
  {Maia}, {Makler}, {Ogando}, {de Simoni}, \& {Mesquita}}]{2011Ramos}
{Ramos}, B.~H.~F., {Pellegrini}, P.~S., {Benoist}, C., {da Costa}, L.~N.,
  {Maia}, M.~A.~G., {Makler}, M., {Ogando}, R.~L.~C., {de Simoni}, F., \&
  {Mesquita}, A.~A. 2011, \aj, 142, 41

\bibitem[{{Rasmussen} {et~al.}(2012){Rasmussen}, {Mulchaey}, {Bai}, {Ponman},
  {Raychaudhury}, \& {Dariush}}]{2012Rasmussen}
{Rasmussen}, J., {Mulchaey}, J.~S., {Bai}, L., {Ponman}, T.~J., {Raychaudhury},
  S., \& {Dariush}, A. 2012, \apj, 757, 122

\bibitem[{{Rees} \& {Ostriker}(1977)}]{1977Rees}
{Rees}, M.~J., \& {Ostriker}, J.~P. 1977, \mnras, 179, 541

\bibitem[{{Rojas} {et~al.}(2004){Rojas}, {Vogeley}, {Hoyle}, \&
  {Brinkmann}}]{2004Rojas}
{Rojas}, R.~R., {Vogeley}, M.~S., {Hoyle}, F., \& {Brinkmann}, J. 2004, \apj,
  617, 50

\bibitem[{{Rosario} {et~al.}(2013){Rosario}, {Mozena}, {Wuyts}, {Nandra},
  {Koekemoer}, {McGrath}, {Hathi}, {Dekel}, {Donley}, {Dunlop}, {Faber},
  {Ferguson}, {Giavalisco}, {Grogin}, {Guo}, {Kocevski}, {Koo}, {Laird},
  {Newman}, {Rangel}, \& {Somerville}}]{2013Rosario}
{Rosario}, D.~J., {Mozena}, M., {Wuyts}, S., {Nandra}, K., {Koekemoer}, A.,
  {McGrath}, E., {Hathi}, N.~P., {Dekel}, A., {Donley}, J., {Dunlop}, J.~S.,
  {Faber}, S.~M., {Ferguson}, H., {Giavalisco}, M., {Grogin}, N., {Guo}, Y.,
  {Kocevski}, D.~D., {Koo}, D.~C., {Laird}, E., {Newman}, J., {Rangel}, C., \&
  {Somerville}, R. 2013, \apj, 763, 59

\bibitem[{{Salim} {et~al.}(2007){Salim}, {Rich}, {Charlot}, {Brinchmann},
  {Johnson}, {Schiminovich}, {Seibert}, {Mallery}, {Heckman}, {Forster},
  {Friedman}, {Martin}, {Morrissey}, {Neff}, {Small}, {Wyder}, {Bianchi},
  {Donas}, {Lee}, {Madore}, {Milliard}, {Szalay}, {Welsh}, \& {Yi}}]{2007Salim}
{Salim}, S., {Rich}, R.~M., {Charlot}, S., {Brinchmann}, J., {Johnson}, B.~D.,
  {Schiminovich}, D., {Seibert}, M., {Mallery}, R., {Heckman}, T.~M.,
  {Forster}, K., {Friedman}, P.~G., {Martin}, D.~C., {Morrissey}, P., {Neff},
  S.~G., {Small}, T., {Wyder}, T.~K., {Bianchi}, L., {Donas}, J., {Lee}, Y.,
  {Madore}, B.~F., {Milliard}, B., {Szalay}, A.~S., {Welsh}, B.~Y., \& {Yi},
  S.~K. 2007, \apjs, 173, 267

\bibitem[{{Santini} {et~al.}(2012){Santini}, {Rosario}, {Shao}, {Lutz},
  {Maiolino}, {Alexander}, {Altieri}, {Andreani}, {Aussel}, {Bauer}, {Berta},
  {Bongiovanni}, {Brandt}, {Brusa}, {Cepa}, {Cimatti}, {Daddi}, {Elbaz},
  {Fontana}, {F{\"o}rster Schreiber}, {Genzel}, {Grazian}, {Le Floc'h},
  {Magnelli}, {Mainieri}, {Nordon}, {P{\'e}rez Garcia}, {Poglitsch}, {Popesso},
  {Pozzi}, {Riguccini}, {Rodighiero}, {Salvato}, {Sanchez-Portal}, {Sturm},
  {Tacconi}, {Valtchanov}, \& {Wuyts}}]{2012Santini}
{Santini}, P., {Rosario}, D.~J., {Shao}, L., {Lutz}, D., {Maiolino}, R.,
  {Alexander}, D.~M., {Altieri}, B., {Andreani}, P., {Aussel}, H., {Bauer},
  F.~E., {Berta}, S., {Bongiovanni}, A., {Brandt}, W.~N., {Brusa}, M., {Cepa},
  J., {Cimatti}, A., {Daddi}, E., {Elbaz}, D., {Fontana}, A., {F{\"o}rster
  Schreiber}, N.~M., {Genzel}, R., {Grazian}, A., {Le Floc'h}, E., {Magnelli},
  B., {Mainieri}, V., {Nordon}, R., {P{\'e}rez Garcia}, A.~M., {Poglitsch}, A.,
  {Popesso}, P., {Pozzi}, F., {Riguccini}, L., {Rodighiero}, G., {Salvato}, M.,
  {Sanchez-Portal}, M., {Sturm}, E., {Tacconi}, L.~J., {Valtchanov}, I., \&
  {Wuyts}, S. 2012, \aap, 540, A109

\bibitem[{{Scannapieco} \& {Oh}(2004)}]{2004Scannapieco}
{Scannapieco}, E., \& {Oh}, S.~P. 2004, \apj, 608, 62

\bibitem[{{Schmidt}(1959)}]{1959Schmidt}
{Schmidt}, M. 1959, \apj, 129, 243

\bibitem[{{Silk}(1977)}]{1977Silk}
{Silk}, J. 1977, \apj, 211, 638

\bibitem[{{Strateva} {et~al.}(2001){Strateva}, {Ivezi{\'c}}, {Knapp},
  {Narayanan}, {Strauss}, {Gunn}, {Lupton}, {Schlegel}, {Bahcall}, {Brinkmann},
  {Brunner}, {Budav{\'a}ri}, {Csabai}, {Castander}, {Doi}, {Fukugita}, {Gy{\H
  o}ry}, {Hamabe}, {Hennessy}, {Ichikawa}, {Kunszt}, {Lamb}, {McKay},
  {Okamura}, {Racusin}, {Sekiguchi}, {Schneider}, {Shimasaku}, \&
  {York}}]{2001Strateva}
{Strateva}, I., {Ivezi{\'c}}, {\v Z}., {Knapp}, G.~R., {Narayanan}, V.~K.,
  {Strauss}, M.~A., {Gunn}, J.~E., {Lupton}, R.~H., {Schlegel}, D., {Bahcall},
  N.~A., {Brinkmann}, J., {Brunner}, R.~J., {Budav{\'a}ri}, T., {Csabai}, I.,
  {Castander}, F.~J., {Doi}, M., {Fukugita}, M., {Gy{\H o}ry}, Z., {Hamabe},
  M., {Hennessy}, G., {Ichikawa}, T., {Kunszt}, P.~Z., {Lamb}, D.~Q., {McKay},
  T.~A., {Okamura}, S., {Racusin}, J., {Sekiguchi}, M., {Schneider}, D.~P.,
  {Shimasaku}, K., \& {York}, D. 2001, \aj, 122, 1861

\bibitem[{{Swinbank} {et~al.}(2012){Swinbank}, {Balogh}, {Bower}, {Zabludoff},
  {Lucey}, {McGee}, {Miller}, \& {Nichol}}]{2012Swinbank}
{Swinbank}, A.~M., {Balogh}, M.~L., {Bower}, R.~G., {Zabludoff}, A.~I.,
  {Lucey}, J.~R., {McGee}, S.~L., {Miller}, C.~J., \& {Nichol}, R.~C. 2012,
  \mnras, 420, 672

\bibitem[{{Tanaka} {et~al.}(2012){Tanaka}, {Finoguenov}, {Lilly}, {Bolzonella},
  {Carollo}, {Contini}, {Iovino}, {Kneib}, {Lamareille}, {Le Fevre},
  {Mainieri}, {Presotto}, {Renzini}, {Scodeggio}, {Silverman}, {Zamorani},
  {Bardelli}, {Bongiorno}, {Caputi}, {Cucciati}, {de la Torre}, {de Ravel},
  {Franzetti}, {Garilli}, {Kampczyk}, {Knobel}, {Kova{\v c}}, {Le Borgne}, {Le
  Brun}, {L{\'o}pez-Sanjuan}, {Maier}, {Mignoli}, {Pello}, {Peng},
  {Perez-Montero}, {Tasca}, {Tresse}, {Vergani}, {Zucca}, {Barnes}, {Bordoloi},
  {Cappi}, {Cimatti}, {Coppa}, {Koekemoer}, {McCracken}, {Moresco}, {Nair},
  {Oesch}, {Pozzetti}, \& {Welikala}}]{2012Tanaka}
{Tanaka}, M., {Finoguenov}, A., {Lilly}, S.~J., {Bolzonella}, M., {Carollo},
  C.~M., {Contini}, T., {Iovino}, A., {Kneib}, J.-P., {Lamareille}, F., {Le
  Fevre}, O., {Mainieri}, V., {Presotto}, V., {Renzini}, A., {Scodeggio}, M.,
  {Silverman}, J.~D., {Zamorani}, G., {Bardelli}, S., {Bongiorno}, A.,
  {Caputi}, K., {Cucciati}, O., {de la Torre}, S., {de Ravel}, L., {Franzetti},
  P., {Garilli}, B., {Kampczyk}, P., {Knobel}, C., {Kova{\v c}}, K., {Le
  Borgne}, J.-F., {Le Brun}, V., {L{\'o}pez-Sanjuan}, C., {Maier}, C.,
  {Mignoli}, M., {Pello}, R., {Peng}, Y., {Perez-Montero}, E., {Tasca}, L.,
  {Tresse}, L., {Vergani}, D., {Zucca}, E., {Barnes}, L., {Bordoloi}, R.,
  {Cappi}, A., {Cimatti}, A., {Coppa}, G., {Koekemoer}, A.~M., {McCracken},
  H.~J., {Moresco}, M., {Nair}, P., {Oesch}, P., {Pozzetti}, L., \& {Welikala},
  N. 2012, \pasj, 64, 22

\bibitem[{{Tanaka} {et~al.}(2004){Tanaka}, {Goto}, {Okamura}, {Shimasaku}, \&
  {Brinkmann}}]{2004Tanaka}
{Tanaka}, M., {Goto}, T., {Okamura}, S., {Shimasaku}, K., \& {Brinkmann}, J.
  2004, \aj, 128, 2677

\bibitem[{{Tanaka} {et~al.}(2007){Tanaka}, {Kodama}, {Kajisawa}, {Bower},
  {Demarco}, {Finoguenov}, {Lidman}, \& {Rosati}}]{2007Tanaka}
{Tanaka}, M., {Kodama}, T., {Kajisawa}, M., {Bower}, R., {Demarco}, R.,
  {Finoguenov}, A., {Lidman}, C., \& {Rosati}, P. 2007, \mnras, 377, 1206

\bibitem[{{Taranu} {et~al.}(2012){Taranu}, {Hudson}, {Balogh}, {Smith},
  {Power}, \& {Krane}}]{2012Taranu}
{Taranu}, D.~S., {Hudson}, M.~J., {Balogh}, M.~L., {Smith}, R.~J., {Power}, C.,
  \& {Krane}, B. 2012, ArXiv e-prints

\bibitem[{{Thomas} {et~al.}(2010){Thomas}, {Maraston}, {Schawinski}, {Sarzi},
  \& {Silk}}]{2010Thomas}
{Thomas}, D., {Maraston}, C., {Schawinski}, K., {Sarzi}, M., \& {Silk}, J.
  2010, \mnras, 404, 1775

\bibitem[{{Tonnesen} \& {Bryan}(2009)}]{2009Tonnesen}
{Tonnesen}, S., \& {Bryan}, G.~L. 2009, \apj, 694, 789

\bibitem[{{Tonnesen} \& {Cen}(2012)}]{2012Tonnesen}
{Tonnesen}, S., \& {Cen}, R. 2012, \mnras, 425, 2313

\bibitem[{{Tripp} {et~al.}(2008){Tripp}, {Sembach}, {Bowen}, {Savage},
  {Jenkins}, {Lehner}, \& {Richter}}]{2008Tripp}
{Tripp}, T.~M., {Sembach}, K.~R., {Bowen}, D.~V., {Savage}, B.~D., {Jenkins},
  E.~B., {Lehner}, N., \& {Richter}, P. 2008, \apjs, 177, 39

\bibitem[{{Turk} {et~al.}(2011){Turk}, {Smith}, {Oishi}, {Skory}, {Skillman},
  {Abel}, \& {Norman}}]{2011Turk}
{Turk}, M.~J., {Smith}, B.~D., {Oishi}, J.~S., {Skory}, S., {Skillman}, S.~W.,
  {Abel}, T., \& {Norman}, M.~L. 2011, \apjs, 192, 9

\bibitem[{{van den Bosch} {et~al.}(2008){van den Bosch}, {Aquino}, {Yang},
  {Mo}, {Pasquali}, {McIntosh}, {Weinmann}, \& {Kang}}]{2008vandenBosch}
{van den Bosch}, F.~C., {Aquino}, D., {Yang}, X., {Mo}, H.~J., {Pasquali}, A.,
  {McIntosh}, D.~H., {Weinmann}, S.~M., \& {Kang}, X. 2008, \mnras, 387, 79

\bibitem[{{van der Burg} {et~al.}(2013){van der Burg}, {Muzzin}, {Hoekstra},
  {Lidman}, {Rettura}, {Wilson}, {Yee}, {Hildebrandt}, {Marchesini},
  {Stefanon}, \& {Kuijken}}]{2013vanderBurg}
{van der Burg}, R.~F.~J., {Muzzin}, A., {Hoekstra}, H., {Lidman}, C.,
  {Rettura}, A., {Wilson}, G., {Yee}, H.~K.~C., {Hildebrandt}, H.,
  {Marchesini}, D., {Stefanon}, M., \& {Kuijken}, K. 2013, ArXiv e-prints

\bibitem[{{Weinmann} {et~al.}(2012){Weinmann}, {Pasquali}, {Oppenheimer},
  {Finlator}, {Mendel}, {Crain}, \& {Macci{\`o}}}]{2012Weinmann}
{Weinmann}, S.~M., {Pasquali}, A., {Oppenheimer}, B.~D., {Finlator}, K.,
  {Mendel}, J.~T., {Crain}, R.~A., \& {Macci{\`o}}, A.~V. 2012, \mnras, 426,
  2797

\bibitem[{{Weinmann} {et~al.}(2006){Weinmann}, {van den Bosch}, {Yang}, \&
  {Mo}}]{2006Weinmann}
{Weinmann}, S.~M., {van den Bosch}, F.~C., {Yang}, X., \& {Mo}, H.~J. 2006,
  \mnras, 366, 2

\bibitem[{{Wetzel} {et~al.}(2012){Wetzel}, {Tinker}, {Conroy}, \& {van den
  Bosch}}]{2012Wetzel}
{Wetzel}, A.~R., {Tinker}, J.~L., {Conroy}, C., \& {van den Bosch}, F.~C. 2012,
  ArXiv e-prints

\bibitem[{{Wetzel} {et~al.}(2013{\natexlab{a}}){Wetzel}, {Tinker}, {Conroy}, \&
  {van den Bosch}}]{2013aWetzel}
---. 2013{\natexlab{a}}, \mnras, 432, 336

\bibitem[{{Wetzel} {et~al.}(2013{\natexlab{b}}){Wetzel}, {Tinker}, {Conroy}, \&
  {van den Bosch}}]{2013bWetzel}
---. 2013{\natexlab{b}}, ArXiv e-prints

\bibitem[{{Whitaker} {et~al.}(2011){Whitaker}, {Labb{\'e}}, {van Dokkum},
  {Brammer}, {Kriek}, {Marchesini}, {Quadri}, {Franx}, {Muzzin}, {Williams},
  {Bezanson}, {Illingworth}, {Lee}, {Lundgren}, {Nelson}, {Rudnick}, {Tal}, \&
  {Wake}}]{2011Whitaker}
{Whitaker}, K.~E., {Labb{\'e}}, I., {van Dokkum}, P.~G., {Brammer}, G.,
  {Kriek}, M., {Marchesini}, D., {Quadri}, R.~F., {Franx}, M., {Muzzin}, A.,
  {Williams}, R.~J., {Bezanson}, R., {Illingworth}, G.~D., {Lee}, K.-S.,
  {Lundgren}, B., {Nelson}, E.~J., {Rudnick}, G., {Tal}, T., \& {Wake}, D.~A.
  2011, \apj, 735, 86

\bibitem[{{Wijesinghe} {et~al.}(2012){Wijesinghe}, {Hopkins}, {Brough},
  {Taylor}, {Norberg}, {Bauer}, {Brown}, {Cameron}, {Conselice}, {Croom},
  {Driver}, {Grootes}, {Jones}, {Kelvin}, {Loveday}, {Pimbblet}, {Popescu},
  {Prescott}, {Sharp}, {Baldry}, {Sadler}, {Liske}, {Robotham}, {Bamford},
  {Bland-Hawthorn}, {Gunawardhana}, {Meyer}, {Parkinson}, {Drinkwater},
  {Peacock}, \& {Tuffs}}]{2012Wijesinghe}
{Wijesinghe}, D.~B., {Hopkins}, A.~M., {Brough}, S., {Taylor}, E.~N.,
  {Norberg}, P., {Bauer}, A., {Brown}, M.~J.~I., {Cameron}, E., {Conselice},
  C.~J., {Croom}, S., {Driver}, S., {Grootes}, M.~W., {Jones}, D.~H., {Kelvin},
  L., {Loveday}, J., {Pimbblet}, K.~A., {Popescu}, C.~C., {Prescott}, M.,
  {Sharp}, R., {Baldry}, I., {Sadler}, E.~M., {Liske}, J., {Robotham},
  A.~S.~G., {Bamford}, S., {Bland-Hawthorn}, J., {Gunawardhana}, M., {Meyer},
  M., {Parkinson}, H., {Drinkwater}, M.~J., {Peacock}, J., \& {Tuffs}, R. 2012,
  \mnras, 423, 3679

\bibitem[{{Willmer} {et~al.}(2006){Willmer}, {Faber}, {Koo}, {Weiner},
  {Newman}, {Coil}, {Connolly}, {Conroy}, {Cooper}, {Davis}, {Finkbeiner},
  {Gerke}, {Guhathakurta}, {Harker}, {Kaiser}, {Kassin}, {Konidaris}, {Lin},
  {Luppino}, {Madgwick}, {Noeske}, {Phillips}, \& {Yan}}]{2006Willmer}
{Willmer}, C.~N.~A., {Faber}, S.~M., {Koo}, D.~C., {Weiner}, B.~J., {Newman},
  J.~A., {Coil}, A.~L., {Connolly}, A.~J., {Conroy}, C., {Cooper}, M.~C.,
  {Davis}, M., {Finkbeiner}, D.~P., {Gerke}, B.~F., {Guhathakurta}, P.,
  {Harker}, J., {Kaiser}, N., {Kassin}, S., {Konidaris}, N.~P., {Lin}, L.,
  {Luppino}, G., {Madgwick}, D.~S., {Noeske}, K.~G., {Phillips}, A.~C., \&
  {Yan}, R. 2006, \apj, 647, 853

\bibitem[{{Woo} {et~al.}(2013){Woo}, {Dekel}, {Faber}, {Noeske}, {Koo},
  {Gerke}, {Cooper}, {Salim}, {Dutton}, {Newman}, {Weiner}, {Bundy}, {Willmer},
  {Davis}, \& {Yan}}]{2013Woo}
{Woo}, J., {Dekel}, A., {Faber}, S.~M., {Noeske}, K., {Koo}, D.~C., {Gerke},
  B.~F., {Cooper}, M.~C., {Salim}, S., {Dutton}, A.~A., {Newman}, J., {Weiner},
  B.~J., {Bundy}, K., {Willmer}, C.~N.~A., {Davis}, M., \& {Yan}, R. 2013,
  \mnras, 428, 3306

\bibitem[{{Xue} {et~al.}(2010){Xue}, {Brandt}, {Luo}, {Rafferty}, {Alexander},
  {Bauer}, {Lehmer}, {Schneider}, \& {Silverman}}]{2010Xue}
{Xue}, Y.~Q., {Brandt}, W.~N., {Luo}, B., {Rafferty}, D.~A., {Alexander},
  D.~M., {Bauer}, F.~E., {Lehmer}, B.~D., {Schneider}, D.~P., \& {Silverman},
  J.~D. 2010, \apj, 720, 368

\bibitem[{{Yao} {et~al.}(2009){Yao}, {Tripp}, {Wang}, {Danforth}, {Canizares},
  {Shull}, {Marshall}, \& {Song}}]{2009Yao}
{Yao}, Y., {Tripp}, T.~M., {Wang}, Q.~D., {Danforth}, C.~W., {Canizares},
  C.~R., {Shull}, J.~M., {Marshall}, H.~L., \& {Song}, L. 2009, \apj, 697, 1784

\bibitem[{{Zehavi} {et~al.}(2011){Zehavi}, {Zheng}, {Weinberg}, {Blanton},
  {Bahcall}, {Berlind}, {Brinkmann}, {Frieman}, {Gunn}, {Lupton}, {Nichol},
  {Percival}, {Schneider}, {Skibba}, {Strauss}, {Tegmark}, \&
  {York}}]{2011Zehavi}
{Zehavi}, I., {Zheng}, Z., {Weinberg}, D.~H., {Blanton}, M.~R., {Bahcall},
  N.~A., {Berlind}, A.~A., {Brinkmann}, J., {Frieman}, J.~A., {Gunn}, J.~E.,
  {Lupton}, R.~H., {Nichol}, R.~C., {Percival}, W.~J., {Schneider}, D.~P.,
  {Skibba}, R.~A., {Strauss}, M.~A., {Tegmark}, M., \& {York}, D.~G. 2011,
  \apj, 736, 59

\bibitem[{{Zheng} {et~al.}(2007){Zheng}, {Bell}, {Papovich}, {Wolf},
  {Meisenheimer}, {Rix}, {Rieke}, \& {Somerville}}]{2007Zheng}
{Zheng}, X.~Z., {Bell}, E.~F., {Papovich}, C., {Wolf}, C., {Meisenheimer}, K.,
  {Rix}, H., {Rieke}, G.~H., \& {Somerville}, R. 2007, \apjl, 661, L41

\end{thebibliography}

\end{document}